\documentclass[10pt,journal,compsoc]{IEEEtran}
\ifCLASSOPTIONcompsoc
  \usepackage[nocompress]{cite}
\else
  \usepackage{cite}
\fi

\ifCLASSINFOpdf
  \usepackage{graphicx}
\else
\fi

\usepackage[cmex10]{amsmath}
\usepackage[linesnumbered]{algorithm2e}
\usepackage{color}
\usepackage{url}
\ifCLASSOPTIONcompsoc
\usepackage[tight,normalsize,sf,SF]{subfigure}
\else
\usepackage[tight,footnotesize]{subfigure}
\fi

\usepackage{caption}
\usepackage{color}
\usepackage{ragged2e}
\usepackage{float}
\usepackage{array}
\usepackage{makecell}
\usepackage{multirow}
\ifCLASSINFOpdf
\else
\fi
\begin{document}
%
\title{
Emotion-Based Crowd Simulation Model Based on Physical Strength Consumption for Emergency Scenarios
}
%
%
%
%


\author{Mingliang~Xu,
        Chaochao~Li,
        Pei~Lv,
        Wei~Chen,
        Zhigang~Deng,
        Bing~Zhou
        and Dinesh Manocha

{
 \IEEEcompsocitemizethanks{
 \IEEEcompsocthanksitem \vfill{Mingliang Xu, Chaochao Li, Pei Lv,  and Bing Zhou are with Center for Interdisciplinary Information Science Research, Zhengzhou University, 450000.}
 \IEEEcompsocthanksitem \vfill{Wei Chen is with State Key Lab of CAD$\&$CG, Zhejiang University, 310058.}
 \IEEEcompsocthanksitem \vfill{Zhigang Deng is with Department of Computer Science, University of Houston, Houston, TX, 77204-3010.}
 \IEEEcompsocthanksitem \vfill{Dinesh Manocha is with Department of Computer Science and Electrical $\&$ Computer Engineering, University of Maryland, College Park, MD, USA.}\protect\\
 \vfill{E-mail: \{iexumingliang, ielvpei, iebzhou\} @zzu.edu.cn;  \hfil\break zzulcc@gs.zzu.edu.cn; chenwei@cad.zju.edu.cn; zdeng4@uh.edu; \hfil\break
dm@cs.umd.edu}}
\thanks{}
}

}

%
%

\markboth{Journal of \LaTeX\ Class Files,~Vol.~X, No.~X, October~2018}%
{Shell \MakeLowercase{\textit{et al.}}: Bare Demo of IEEEtran.cls for Computer Society Journals}
%



\IEEEtitleabstractindextext{%
\begin{abstract}
\justifying
Increasing attention is being given to the modeling and simulation of traffic flow and crowd movement, two phenomena that both deal with interactions between pedestrians and cars in many situations. In particular, crowd simulation is important for understanding mobility and transportation patterns.
In this paper, we propose an emotion-based crowd simulation model integrating physical strength consumption.
Inspired by the theory of ``the devoted actor," the movements of each individual in our model are determined by modeling the influence of physical strength consumption and the emotion of panic. In particular, human physical strength consumption is computed using a physics-based numerical method. Inspired by the James-Lange theory, panic levels are estimated by means of an enhanced emotional contagion model that leverages the inherent relationship between physical strength consumption and panic.
To the best of our knowledge, our model is the first method integrating physical strength consumption into an emotion-based crowd simulation model by exploiting the relationship between physical strength consumption and emotion.
We highlight the performance on different scenarios and compare the resulting behaviors with real-world video sequences. Our approach can reliably predict  changes in physical strength consumption and panic levels of individuals in an emergency situation.

\end{abstract}

}

\maketitle

\IEEEdisplaynontitleabstractindextext

%
\IEEEpeerreviewmaketitle

\IEEEraisesectionheading{\section{Introduction}\label{introduction}}

\IEEEPARstart{E}{fficient} and accurate crowd simulation is useful for intelligent transportation systems 
since it can help improve emergency planning and prevent congestion in transit hubs such as train stations and airports \cite{102}.
One can also analyze human mobility through the trajectories obtained by crowd simulation models to get more knowledge 
of pedestrian mobility behaviors in both qualitative and quantitative ways. 
Because of various complex factors, it is challenging to model realistic crowd behaviors in emergency scenarios. 
At a broad level, crowd behavior in emergencies is governed by panic and physical strength consumption \cite{15}. 

The main purpose of crowd simulation algorithms is to model the movements (in terms of speed and direction) of individuals in a crowd \cite{79}. We basically 
deal with two aspects of human motivations: physiological and psychological factors. 
Physical strength consumption and emotion \cite{100} are two representative physiological and psychological factors, respectively. 
Both have a great influence on individual movements.
These two factors influence each other and evolve dynamically. 
It is important to describe the inherent relationship between these two factors, which is more obvious in emergency or evacuation situations \cite{4}. 
Many approaches incorporate emotions of individuals in crowd simulations, 
making it one of the most commonly used psychological factors \cite{11}. 
Panic can prevent an individual from taking proper actions in emergency situations \cite{100}.
Researchers have observed that external dangers can directly cause changes in panic levels in an individual, thereby further determining his or her 
movements \cite{2}. 
We mainly focus on the emotion of panic in emergency situations. 
Most of the previous studies don't consider the effect of physical strength consumption on panic \cite{60}.
Physical strength is a person's or animal's ability to exert force on physical objects using muscles \cite{90}. 
Physical strength consumption is defined as the energy expenditure \cite{8} of a human, which directly affects that human's moving speed \cite{26}.  
However, it is difficult to describe the inherent relationship between physical strength consumption and panic  
and to then combine these factors to determine the movement of each individual \cite{4}. 
Therefore, an emotion-based crowd simulation model integrating physical strength consumption is challenging due to the following reasons:

(1) It is difficult to model the physical strength consumption of an individual in a crowd accurately \cite{89}. 
This task involves considering many factors that are needed to quantify the influence of 
physical strength consumption on crowd movement \cite{5}.

(2) Accurately modeling an individual's panic level in a crowd is difficult because of its constant and dynamic changes \cite{85}. Various factors such as physical strength consumption and individual movement affect panic levels.

\begin{figure}[t] \begin{centering} 
  \centering 
  \includegraphics[scale=0.1]{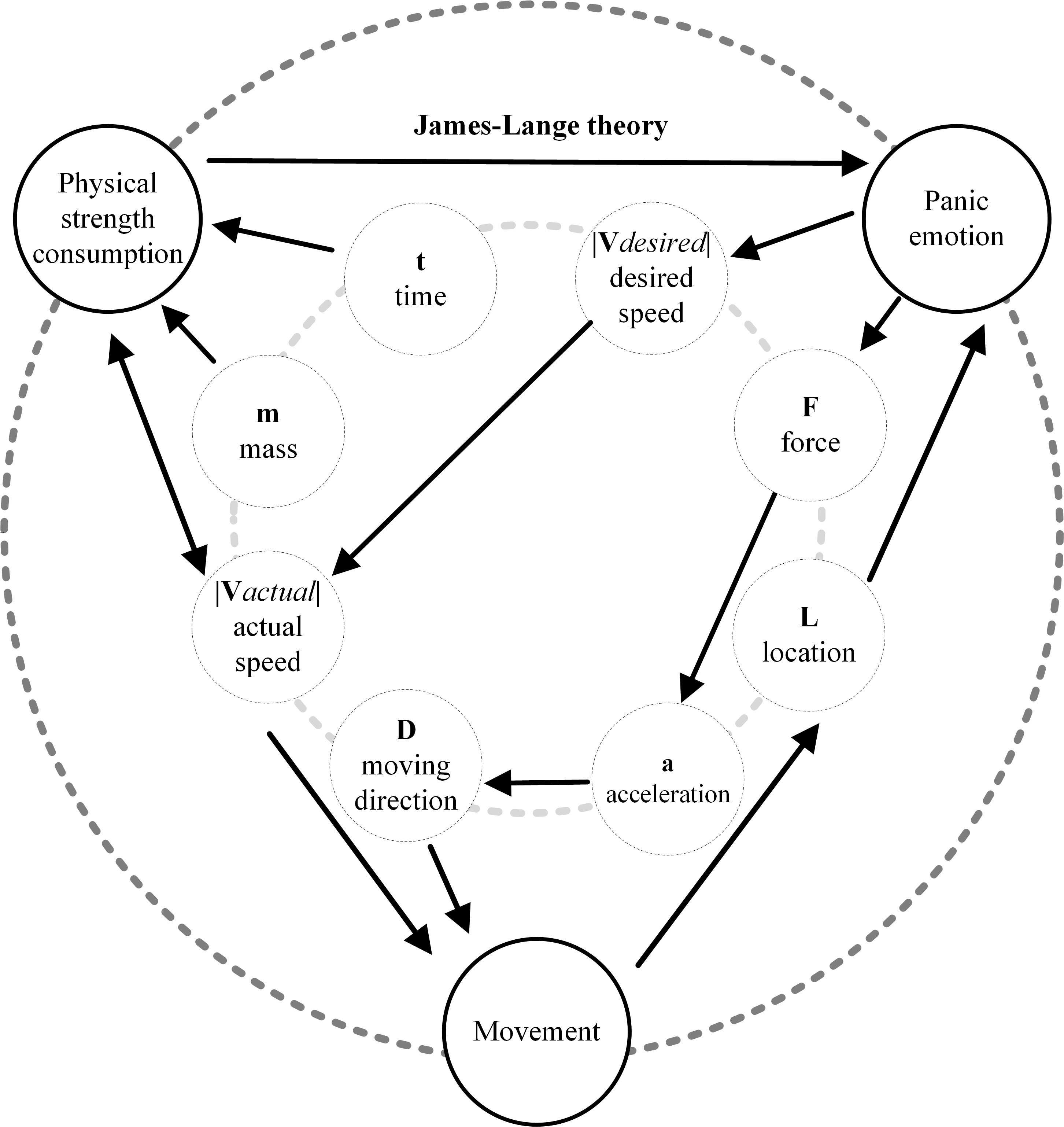}
  \centering
  \caption{The relationships among physical strength consumption, panic, and movement of individuals in a crowd.
Physical strength consumption is calculated according to the actual speed, mass, and moving time of individuals. 
The panic levels are determined by the location and physical strength consumption of the individual, and they further affect the individual's desired speed. 
Moreover, each individual's current panic level will affect his or her moving direction by changing the acceleration based on the inferred force. 
}
  \label{fig:1}
  \end{centering} 
\end{figure}

Inspired by the theory of ``the devoted actor" \cite{15}, 
which shows that an individual's physiological state has an effect on his or her psychological state, 
we propose the first (to the best of our knowledge) emotion-based crowd simulation model based on physical strength consumption 
(illustrated in Figure \ref{fig:1}). 
The main contributions of our work include:

\begin{itemize}
\item We introduce a physical strength consumption calculation method based on how individuals work under the laws of physics  
and quantitively characterize their dynamic changes during the crowd movement.

\item We present a comprehensive emotion calculation method for physical strength consumption based on the James-Lange theory. 
Our new proposed model is used to derive the relationship between physical strength consumption and panic and examines how both of them 
govern the movement. 

\end{itemize}

The rest of this paper is organized as follows. Background and related work are reviewed in 
Section \ref{related work}. The definition of our proposed crowd simulation model is introduced in Section \ref{3p model}. 
Different benchmark scenarios and results are presented in Section \ref{experiments}.

\section{Related Work}
\label{related work}

In this section, we provide a brief overview of prior work on crowd simulation. We divide the summaries 
based on whether the works involve physical, psychological, or physiological factors.

\subsection{Simple crowd simulation models}
In this subsection, we summarize representative crowd simulation models that do not consider psychological or physiological factors \cite{64,57,72,102}.

In the real world, many environmental factors influence an individual's 
movement, i.e. scene layout, moving pedestrians, and stationary groups
\cite{30, 75,101}. During the evacuation of a crowd, the behavioral choice of 
an individual is highly dependent on the moving directions of nearby 
individuals, the hazard location, and obstacles \cite{76}. 
Moussaid et al. \cite{107} propose a cognitive science approach based on behavioral heuristics.
Guided by visual information, pedestrians apply two simple cognitive procedures to adapt their moving directions and speed.
Zhou et al. \cite{108} propose a fuzzy logic approach to model and simulate pedestrian dynamic behaviors, which are based on human experiences 
and human knowledge, and perceptual information obtained from interactions with the surrounding environment. 
Zhou et al. \cite{109} focus on the role of leaders who can guide the movements of passengers during the evacuation. 
Cassol et al. \cite{87} focus on global path planning with the main goal of identifying the best evacuation routes for a specific population 
when leaving a certain building. 
To realize better 
behavioral choices, most approaches calculate the position of 
each individual at the next time step to obtain a conflict-free moving path in a 
global scenario \cite{33}. However, these approaches are not 
applicable to highly complex scenes with dense crowds. 
Other approaches use local obstacle 
avoidance methods. Namely, once the movement state of an individual is 
determined, the movement states of other individuals are updated by 
using local collision avoidance techniques \cite{34}.

In practice, these approaches face many difficulties in terms of accurately controlling the individual movements.
Researchers in this field are increasingly focusing on 
integrating global path planning and local obstacle avoidance
\cite{35,103}. Weiss et al. \cite{36} model collision avoidance 
constraints both in terms of short and long-term ranges to deal with sparse and dense 
crowds. In \cite{29}, intergroup- and intragroup-level techniques are 
presented to perform coherent and collision-free navigation using reciprocal 
collision avoidance. 
Mutual information about the dynamic crowd is used to guide agents' movements by 
combining both macroscopic and microscopic controls \cite{81}.
By constructing a visual tree, the shortest path without collisions is 
obtained in \cite{37}. In addition, in \cite{40,41,42}, and \cite{43}, path 
planning and navigation algorithms are described for crowd simulation in complex contexts. 
Furthermore, in \cite{44}, an effective long-range collision 
avoidance algorithm is proposed.

In contrast to these works, our model enhances the traditional social 
force model to avoid collisions with surrounding individuals and obstacles 
by combining panic and physical strength consumption calculations. 
Traditional crowd simulation models are not concerned with this approach. 
In our model, we mainly deal with moving directions and moving speeds, which are largely influenced by panic and 
physical strength consumption during a relatively short period of time.

\subsection{Crowd simulation with psychological factors}

The psychological state of an individual plays a vital role in his or her 
decision-making process \cite{82,88,94}. Stress and panic are typical 
psychological factors and have a great influence on the movement of individuals in a 
crowd. In this subsection, we introduce representative works 
on them.

In \cite{85}, authors focus on stress, which is defined as any change 
caused by interactions between the environment and individuals. 
Generally, stress is caused by a discrepancy between 
environmental demands and the abilities of individuals. 
Stress can have positive effects on individual behavior. In emergency or evacuation situations, stress improves the performance of individuals \cite{85}.
It can be chronic and long-term \cite{85}. 
However, stress and panic are inherently different. 
Panic is short-term and changeable \cite{91} and usually leads to negative effects on individuals \cite{98}. 
One of the most disastrous forms of collective human behavior is the kind of crowd stampede induced by panic, 
often leading to fatalities as people are crushed or trampled \cite{2}.

An individual's stress and panic are mirrored by others and they are disseminated within the crowd \cite{11}. 
There are two separate lines of emotional contagion research: epidemiological-based and thermodynamics-based.

The epidemiological SIR model \cite{20} divides the individuals in a crowd 
into three categories: infected, susceptible, and recovered. 
At first this model is used to simulate the spread of rumor \cite{21}. 
Then the epidemiological SIR model is used to describe 
emotion propagation. 
In \cite{11}, the epidemiological SIR model is combined with the OCEAN personality model \cite{14}. 
The phenomenon of emotional contagion occurs more obviously in a panicked 
crowd. In \cite{10}, the cellular automata model is combined with the SIR model (CA-SIRS) to describe emotional contagion 
in an emergency situation. 
In \cite{22}, a qualitatively simulated approach is proposed to model emotional contagion process in a large-scale emergency evacuation siutation, 
which confirms that the effectiveness of rescue guidance is influenced by the leading emotion of the crowd.
There is another kind of emotional contagion models based on thermodynamics \cite{047}. 
Bosse et al. define emotional contagion within groups based on a multi-agent approach. 
They focuses mostly on emotions of groups rather than those of single individuals. 
Neto et al. \cite{88} improve this model adapting it into BioCrowds and coping with emotional contagion within different groups of agents.
Some researchers combine these two kinds of emotional contagion models to describe dynamic emotion propagation 
from the perspective of social psychology \cite{24}.

Because panic has a great influence on individual movement and often leads to serious consequences, we focus on panic in emergency situations. 
Inspired by the James-Lange theory in biological psychology, we improve the 
Durupinar model \cite{11} by considering the influence of 
physical strength consumption on panic levels. In contrast to previous methods considering only panic, 
we further demonstrate the relationship between physical 
strength consumption and panic.

\subsection{Crowd simulation with physiological factors} 
To complete a comprehensive analysis of crowd movement, we must consider not only 
psychological factors, but also physiological factors of individuals as other important factors 
in determining the crowd movement \cite{26}. 

Physical strength is one of the 
most important physiological parameters that affects individual movement. 
Bruneau et al. \cite{28} apply the principle of minimum energy (PME) 
on groups of different sizes and densities. 
In \cite{8,9}, some 
physiological indicators (such as physical strength consumption and heart 
rate) are described. Furthermore, the relationship between 
physical strength consumption and heart rate is revealed, which is also a method for 
predicting physical strength consumption based on heart rate during 
moderate and vigorous exercise.
Work in \cite{26} shows that the relationship between physical strength consumption and 
speed is nonlinear. 
In \cite{5}, researchers investigate how the cumulative consumption of physical 
strength affects the evacuation time of individuals. Guy et al. 
\cite{16} propose the principle of least effort (PLE) to compute the physical 
strength consumption required by various movements. Furthermore, Guy et al. \cite{38} propose a less 
energy-consuming, conflict-free crowd movement method based on the criterion 
of minimal physical strength consumption \cite{16}. 
These approaches are focused on the relationship 
between physical strength and other physiological parameters (heart rate and oxygen uptake, for example) or individual movement. 
In \cite{89}, the authors choose four other basic physiological 
characteristics, including gender, age, health, and body shape, and map them to a navigation method.

Inspired by prior approaches, we focus on physical strength consumption, which is a very important physiological factor. 
Physical strength consumption is central to research in human biology and biological 
anthropology \cite{97} and is closely related to a variety of factors such as heart rate, oxygen consumption, etc. \cite{9}. 
It directly affects the moving speed of an individual \cite{5}. 
Other physiological factors (such as gender, age, health, and body shape) can influence movement through physical strength consumption.
We analyze the relationship between physical strength consumption and panic. We also describe the effects of 
physical strength consumption on the physical movements of individuals. 

\begin{figure*}[htb] \begin{centering} 
  \centering 
  \includegraphics[scale=0.38]{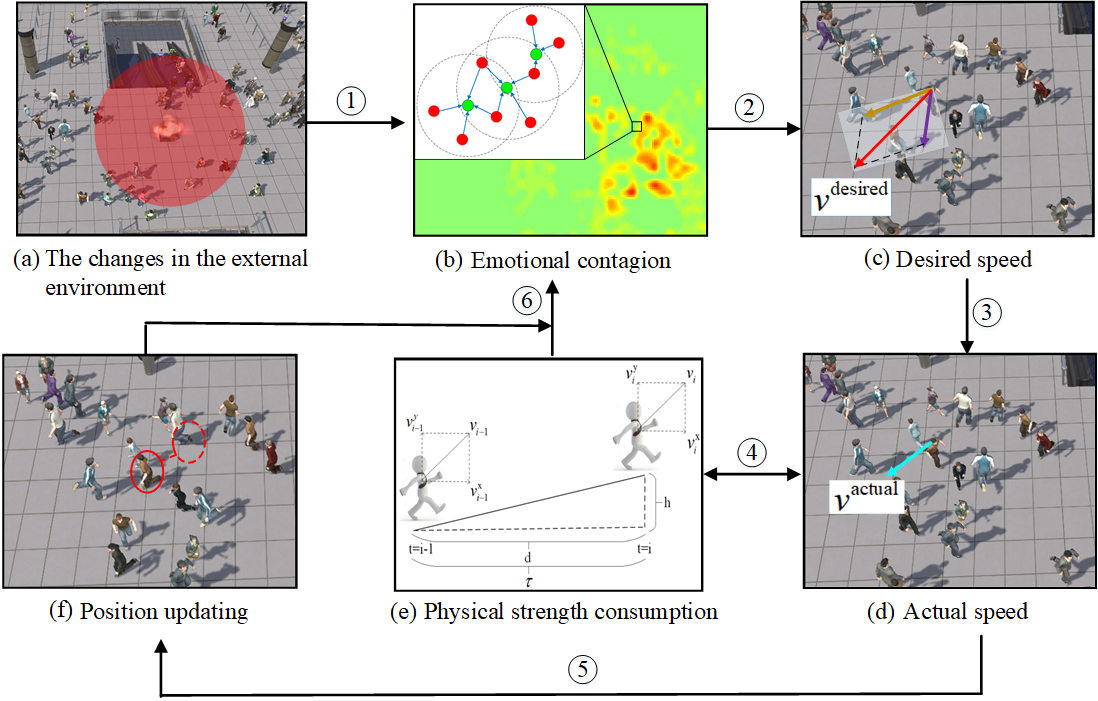}
  \centering
  \caption{The flowchart of our model. 
(a) Changes in the external environment can cause emotional fluctuations. For example, a hazard occurs and the red area represents the range of influence of the hazard.
(b) The emotional changes of one individual are calculated according to the direct impact of the hazard and emotional contagion of his or her neighbors (Section \ref{emotion-calcul}).
(c) The desired speed and direction of each individual are calculated based on an updated panic assessment (Section \ref{move}).
(d) Using the limit of physical strength consumption, the actual speed is determined \cite{26} (Section \ref{move}).
(e) 
The calculation of physical strength consumption affected by the actual speed (Section \ref{psm}). In contrast,  
the cumulative physical strength consumption also determines the actual maximal speed of the individual at the next timestep (Section \ref{move}). 
The current physical strength consumption reflects the emotional experience of an individual (Section \ref{emotion-calcul}).
(f) The position of the individual is updated according to its actual speed. If the individual is panicked, we return to step (b); 
otherwise, the flowchart ends (Section \ref{emotion-calcul}). }
  \label{fig:2}
  \end{centering} 
\end{figure*}

\section{Our Model}
\label{3p model}

Our emotion-based crowd simulation model comprehensively considers physical strength consumption in emergencies to influence crowd movement.
The flowchart of our model is presented in Figure \ref{fig:2}.
Strenuous movements are often observed in individuals in emergency or evacuation situations, 
and the relationship among them is more obvious in such situations \cite{104}. 
Therefore, we mainly focus on simulating crowd movements in such emergency situations.

Our model consists of three important components: physical strength consumption, panic, and individual movement.
Human physical strength consumption is computed with a physics-based method (Section \ref{psm}). 
Panic levels are determined through an enhanced emotional contagion model that leverages the inherent relationship 
between physical strength consumption and panic (Section \ref{emotion-calcul}). 
Our model computes the movement of an individual by modeling the physical influence of strength consumption and panic (Section \ref{move}).

\subsection{Symbols and notations} \label{Symbols and Notation} 
For convenience, the important parameters and their descriptions used in our model are listed in Table \ref{terminology and parameters}.

\begin{table}[htbp]
\setlength{\belowcaptionskip}{10pt}
\renewcommand{\arraystretch}{1.3}
\caption{The parameters used in our model.}
\label{terminology and parameters}
\centering
\begin{tabular}{m{1.3cm}<{\centering}|m{6.5cm}}
\hline
\bfseries Notation & \bfseries Description\\
\hline\hline
 $P\left( t \right)$ &  Physical strength consumption at time $t$ \\
\hline
 $P_{hor}\left( t \right)$ & Physical strength consumption along the horizontal direction at time $t$\\
\hline
 $P_{ver}\left( t \right)$  & Physical strength consumption along the vertical direction at time $t$\\
\hline
 $F^{x}$  & Driving force of individual along the horizontal direction \\
\hline
 $F^{y}$  & Pulling force of individual along the vertical direction \\
\hline
 $E$  & Panic emotion \\
\hline
 $E_o$ & Emotional cognitive component \\
\hline
 $E_p$ & Emotional experience component \\
\hline
 $E_o^h$ & The emotion affected by hazards \\
\hline
 $E_o^c$ & Emotional contagion \\
\hline
 $V_i \left(t\right)$ & Moving direction of the individual $i$ at time $t$ \\
\hline
 $V_i^s \left(L,t\right)$ & Safety evacuation direction of the individual $i$ at location $L$ and at time $t$ \\
\hline
 $V_i^{round}\left(t\right)$  & Combined moving directions of individuals who are in the perceived range of the individual $i$ at time $t$ \\
\hline
 $v_i^{desired}$  & The desired speed $v_i^{desired}$ of the individual $i$ considers only the emotion factor. \\
\hline
 $v_i^{actual}$ & The actual speed $v_i^{actual}$ of the individual $i$ is limited by his own physical strength consumption. \\
\hline
 $v^p$ & Maximum speed $v^p$ according to current physical strength consumption\\
\hline
 $v_i^{\textit{MAX}}$ & Maximum speed that the individual $i$ can run \\
\hline
 $v_i^{\textit{NOR}}$ & Speed of the individual $i$ in the normal case (emotion value is equal to zero) \\
\hline
PR & The radius of perceived range\\
\hline
Num & The number of individuals in a scene\\
\hline
$R_a$  &  The radius of an individual\\
\hline
\end{tabular}
\end{table}

\subsection{Physical strength consumption calculation} \label{psm} 

Physical strength consumption is one of the most commonly used physiological indicators and is closely related to individual movement. It is defined by the following equation:

\begin{equation}\label{eq1} 
P\left( t \right)=P_{hor}\left( t \right)+P_{ver}\left( t \right)
\end{equation} 
where $P\left( t \right)$ denotes the total physical strength consumption at time $t$ and $P_{hor}\left( t \right)$, $P_{ver}\left( t \right)$ 
denote the physical strength consumption along the horizontal and the vertical directions, respectively. They are defined as follows:

\begin{equation}\label{eq3} 
P_{hor}\left( t \right)=\sum\limits_{i=1}^{t}{F^{x}_{i}\cdot d_{i}}
\end{equation} 
\begin{equation}\label{eq2} 
P_{ver}\left( t \right)=\sum\limits_{i=1}^{t}{F^{y}_{i}\cdot h_{i}}
\end{equation} 
$F^{x}_{i}$ is the driving force of the individual along the horizontal direction. This 
force overcomes friction. $d_{i}$ is the moving distance of the individual at time $t$.  $F^{x}_{i}\cdot d_{i}$ represents the work done by the individual along the horizontal direction.
$F^{y}_{i}$ is the pulling force of the individual along the vertical direction. This force overcomes gravity. $h_{i}$ is the rising height of the individual at time $t$, and $F^{y}_{i}\cdot h_{i}$ represents 
the work done by the individual along the vertical direction.

\begin{figure}[t] \begin{centering} 
  \centering 
  \includegraphics[width=7.8cm]{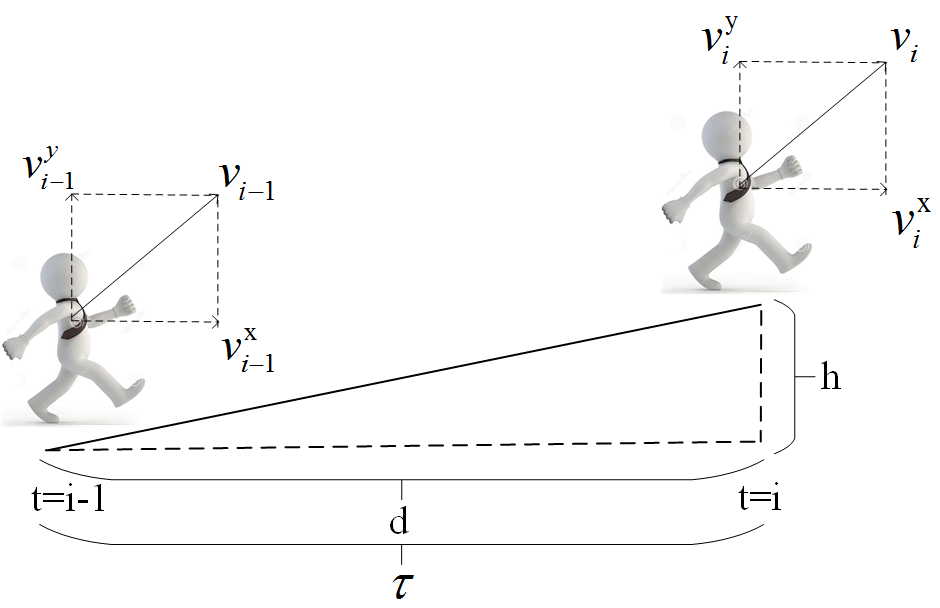}
  \centering
  \caption{Schematic of the physical strength consumption calculation. 
$v_i^x$ is the velocity component of an individual in the horizontal direction at time $i$, and the length of each time step is $\tau$. 
The horizontal speed of the individual changes from $v_{i-1}^x$ to $v_i^x$ in time interval $\tau$. 
} 
  \label{fig:3}
  \end{centering} 
\end{figure}

According to the laws of physics, $F^x_{i}$ is defined as follows:
\begin{equation}\label{eq4} 
F^x_{i}=f_{i}+\frac{(v_i^x-v_{i-1}^x)m}{\tau} 
\end{equation} 
A  diagram of the physical strength consumption calculation is 
shown in Figure \ref{fig:3}.

The friction $f_{i}$ is defined in Equation \ref{eq5}, $k_{i}$ is defined in Equation \ref{eq51}, and 
$t_{i}$ is defined in Equation \ref{eq52} according to \cite{46,47}. 
$\mu$ is the friction factor, which is related to the shoes and the ground. In our implementation, $\mu$=0.58 is adopted, which is also recommended in \cite{73}. 
$v_{i}$ is the current velocity magnitude, $v_{min}$ is the minimal velocity magnitude, 
and $v_{max}$ is the maximal velocity magnitude.

\begin{equation}\label{eq5} 
f_{i}=t_{i} \cdot \mu \cdot mg \cdot k_{i}
\end{equation} 
\begin{equation}\label{eq51} 
k_{i}=1.5+0.5 \cdot \frac{v_{i}-v_{min}}{v_{max}-v_{min}}
\end{equation} 
\begin{equation}\label{eq52} 
t_{i}=0.6-0.2 \cdot \frac{v_{i}-v_{min}}{v_{max}-v_{min}}
\end{equation} 
where $k_{i}$ is the coefficient of the weight, $t_{i}$ is the time of the individual's 
foot touching the ground, $k_{i} \propto v_{i}$, $t_{i} \propto^{-1} v_{i}$, $f_{i} \propto k_{i}$, and $f_{i} \propto t_{i}$.
If one stands with both feet on a force plate, $t_i = k_i = 1$.

The physical strength consumption in the horizontal direction is defined by:
\begin{equation}\label{eq8} 
\scalebox{0.8}
{$
P_{hor}\left( t \right)= \frac{1}{2} \cdot \sum\limits_{i=1}^{t}{ \left\{  \left( \left({v_i^x}\right)^2 - \left({v_{i-1}^x}\right)^2 \right) m +\\ 
                         t_{i} \cdot \mu \cdot mg \cdot k_{i} \left( v_i^x + v_{i-1}^x \right) \tau  \right\}  }
$}
\end{equation} 

According to the laws of physics, $F^y_{i}$ is defined by the following equation:
\begin{equation}\label{eq9} 
F^y_{i}=mg+\frac{(v_i^y-v_{i-1}^y)m}{\tau}  
\end{equation} 
where $v_i^y$ is the velocity component in the vertical direction at time $i$.

The physical strength consumption in the vertical direction is defined by:
\begin{equation}\label{eq12} 
\scalebox{0.8}
{$
P_{ver}\left( t \right) =\frac{1}{2} \cdot
\sum\limits_{i=1}^{t}{ \left\{  \left( \left({v_i^y}\right)^2 - \left({v_{i-1}^y}\right)^2 \right) m +  
 \left( v_i^y + v_{i-1}^y \right)mg \tau  \right\}  }
$}
\end{equation} 

\subsection{Panic calculation considering physical strength consumption} \label{emotion-calcul}
This section presents the calculation method for the panic level of an individual. 
$E \in [0,1]$ indicates the approximate level of panic. 
The panic level $E$ consists of two components. The first is 
the emotional cognitive component $E_o$, 
which relates to the hazard and encompasses emotional contagion. The second is the emotional experience component $E_p$, 
which is calculated using physical strength consumption and heart rate. Therefore, the final emotion value is defined as follows:

\begin{equation}\label{eq14-2} 
 E=w \cdot E_o+ (1-w) \cdot E_p
\end{equation} 
where $w$ is a weighting parameter, and $0<w<1$.

\subsubsection{The emotional cognitive component}
In this section, we present the calculation method of $E_o$. $E_o$ consists of three terms: 
effect from hazard $E_o^h$, emotional contagion $E_o^c$, and emotional attenuation $E_o^d$. 

\textbf{Effect from hazard} $\pmb{E_o^h}$ \textbf{:}
When individuals are able to perceive a hazard, they may become panicked. $E_o^h$ is 
defined as follows \cite{105}:
\begin{equation}\label{eq19} 
E_o^h \left( L,t \right)= \sum\limits_{s=0}^{n-1}{{\Gamma }_{s}\left( L,t \right)}
\end{equation} 
\begin{equation} 
\label{eq18} 
\scalebox{0.9}{$
{{\Gamma }_{s}}\left( L,t \right) =\left \{ 
\begin{array}{lll}\!\! 
 {\!\!\!\!}         \frac{1}{\sqrt{2\pi }\cdot {{r}_{s}}}{{e}^{-\frac{{{\left( L-{{L}_{s}}
   \right)}^{2}}}{2{{r}_{s}}^{2}}}}                         
 {\!\!\!\!}           & \mbox{if} \left\|\left. L-{{L}_{s}} \right\|<{{r}_{s}}\text{ }and\text{ }t\in U \right. \\
           \\
                0
                 &  \mbox{otherwise}\\
\end{array} \right.
$}
\end{equation} 
where $L$ is the location of an individual, $L_s$ is the location of a hazard, $r_s$ is the radius of the influence range of the hazard, and
$U$ is the duration of the hazard.

\textbf{Effect from emotional contagion} $\pmb{E_o^c}$ \textbf{:}
There are two kinds of representative models of emotional contagion: the Neto model \cite{88} and the Durupinar model \cite{11}. 
They use fundamentally different mechanisms, but both can generate good results. 
However, the Neto model defines too many parameters for each pairwise 
interaction \cite{93} and it is hard to compute these parameters automatically. 
Moreover, personality is also a very important, long-term, stable psychological factor 
and it is vital for simulating heterogeneous crowd behavior \cite{11}. 
The Neto model simplifies the personality factor while 
the Durupinar model pays more attention to that factor and is effective at capturing the differences between individuals. 
Personality is an important part of our model. 
We consider the effect of personality on panic. 
According to the above analysis, the Durupinar model is more suitable. 
In the virtual scenario of Section \ref{comp-real}, we implement a comparable experiment to verify our motivation.
Next, we present the emotional contagion method in our model.

During evacuation, individuals can be in one of two states: susceptible or infected. When the panic level of an individual exceeds a 
certain threshold $T_1$, the individual will be infected. If the intensity of an individual's panic surpasses another threshold $T_2$, 
then the individual can spread the panic to his or her neighbors. In a general case, $T_1<T_2$. $T_1$ and $T_2$ are 
correlated with individual personalities. Here we represent the personalities of individuals using the OCEAN personality model \cite{11}. 
The personality of an individual is represented by a five-dimensional vector $<O, C, E, A, N>$.  
Each factor is randomly distributed with a Gaussian distribution $N(0,0.25)$ \cite{11}.
$T_1$ $\propto^{-1}$ N, $T_1$ $\propto$ C \cite{14}. $T_2$ $\propto^{-1}$ E \cite{11,14}. 
$T_1$ and $T_2$ are defined by the following:

\begin{equation}\label{eq16} 
T_1=\alpha \cdot C -\beta \cdot N + \gamma
\end{equation} 
where $\alpha=0.1$, $\beta=0.1$, and $\gamma=0.15$.
\begin{equation}\label{eq17} 
T_2=\delta - \xi \cdot E
\end{equation} 
where $\delta=0.35$, and $\xi=0.1$. 
These parameters are determined according to the methods in \cite{95,96}.

%

Within the perceived range, when a susceptible individual $i$ sees an expressive individual $j$ (the panic value is higher 
than threshold $T_2$), $i$ gets exposed by receiving a random dose $d_i$ from a specified probability distribution 
multiplied by the panic intensity of $j$. 
The dose values $d_i$ are randomly distributed with a Gaussian distribution $N(0.1,0.01)$. 
We denote the panic value 
of individual $j$ at time $t'$ as $e_j\left( t'\right)$. The panic value of individual $i$ due to emotional 
contagion is defined in Equation \ref{eq20} \cite{11}.
\begin{equation}\label{eq20}
\scalebox{0.9}
{$
E_{i,o}^c \left( L,t \right) = \sum\limits_{t' = 0}^t \,  {\sum\limits_{\forall j \mid j \in \textit{Visibility} \left( i \right) \land j \,\,\textit{is}\,\, \textit{expressive}} {{d_i}\left( {t'} \right)e_{j}\left( t' \right)} }
$}
\end{equation}

\textbf{Effect from emotional attenuation} $\pmb{E_o^d}$ \textbf{:}
Emotional attenuation is defined in Equation \ref{eq21} \cite{105}.

\begin{equation}\label{eq21}
E_{o}^d \left( L,t \right) =  E_{o} \left( L^{pre},t-1 \right) \cdot \eta(t)
\end{equation}
where $E_o^d \left(L,t\right)$ is an emotion decay function and $\eta(t)$ is the decay rate. $\eta(t)$ is positively related to 
the individual personality factor $N$ \cite{11}. Inspired by \cite{11,5}, it is defined as follows:

\begin{equation} 
\label{eq22} 
\scalebox{0.9}{$
{{\eta }}\left( t \right) =\left \{ 
\begin{array}{lll}\!\! 
 {\!\!\!\!}
 \\ 0                             
 {\!\!\!\!}         & t<t_1 \\
           \\
                \frac{e^ {\beta_2 \left( t-t_2 \right)} - e^ {\beta_2 \left( t-1-t_2 \right) } }{1+ e^{ {\beta_2 \left( t-t_2 \right)} }} + \alpha \cdot N
                 &  t \ge t_1\\
\end{array} \right. 
$}
\end{equation} 
where $\beta_2=0.1$, $\eta \propto N$, and $\alpha=0.1$.

The change of the emotional cognitive component $\Delta E_o \left( L,t \right)$ is defined in Equation \ref{eq24} \cite{105}. The $E_o$ is defined in Equation \ref{eq25} \cite{105}.
\begin{equation}\label{eq24}
\Delta E_o \left( L,t \right) = E_o^h \left(L,t \right) + E_o^c \left( L,t \right) - E_o^d \left( L,t \right)
\end{equation}
\begin{equation}\label{eq25}
E_o \left( L,t \right) = E_o \left(L^{pre},t-1 \right) + \Delta E_o \left( L,t \right)
\end{equation}

\subsubsection{The emotional experience component}

In this section, we present the calculation method of $E_p$. Individual emotions undergo three stages: cognition, action, and experience. 
First, an event occurs, and the individual perceives the current 
scene (emotional cognitive stage). Subsequently, 
the individual acts in a way that corresponds with physiological changes (action stage). 
Finally, the individual has the emotional experience (experience stage) \cite{4}.

Under emergency situations, once a hazard occurs, the individuals around it immediately take different actions, 
requiring physical strength consumption. 
Physical strength consumption in one minute is chosen as the measure of physiological changes. 
The current heart rate is calculated using physical strength consumption. Then, the increment of the emotional 
experience value is calculated based on the heart rate increment. Thereafter, the current 
emotional experience value $E_p$ is obtained. 
The details of the calculation method are as follows.

Equation \ref{eq26} describes the relationship between physical strength consumption in a minute (KJ/min) and heart rate (beat/min) when individuals experience panic and attempt to escape from the hazard \cite{9}. 
According to Equation \ref{eq26}, we can calculate the current heart rate (\textit{HR}) based on physical strength consumption in a minute ($\bigtriangleup P$).
\begin{equation}\label{eq26}
\scalebox{0.75}
{$
\textit{HR}(t)=\left\{ \begin{matrix}
   \begin{matrix}
   87.3306+\text{1}\text{.5850}\Delta P(t)-0.3151weight-0.3197age & gender=1  \\
\end{matrix}  \\
   \begin{matrix}
   45.6221+2.2361\Delta P(t)+0.2824weight-0.1655age & gender=0  \\
\end{matrix}  \\
\end{matrix} \right.
$}
\end{equation}
where gender=1 for males and 0 for females, age (year) $\in$ [19,45], weight (kg) $\in$ [47,116], $\Delta P(t)=P(t)-P(t-\tau)$, and $\tau=60s$ .

Furthermore, according to \cite{53}, heart rate $\left(\textit{HR}\right)$ and intensity of anxiety or fear (emotional experience) are positively correlated.  
In \cite{53}, the heart rate per minute is recorded before and after an electric shock, and emotional experience is reported once per minute. 
${\bigtriangleup E_p}$ and ${\bigtriangleup \textit{HR}}$ are the increments of emotional experience and heart rate, respectively, compared with the values when 
individuals are not panicked. The ${\bigtriangleup \textit{HR}}$ is defined as follows:
\begin{equation}\label{eq201965}
\scalebox{1}
{$
\bigtriangleup \textit{HR}(t)=\textit{HR}(t)-\textit{HR}(0)
$}
\end{equation}
where $\textit{HR}(t)$ is the heart rate at time $t$ and $\textit{HR}(0)$ is the heart rate when individuals are not panicked.

Using a linear curve fitting method, we can obtain the relationship between  $\bigtriangleup \textit{HR}$ and $\bigtriangleup E_p$.
\begin{equation}
\bigtriangleup E_p(t)=0.03669 \cdot \bigtriangleup \textit{HR}(t)-0.0724
\end{equation}
$E_p$ is defined in Equation \ref{eq251} and $E_p \left( 0 \right)=0$.
\begin{equation}\label{eq251}
E_p \left( t \right) = E_p \left( t-1 \right) + \Delta E_p \left( t \right)
\end{equation}



\subsection{Individual movement model} \label{move} 

Based on the results of physical strength consumption and panic, the movement of each individual can be determined accurately through moving direction and 
moving speed.

\subsubsection {Moving direction}

When a hazard occurs, individuals who can perceive the hazard directly will be panicked and calculate their own 
safety evacuation directions $V_i^s \left(L,t\right)$ \cite{105}. 
$V_i^{round}\left(t\right)$ is the combined moving directions of individuals who are in the perceived range of the individual $i$.  
\begin{equation} \label{eq27} \scalebox{0.9}{$\mathop {V_i^s\left( {L,t} 
\right)}\limits^ \to = \!\! \left \{ \begin{array}{lll} 
\!\!\!          \sum\limits_{s = 0}^{n - 1}{{\Gamma _s}\left( {L,t} \right)} \cdot \mathop{{L_{{s}}}L}\limits^ \to                        
          
           & \mbox{if} \! \left\| \left. L-{{L}_{s}}  \right\|<{{r}_{s}}\text{ }and\text{ }t\in U \right. \\
\!\!\!          \mathop V\limits^ \to 
                 &  \mbox{otherwise}\\

\end{array} \right. 
$}
\end{equation} 
\begin{equation}\label{eq28}
\scalebox{0.9}
{$
V_{i}^{round} \left( t \right) = \sum\limits_{\forall j \mid j \in \textit{Visibility} \left( i \right) \land j \,\,\textit{is}\,\, \textit{expressive}}V_j\left(t\right)  
$}
\end{equation}
Finally, the moving direction $V_i \left(t\right)$ of actual velocity of an individual who directly perceives the hazard is defined as follows:
\begin{equation}\label{eq29}
V_i\left(t\right)=E \cdot V_i^s\left(L,t\right)+\left(1-E\right) \cdot V_i^{round} \left(t\right)
\end{equation}
where $E$ is the panic emotion value. 
The moving direction of an individual $i$ is influenced by panic level, safety evacuation direction, and other neighboring 
panicked individuals.

Individual $i$ can perceive the hazard indirectly through the surrounding panicked individuals. 
The individual $i$ moves in the direction of $V_i \left(t\right)$, as shown 
in Equation \ref{eq30}. $V_i^{old} \left(t\right)$ is the moving direction of the individual $i$ at the last moment 
when he is not panicked. 
The more panicked the individual is, the more easily he moves with other neighboring 
panicked individuals. Nonetheless, if the individual $i$ is not panicked, he or she still moves in his or her original direction. 
\begin{equation}\label{eq30}
\scalebox{0.9}
{$
V_i\left(t\right)=\left(1-E\right) \cdot V_i^{old}\left(t\right)+E \cdot V_i^{round} \left(t\right)
$}
\end{equation}

\subsubsection {Moving speed}

In a panic situation, the speed of an individual $i$ is expressed by the following equation \cite{2}:
\begin{equation}\label{eq31}
\scalebox{0.9}
{$
v_i^{desired}=\left(1-E\right) \cdot v_i^{\textit{NOR}} + E \cdot v_i^{\textit{MAX}}
$}
\end{equation}
where $v_i^{desired}$ is the speed considering only the emotion factor, and $0 \leq E \leq 1$. The speed 
of an individual in the normal case (the panic value is equal to zero) is $v_i^{\textit{NOR}}$ , and 
the maximal speed is $v_i^{\textit{MAX}}$. The more panicked an individual is, the faster his or her speed.

However, an individual is limited by his or her own physical strength consumption. In some cases, the 
moving speed of an individual cannot reach the desired speed due to the maximum limit dictated by current 
physical strength consumption. 
The actual speed cannot exceed the maximal speed $v^p$.
\begin{equation}\label{eq32}
v_i^{actual}=min\left(v_i^{desired} , v^p\right)
\end{equation}

The dependence 
of the decay rate \cite{5} and maximal speed on physical strength consumption is presented 
in Table \ref{The dependence of speed decay rate and maximal-limit speed on physical strength}.

\begin{table}[htb]
\scriptsize
\setlength{\belowcaptionskip}{10pt}
\renewcommand{\arraystretch}{1.3}
\caption{Dependence of speed decay rate and maximal-limit speed on physical strength consumption. 
As the physical strength consumption increases, the maximal-limit speed decreases.}
\label{The dependence of speed decay rate and maximal-limit speed on physical strength}
\centering
\begin{tabular}{c|c|c}
\hline
\makecell[tl]{\bfseries Physical strength \\ \bfseries consumption $p$ (J)} &\bfseries Decay rate $\xi$  $\left(\%\right)$ & \makecell[tl]{ \bfseries Maximal-limit \\ \bfseries speed  $v^p$ $\left(m/s\right)$}\\
\hline\hline
0.0000 -- 20154.0000 & 100.0000 & $v_i^{\textit{MAX}}$ \\
\hline
20154.0000 -- 40279.6713 & 99.8500 & $v_i^{\textit{MAX}} \cdot 0.9985$ \\
\hline
40279.6713 -- 81121.0042 & 89.4200 & $v_i^{\textit{MAX}} \cdot 0.8942$ \\
\hline
81121.0042 -- 166258.8920 & 75.8000 & $v_i^{\textit{MAX}} \cdot 0.7580$ \\
\hline
166258.8920  -- 181569.6090 & 69.8200 & $v_i^{\textit{MAX}} \cdot 0.6982$ \\
\hline
181569.6090 --  196355.1760 & 65.7200 & $v_i^{\textit{MAX}} \cdot 0.6572$ \\
\hline
\end{tabular}
\end{table}

The actual speed can be calculated using Equation \ref{eq33}.
\begin{equation}\label{eq33}
{
v_i^{actual}=min\left(\left(1-E\right) \cdot v_i^{\textit{NOR}} + E \cdot v_i^{\textit{MAX}} , v_i^{\textit{MAX}} \cdot \xi  \right)
}
\end{equation}





\section{Experiments} 
\label{experiments}

\begin{figure*}[!t]
\centering
\begin{tabular}{cccc}
 \includegraphics[width=4cm,height=4cm]{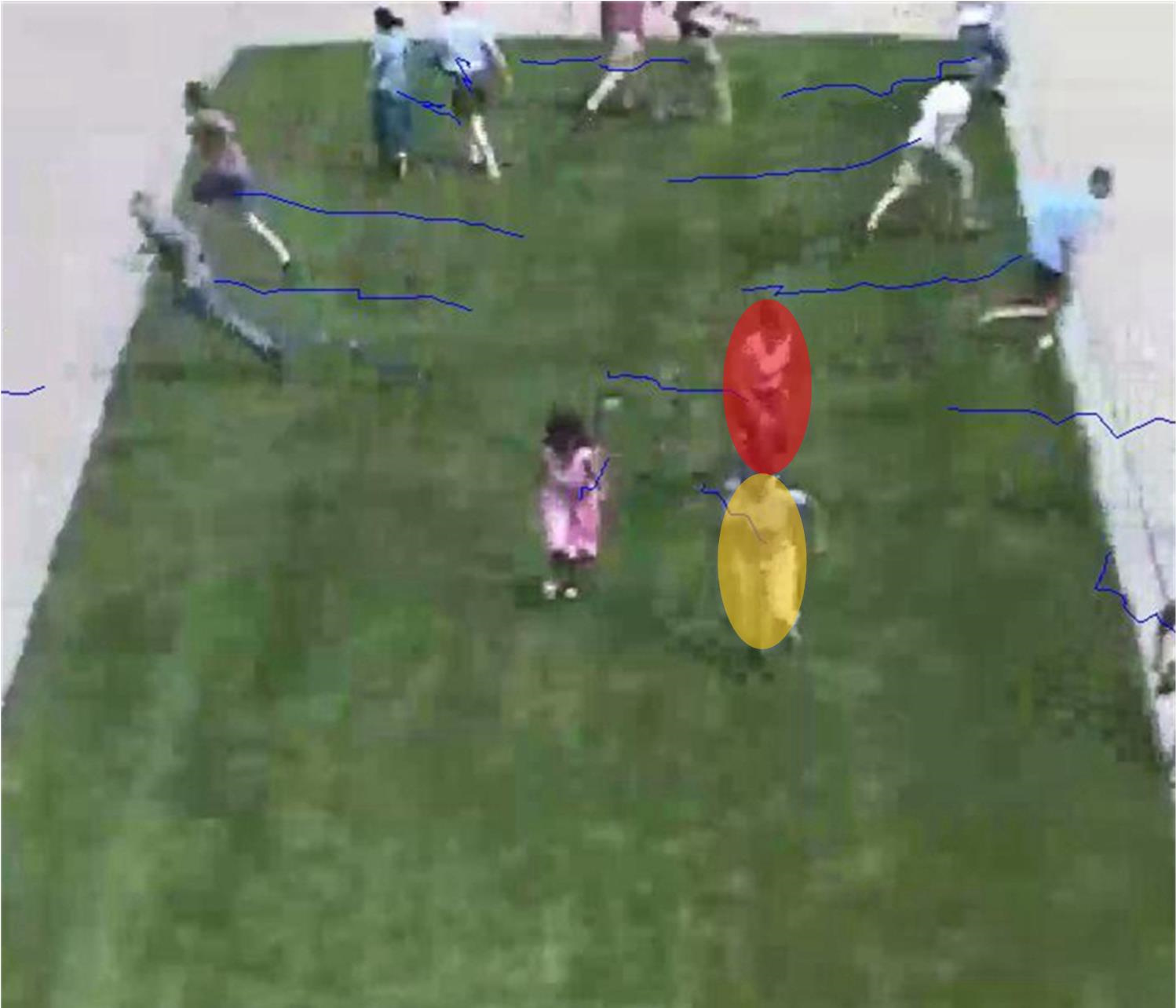} &  \includegraphics[width=4cm,height=4cm]{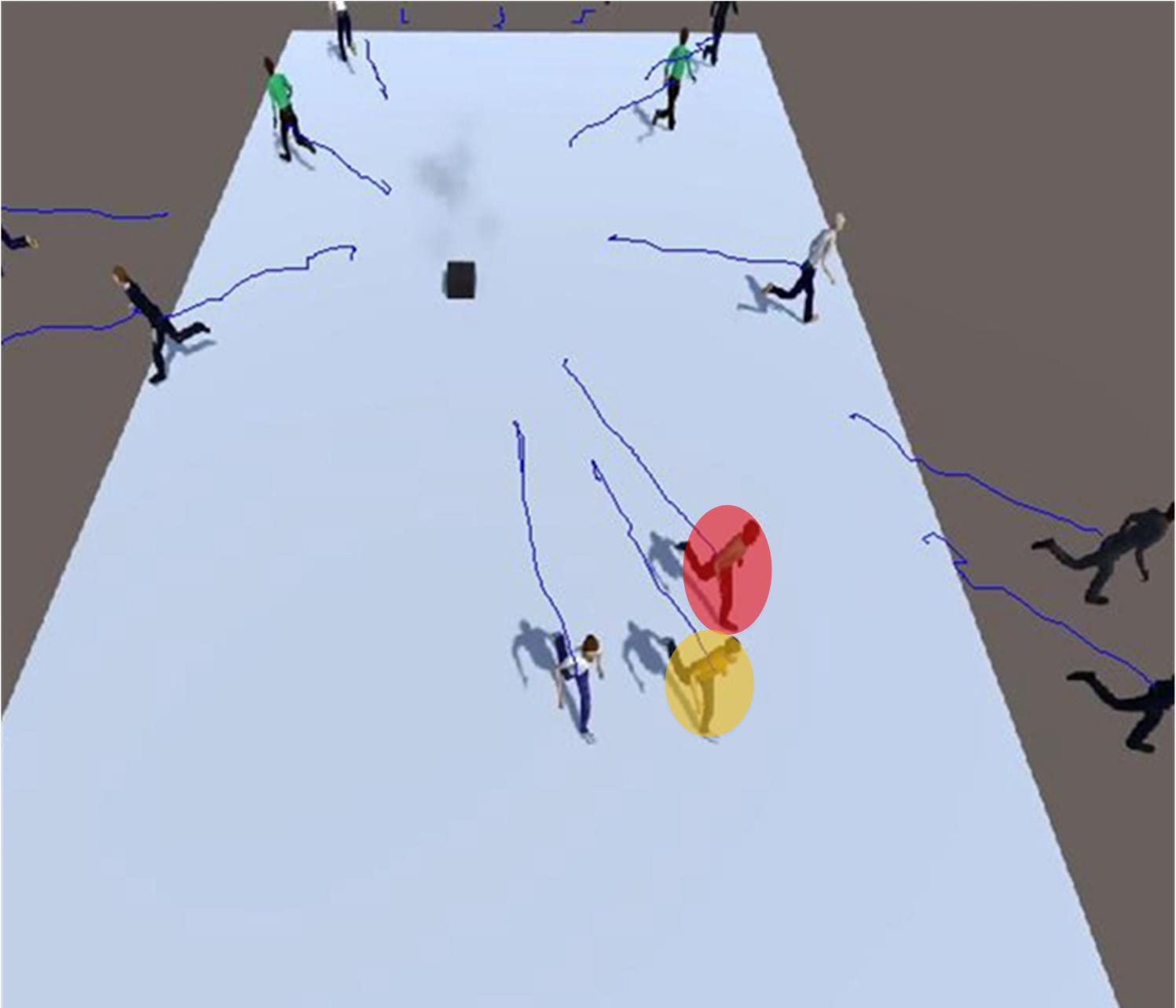}&  \includegraphics[width=4cm,height=4cm]{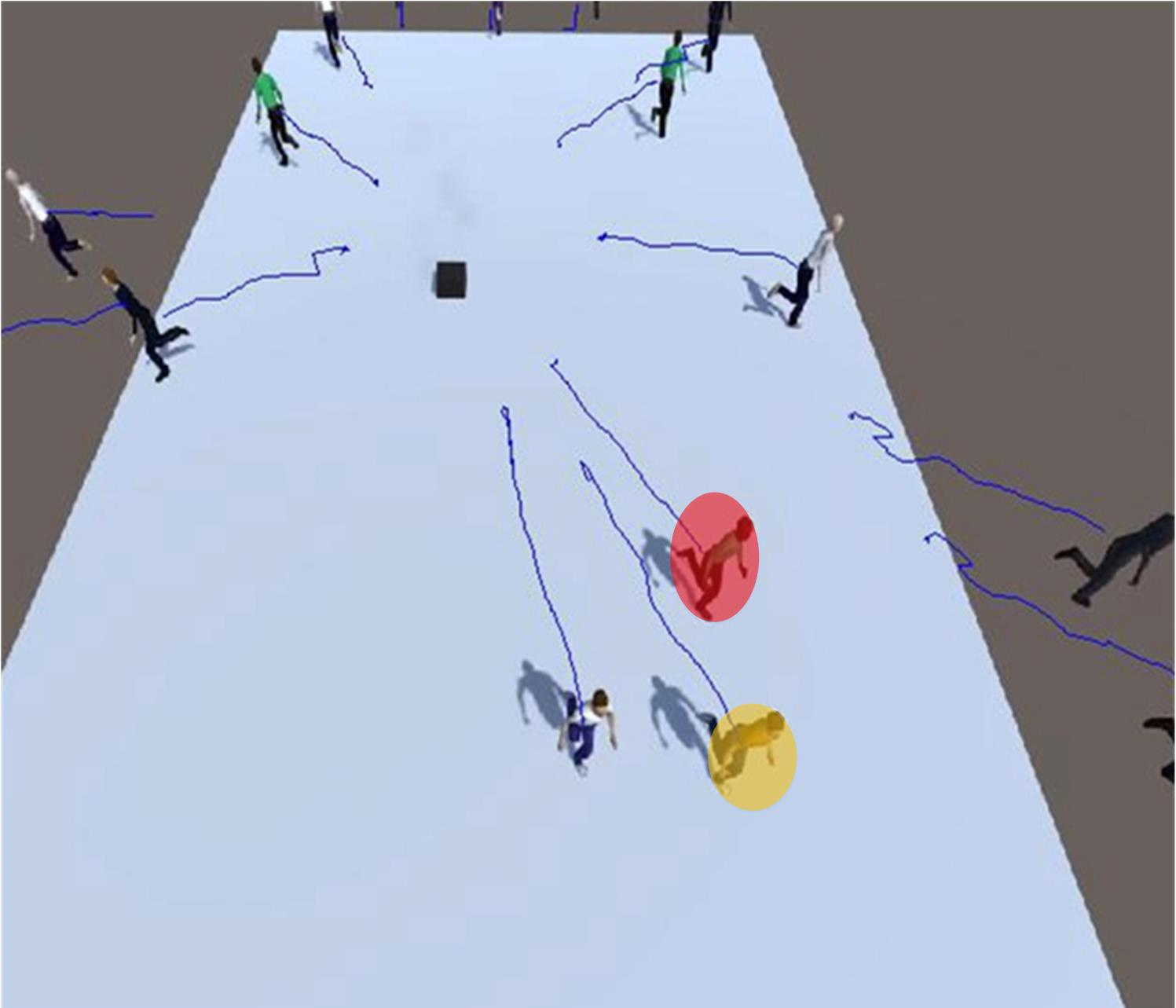}&  \includegraphics[width=4cm,height=4cm]{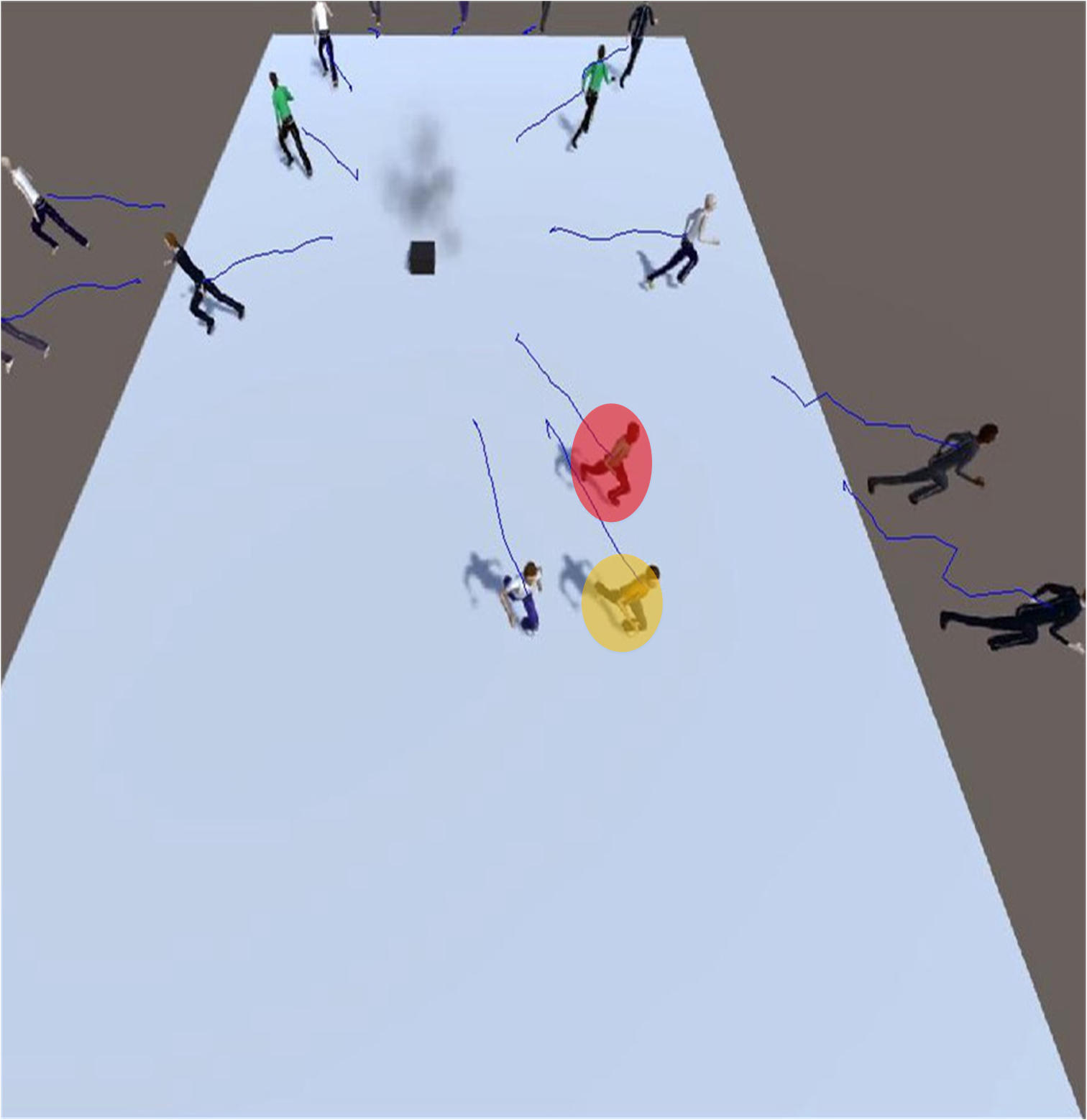}  \\
  (a) &  (b)  &  (c) &  (d)  \\
 \includegraphics[width=4cm,height=4cm]{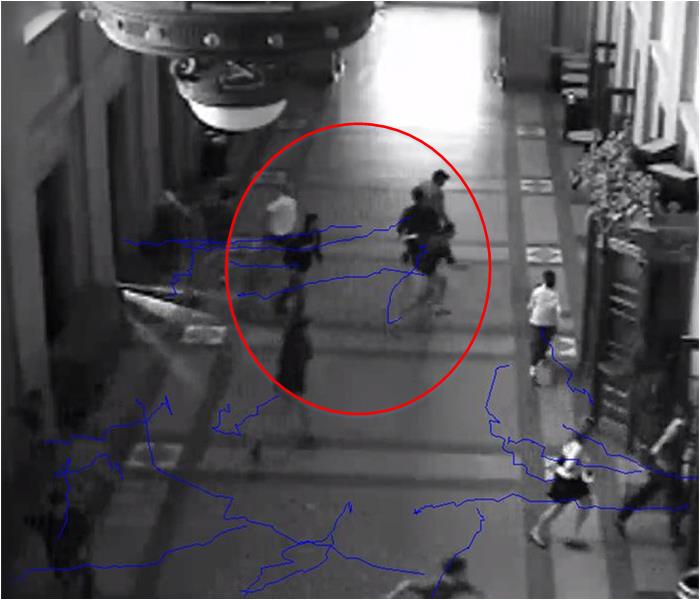} &  \includegraphics[width=4cm,height=4cm]{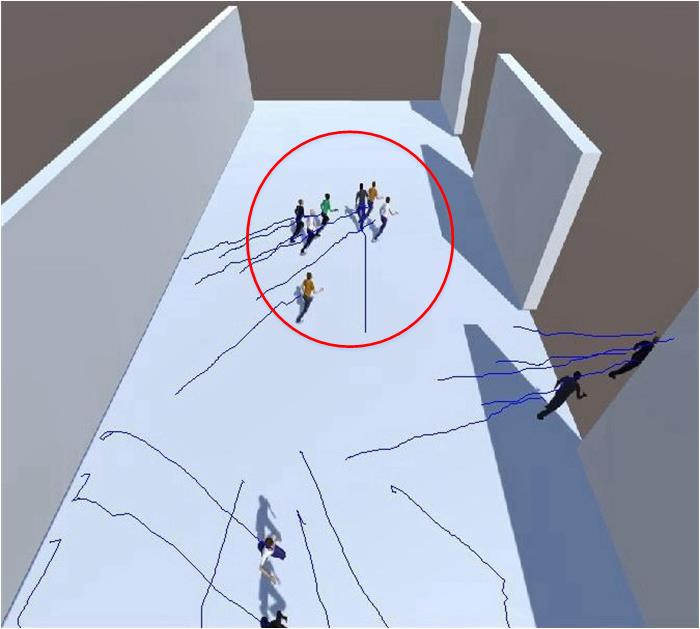}&  \includegraphics[width=4cm,height=4cm]{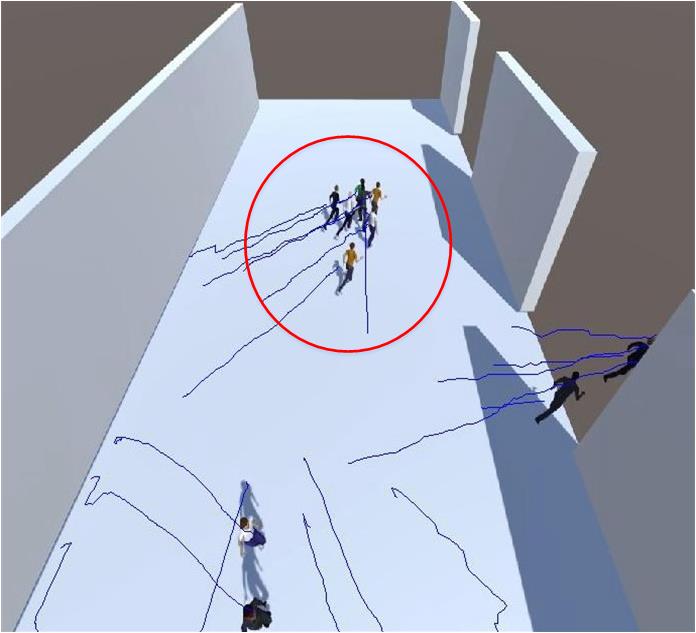}&  \includegraphics[width=4cm,height=4cm]{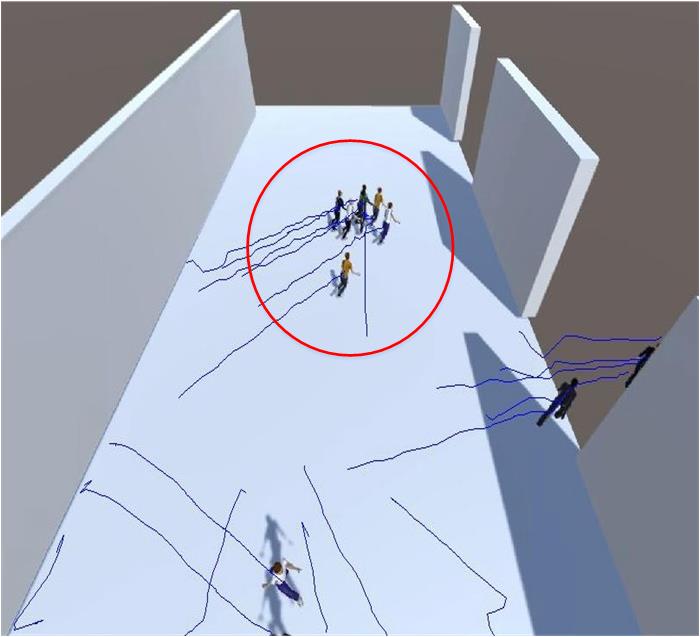} \\
  (e)  &  (f)   &  (g)  & (h)  \\
\end{tabular} 
\caption{Comparisons between real scenes and simulation results by different models: (a) and (e) are real-world videos, 
(b) and (f) are simulated by our model, (c) and (g) are simulated by the Durupinar model, (d) and (h) are simulated by the Neto model.
Each row represents one scene. 
(a) The red ellipse is Individual No. 1 and the yellow one is Individual No. 2. Individual No. 1 gets closer to Individual No. 2.
(b) As the speed is influenced by physical strength consumption, simulating the situation where Individual No. 1 
gets closer to Individual No. 2 is easier using our model.
In (g) and (h), the simulation trajectories of different individuals in the red circle by the Durupinar and Neto model 
are similar and individuals easily get together, which is different from the real-world video. 
} 
\label{fig:1415} 
\end{figure*}

Our proposed algorithm is used to simulate crowd movement in various scenarios and 
we demonstrate the benefits of it in these different scenarios. 
We evaluate our model on the public UMN dataset \cite{83} (Figures \ref{fig:1415}). 
The dataset comprises the videos of 11 different 
scenarios of an escape event in 3 different indoor and outdoor scenes. Each video begins with initial normal behavior 
and ends with sequences of abnormal behavior. In addition, some real-world scenes (Figures \ref{fig2019060501}, \ref{fig2019070601}, and \ref{fig:25}) are chosen from 
real emergency incidents to evaluate our model. 
The simulation results show that our proposed method can generate realistic group behaviors. 
It can also reliably predict the changes of physical strength consumption and panic levels of a crowd in an
emergency. 
We also use our proposed model in different virtual scenarios, such as a subway station and a crosswalk. These scenes have dense crowds 
and the probability of hazard occurrence in these scenarios is high. We simulate the crowd movement in these scenarios after one hazard.

We have implemented the proposed model using Visual C++ to simulate crowd movements in emergencies. 
The Unity3D game engine has been used to visualize our crowd simulation results.
The computing platform corresponds a PC with a quadcore 2.50 GHz CPU,16 GB memory, and an Nvidia GeForce GTX 1080 Ti graphics card. 
The parameter values in different scenarios used in the simulation are listed in Table \ref{2019318}. The mass of each individual is set to 60kg 
on average and the 
radius to 0.3m (in Table \ref{2019318}) \cite{106}. Each factor of the vector $<O, C, E, A, N>$ is randomly distributed 
with a Gaussian distribution $N(0,0.25)$ \cite{11}. In most scenes, the dose values $d_i$ are randomly distributed with a Gaussian 
distribution $N(0.1,0.01)$ \cite{11}. In the Neto model, $\epsilon=0.5$, $\delta=0.5$, $\eta=0.5$, and $\beta=1$ \cite{88}. 
The parameter values are obtained by comparing the simulations with real-world videos. 
A combination of genetic and greedy strategies 
are used to sample plausible parameters for our model, maximizing the match of the simulation algorithm to real data \cite{96}. 
\begin{table*}[]\tiny
\setlength{\belowcaptionskip}{10pt}
\renewcommand{\arraystretch}{1.3}
\caption{List of parameter values used in our simulations}
\begin{tabular}{|c|c|c|c|c|c|c|c|c|c|c|c|c|c|c|c|c|c|c|c|c|}
\hline
\multicolumn{1}{|c|}{\multirow{2}{*}{Scenarios}}                               & \multicolumn{1}{c|}{\multirow{2}{*}{Model}} & Num & $R_a$ & \multicolumn{5}{c|}{T1}                                                                                                                    & \multicolumn{3}{c|}{T2}                                                               & PR & \multicolumn{1}{c|}{\multirow{2}{*}{$d_i$}} & \multirow{2}{*}{$v^{\text{NOR}}$}            & \multirow{2}{*}{$v^{\text{MAX}}$}              & \multirow{2}{*}{\begin{tabular}[c]{@{}c@{}}size of \\ scene\end{tabular}} & \multirow{2}{*}{$\epsilon$} & \multicolumn{1}{l|}{\multirow{2}{*}{$\delta$}} & \multicolumn{1}{l|}{\multirow{2}{*}{$\eta$}} & \multicolumn{1}{l|}{\multirow{2}{*}{$\beta$}} \\ \cline{5-12}
\multicolumn{1}{|c|}{}                                                         & \multicolumn{1}{l|}{}                       &                                                                                   &                                                                                  & $\alpha$                      & $\beta$                      & $\gamma$                     & C                              & N                              & $\delta$                        & $\xi$                        & E                              &                                                                                & \multicolumn{1}{c|}{}                    &                                  &                                    &                                                                           &                          & \multicolumn{1}{l|}{}                       & \multicolumn{1}{l|}{}                     & \multicolumn{1}{l|}{}                      \\ \hline
\begin{tabular}[c]{@{}l@{}} Grass\end{tabular}               & Ours                                       & 16                                                                                & 0.3                                                                              & 0.1                    & 0.1                    & 0.15                   & N(0,0.25)                      & N(0,0.25)                      & 0.35                      & 0.1                      & N(0,0.25)                      & 10                                                                             & \multicolumn{1}{c|}{N(0.1,0.01)}         & 2                                & 0.8                                & 230*111                                                                   & -                        & -                                           & -                                         & -                                          \\ \hline
\begin{tabular}[c]{@{}l@{}} Grass\end{tabular}               & \multicolumn{1}{l|}{Durupinar}              & 16                                                                                & 0.3                                                                              & -                      & -                      & -                      & -                              & -                              & -                         & -                        & -                              & 10                                                                             & N(0.1,0.01)                              & 2                                & 0.8                                & 230*111                                                                   & -                        & -                                           & -                                         & -                                          \\ \hline
\begin{tabular}[c]{@{}l@{}} Grass\end{tabular}             & Neto                                        & 16                                                                                & 0.3                                                                              & -                      & -                      & -                      & -                              & -                              & -                         & -                        & -                              & 10                                                                             & \multicolumn{1}{c|}{-}                   & 2                                & 0.8                                & 230*111                                                                   & 0.5                      & 0.5                                         & 0.5                                       & 1                                          \\ \hline
\begin{tabular}[c]{@{}l@{}} Room\end{tabular}                & Ours                                       & 19                                                                                & 0.3                                                                              & 0.1                    & 0.1                    & 0.15                   & \multicolumn{1}{l|}{N(0,0.25)} & \multicolumn{1}{l|}{N(0,0.25)} & \multicolumn{1}{l|}{0.35} & \multicolumn{1}{l|}{0.1} & \multicolumn{1}{l|}{N(0,0.25)} & 10                                                                             & N(0.1,0.01)                              & 2                                & 0.8                                & 25.6*53.5                                                                 & -                        & -                                           & -                                         & -                                          \\ \hline
\begin{tabular}[c]{@{}l@{}} Room\end{tabular}                & \multicolumn{1}{l|}{Durupinar}              & 19                                                                                & 0.3                                                                              & -                      & -                      & -                      & -                              & -                              & -                         & -                        & -                              & 10                                                                             & N(0.1,0.01)                              & 2                                & 0.8                                & \multicolumn{1}{l|}{25.6*53.5}                                            & -                        & -                                           & -                                         & -                                          \\ \hline
\begin{tabular}[c]{@{}l@{}} Room\end{tabular}                & Neto                                        & 19                                                                                & 0.3                                                                              & -                      & -                      & -                      & -                              & -                              & -                         & -                        & -                              & 10                                                                             & \multicolumn{1}{c|}{-}                   & 2                                & 0.8                                & 25.6*53.5                                                                 & 0.5                      & 0.5                                         & 0.5                                       & 1                                          \\ \hline
\begin{tabular}[c]{@{}c@{}} phone\\explosion\end{tabular}                 & Ours                                       & 152                                                                               & 0.3                                                                              & 0.1                    & 0.1                    & 0.15                   & \multicolumn{1}{l|}{N(0,0.25)} & \multicolumn{1}{l|}{N(0,0.25)} & \multicolumn{1}{l|}{0.35} & \multicolumn{1}{l|}{0.1} & \multicolumn{1}{l|}{N(0,0.25)} & 8                                                                              & N(0.1,0.01)                              & 2                                & 0.8                                & 600*600                                                                   & -                        & -                                           & -                                         & -                                          \\ \hline
\begin{tabular}[c]{@{}c@{}} British \\Parliament\\building\end{tabular}   & Ours                                       & 37                                                                                & 0.3                                                                              & 0.1                    & 0.1                    & 0.15                   & \multicolumn{1}{l|}{N(0,0.25)} & \multicolumn{1}{l|}{N(0,0.25)} & 0.35                      & 0.1                      & \multicolumn{1}{l|}{N(0,0.25)} & 15                                                                             & N(0.4,0.01)                              & {2/2.5}       & {0.8/1.2}       & 600*600                                                                   & -                        & -                                           & -                                         & -                                          \\ \hline
\begin{tabular}[c]{@{}c@{}} Virtual\\ scenario\end{tabular} & Ours                                       & 300                                                                               & 0.3                                                                              & 0.1                    & 0.1                    & 0.15                   & \multicolumn{1}{l|}{N(0,0.25)} & \multicolumn{1}{l|}{N(0,0.25)} & \multicolumn{1}{l|}{0.35} & \multicolumn{1}{l|}{0.1} & \multicolumn{1}{l|}{N(0,0.25)} & 10                                                                             & N(0.1,0.01)                              & {{[}2,4.5{]}} & {{[}0.8,1.2{]}} & 600*600                                                                   & -                        & -                                           & -                                         & -                                          \\ \hline
\begin{tabular}[c]{@{}c@{}} Virtual\\ scenario\end{tabular} & \multicolumn{1}{l|}{Cube-Neto}              & {300}                                                          & {0.3}                                                         & {-} &{-} & {-} & {-}         & {-}         & {-}    & {-}   & {-}         & {10}                                                        & -                                        & {{[}2,4.5{]}} & {{[}0.8,1.2{]}} & {600*600}                                              & 0.5                      & 0.5                                         & 0.5                                       & 1                                          \\ \hline
\begin{tabular}[c]{@{}c@{}} Virtual\\ scenario\end{tabular}  & \multicolumn{1}{l|}{Durupinar}              & {300}                                                          & {0.3}                                                         & {-} & {-} & {-} & {-}         & {-}         & {-}    & {-}   & {-}         & {10}                                                        & N(0.1,0.01)                              & {{[}2,4.5{]}} & {{[}0.8,1.2{]}} & {600*600}                                              & -                        & -                                           & -                                         & -                                          \\ \hline
\begin{tabular}[c]{@{}c@{}} Virtual\\ Scenario\end{tabular} & Neto                                        & 300                                                                               & 0.3                                                                              & -                      & -                      & -                      & -                              & -                              & -                         & -                        & -                              & 10                                                                             & \multicolumn{1}{c|}{-}                   & {[}2,4.5{]}                      & {[}0.8,1.2{]}                      & 600*600                                                                   & 0.5                      & 0.5                                         & 0.5                                       & 1                                          \\ \hline
\end{tabular}
\label{2019318}
\end{table*}

%
%
%
%

\subsection{Comparisons} 
\label{comp-real}



To validate our approach, we compare the simulation results obtained by different methods with real-world crowd 
evacuation videos. The trend in the simulation results obtained by our model is that they are more similar to real-world videos than results from other approaches.

\subsubsection {Comparisons with scenarios from public UMN dataset}

Comparisons between real scenes (chosen from the public UMN dataset \cite{83}) and the corresponding simulation results are presented in Figure \ref{fig:1415}. 
We take two different real-world scenarios as examples, and 
detailed results can be seen in the supplementary video. 
Our model is compared with two other representative emotion models: the Durupinar model \cite{11} and the Neto model \cite{88}.
We annotate the trajectories of all the individuals in the 
real-world video using the video annotation tool in \cite{86} and assign initial movement states to these models. 
Therefore, we can predict the trajectories of these individuals and compare 
them with the actual ones.

In the Grass scenario, Individual No. 1 moves faster than Individual No. 2, and Individual No. 1 moves closer to Individual No. 2 
(Figure \ref{fig:1415}a). 
The simulation result obtained by our model in the Grass scenario is more realistic than those obtained by the Durupinar and Neto models 
because the speed is influenced by physical strength consumption in our model. 
If an individual has consumed more physical strength than other individuals, his moving speed decreases and  
other individuals move faster than he does. 
Thus, simulating the situation is easier when one individual gets closer to another individual.

In the Room scenario, some individuals are marked with red circles in the simulation results obtained by the Durupinar and Neto models (Figures \ref{fig:1415}g and \ref{fig:1415}h). 
The moving directions and moving speeds of these individuals are almost the same. 
The simulation result by our model conforms to the real-world video. 
This is because the emotion mechanism of our model changes the moving directions of individuals and drives them to move away from the hazard. 
Meanwhile, the physical strength consumption influences the individual's speed.

We use the entropy metric \cite{67} to evaluate the trajectories of different simulation algorithms on different 
scenarios. 
Entropy metric is used to measure the similarity between real-world data and simulation results. 
A lower value of the entropy metric means a smaller error and better similarity with the real-world data. 
Its calculation method is described as follows. The real-world crowd state is denoted as $({{x}_{1}}\ldots {{x}_{t}})$, 
which includes the positions of all the agents at different timesteps. $({{y}_{1}}\ldots {{y}_{t}})$ is the corresponding calculation result of our model. 
$M$ is the estimated error variance. 
\begin{equation}\label{eq20191120}
\scalebox{1}
{$
M=\frac{1}{t\cdot n}\cdot \sum\limits_{k=0}^{t}{\sum\limits_{j=1}^{n}{({{x}_{k}}[j]-{{y}_{k}}[j])}}{{({{x}_{k}}[j]-{{y}_{k}}[j])}^{T}}
$}
\end{equation}
where $t$ is the total number of timesteps and $n$ is the number of agents in the scenario. 
The Entropy metric is given by:
\begin{equation}\label{eq201911201}
{
e(\mu )=\frac{1}{2}n\log ({{(2\pi e)}^{d}}\det (M))
}
\end{equation}
where $d$ is the dimension of the state of a single agent. In this paper, we mainly discuss the 2D locations of agents. So, $d=2$ in this paper.

\begin{figure}[htbp] \begin{centering} 
  \centering 
  \includegraphics[width=7cm]{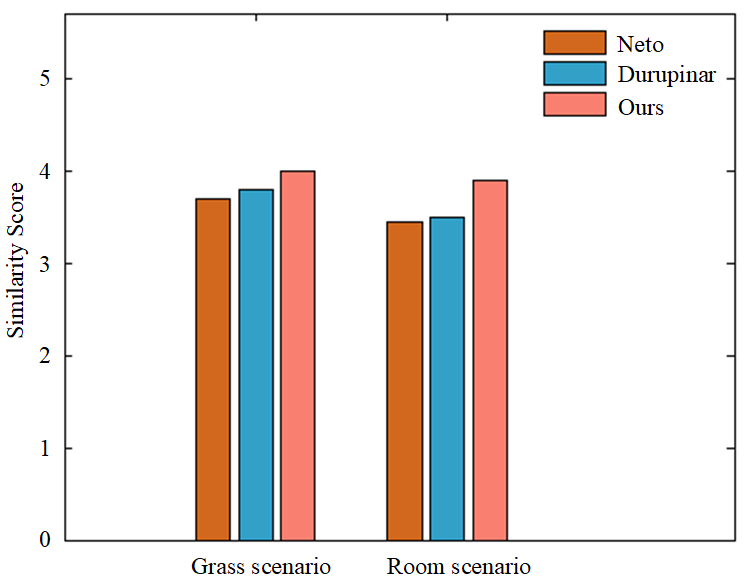}
  \centering
  \caption{ Comparison of similarity scores for movement states
(higher values indicate greater similarity). We compare the movement states in the
original videos with those in crowd simulation results achieved by different algorithms. }
  \label{fig:27}
  \end{centering} 
\end{figure}

For each scenario, a user study is performed. 
There are 39 participants (51.28$\%$ female, 66.67$\%$ in the age group of 20-30) in this study 
and participants are asked to compare the movement states
in the original video clips with the movement states in the crowd simulation results (Figure \ref{fig:27}). 
These similarity scores are computed from the user studies. 
A score of 1 indicates most dissimilar and a score of 5 indicates most similar movement. 
Higher values indicate greater similarity. 
We also calculate average spatial distance between the simulated results and the ground truth over all the timesteps and individuals.
Tables \ref{Entropy metric and spatial distance Grass and Room}
and Figure \ref{fig:27} show that 
the simulated moving trends of our model are closer to those in the real-world videos than the results of other models. 
A rational approach is to combine physical strength consumption and panic to determine the movement of each individual.


\begin{table}[]
\setlength{\belowcaptionskip}{10pt}
\renewcommand{\arraystretch}{1.3}
\caption{Entropy metric and spatial distance for different simulation algorithms on scenarios of Grass and Room. A lower value implies 
higher similarity with respect to the real-world crowd videos.   }
\label{Entropy metric and spatial distance Grass and Room}
\centering
\begin{tabular}{|c|l|c|c|}
\hline
Scenario               & \multicolumn{1}{c|}{Model} & Entropy metric & Spatial distance \\ \hline
\multirow{3}{*}{Grass} & Ours                       & 3.386800       & 1.042990         \\ \cline{2-4} 
                       & Durupinar                  & 3.429000       & 1.105531         \\ \cline{2-4} 
                       & Neto                       & 3.409700       & 1.046776         \\ \hline
\multirow{3}{*}{Room}  & Ours                       & 5.393900       & 1.481327         \\ \cline{2-4} 
                       & Durupinar                  & 5.463200       & 1.484492         \\ \cline{2-4} 
                       & Neto                       & 5.493300       & 1.527570         \\ \hline
\end{tabular}
\end{table}

%

\begin{figure}[htbp]
\centering
\begin{tabular}{cc}
 \includegraphics[width=3.5cm]{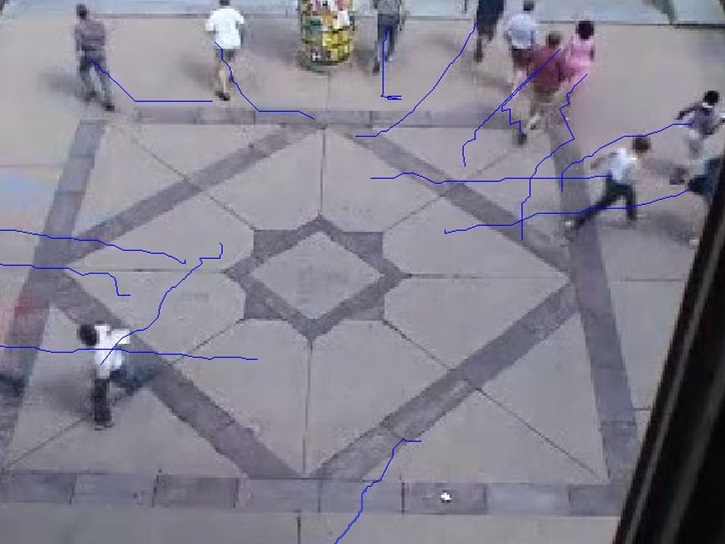} &  \includegraphics[width=3.5cm]{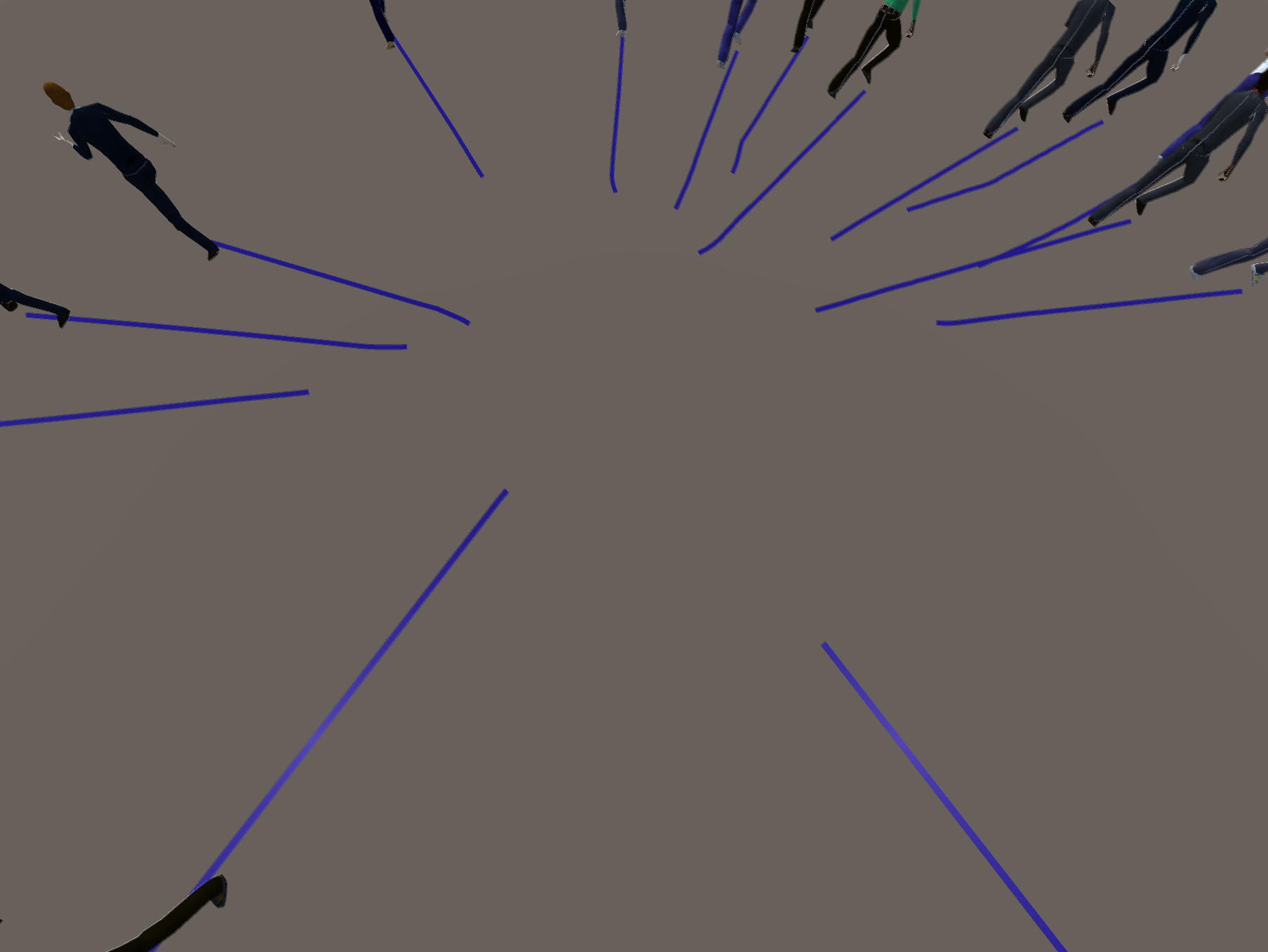}    \\
 (a)  & (b)    \\
 \includegraphics[width=3.5cm]{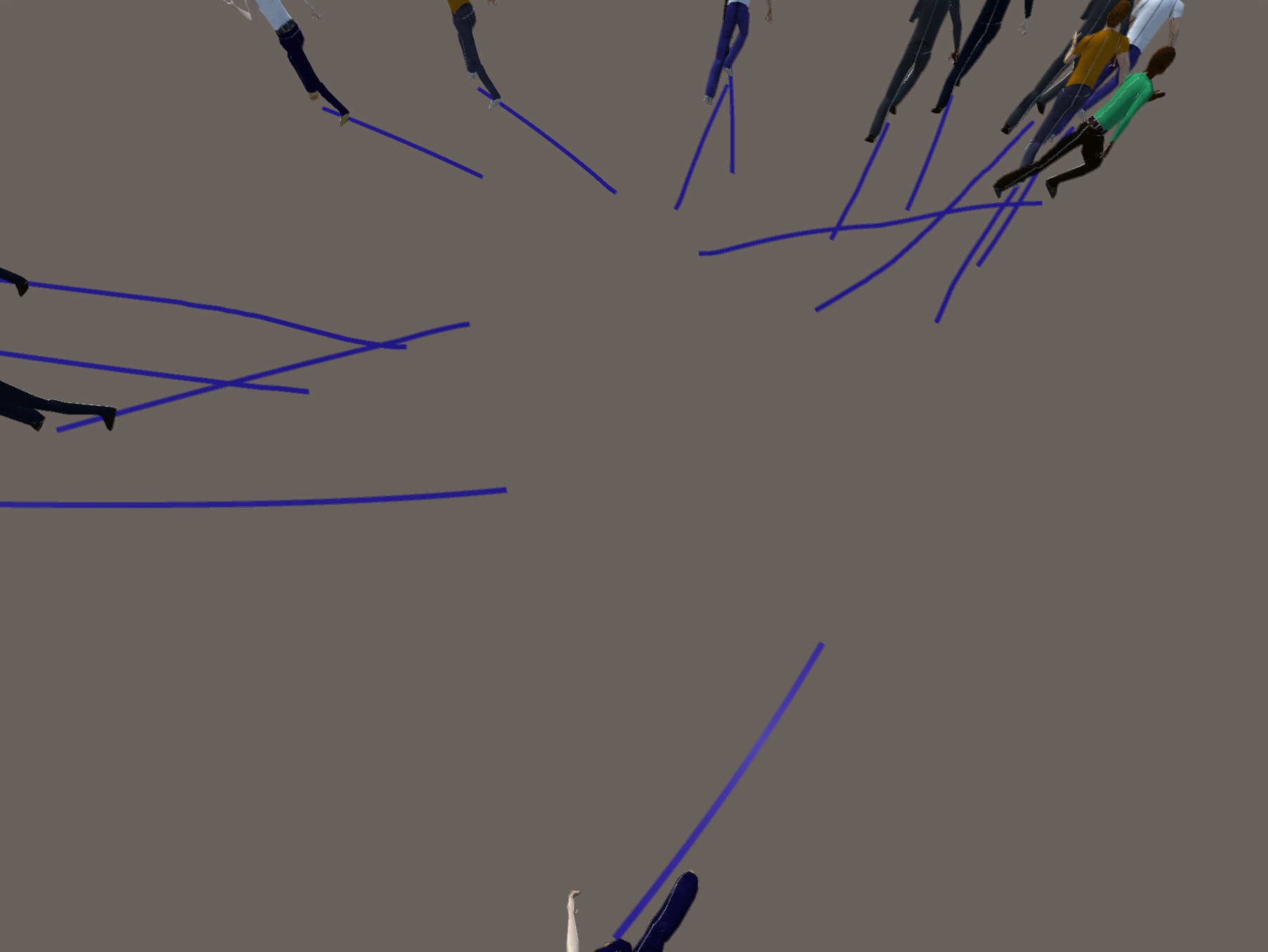} & \includegraphics[width=3.5cm]{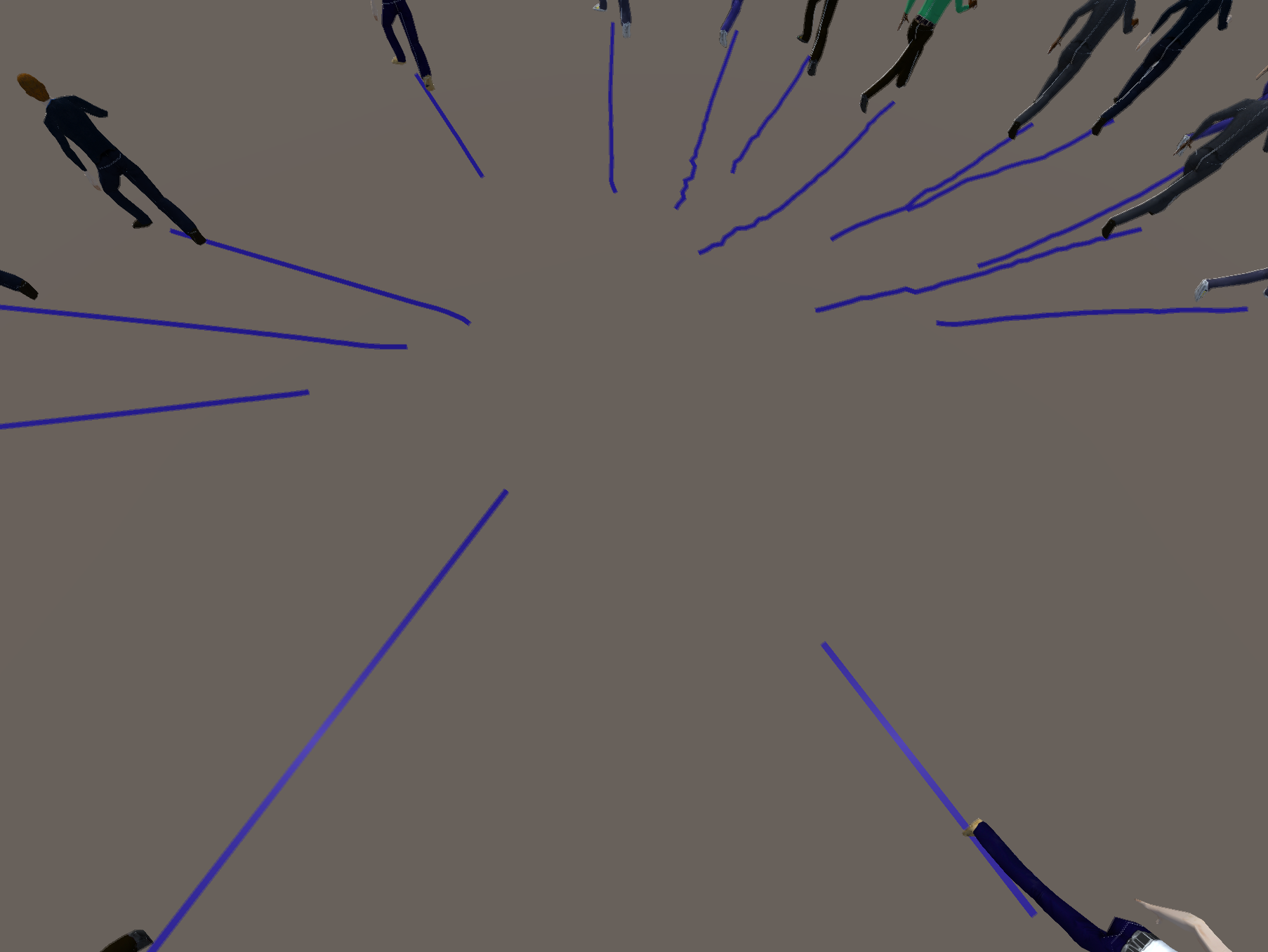}\\
 (c) & (d) \\
 \includegraphics[width=3.5cm]{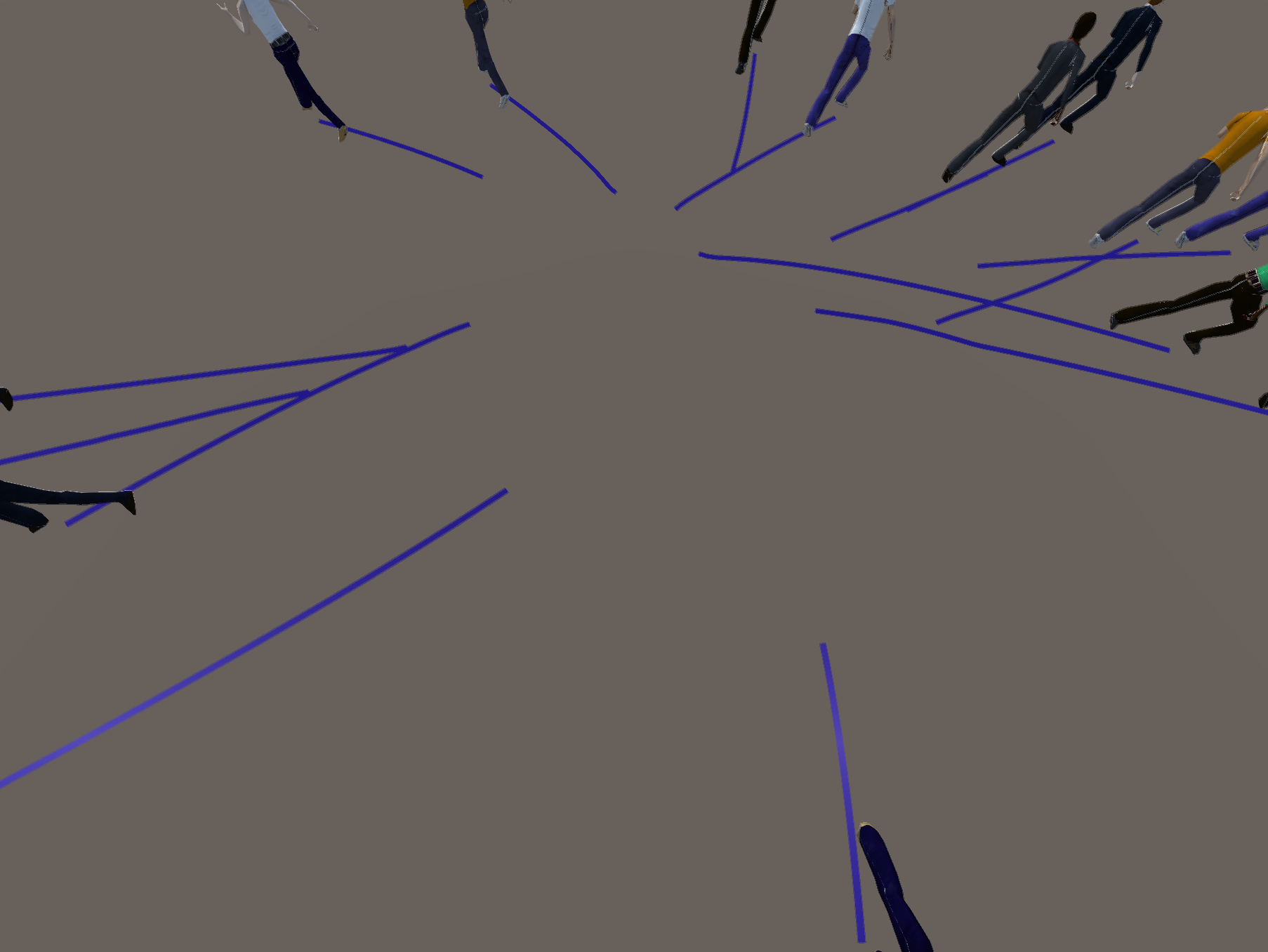} &  \includegraphics[width=3.5cm]{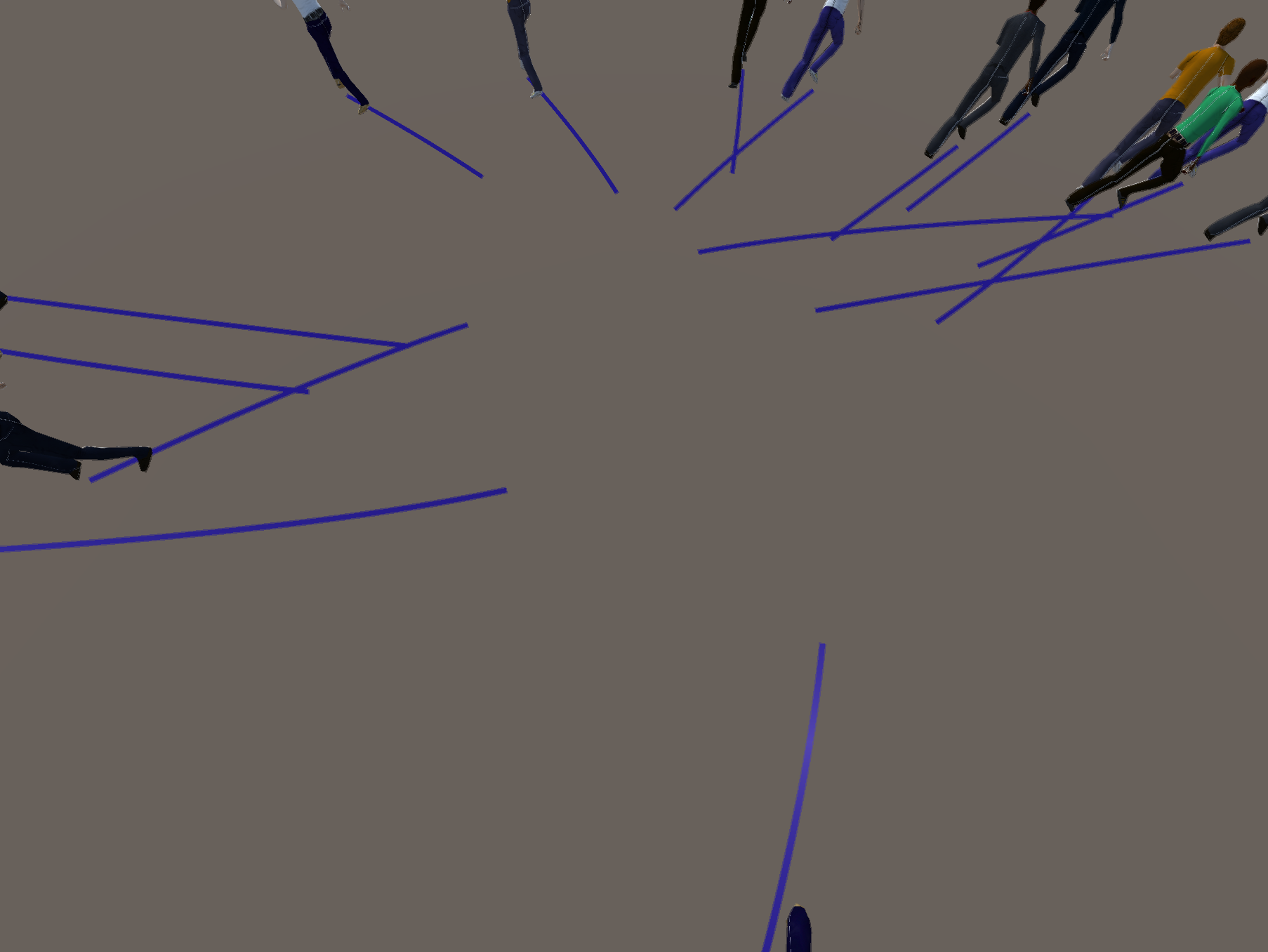} \\
 (e) & (f)  \\
\end{tabular} 
\caption{
Crowd simulation results generated by different models in the Square scenario at the $98th$ frame: 
(a) Real-world scenario, (b) Neto model, (c) Neto-PS model, (d) Durupinar model, (e) Durupinar-PS model, and (f) our model.
} 
\label{fig2019052101} 
\end{figure} 

\begin{figure}[htbp]
\centering
\begin{tabular}{cc}
 \includegraphics[width=3.5cm]{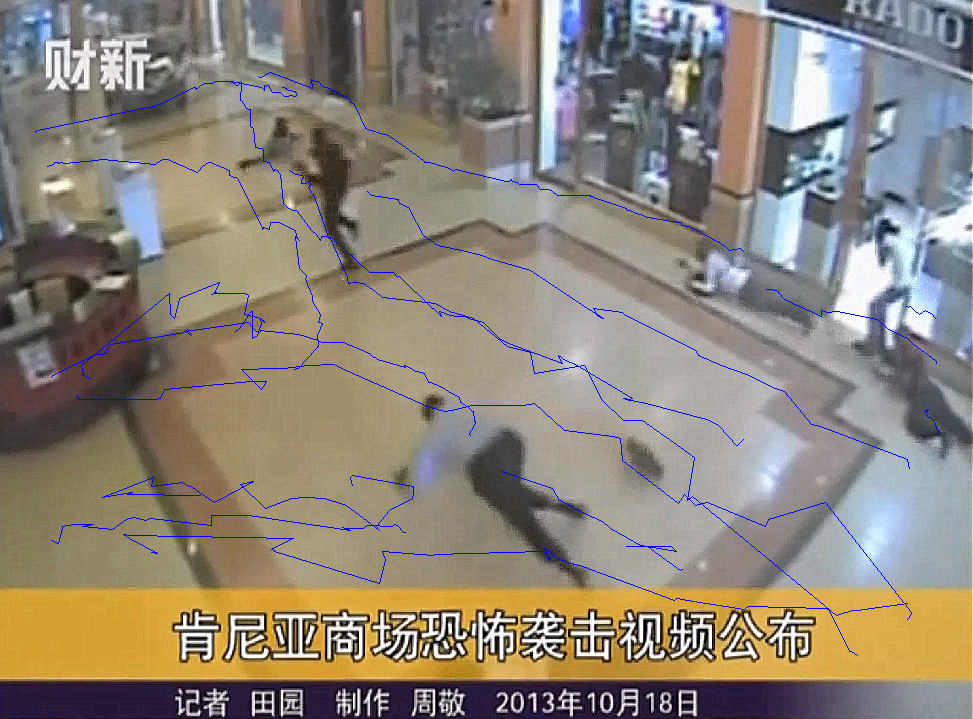} &  \includegraphics[width=3.5cm]{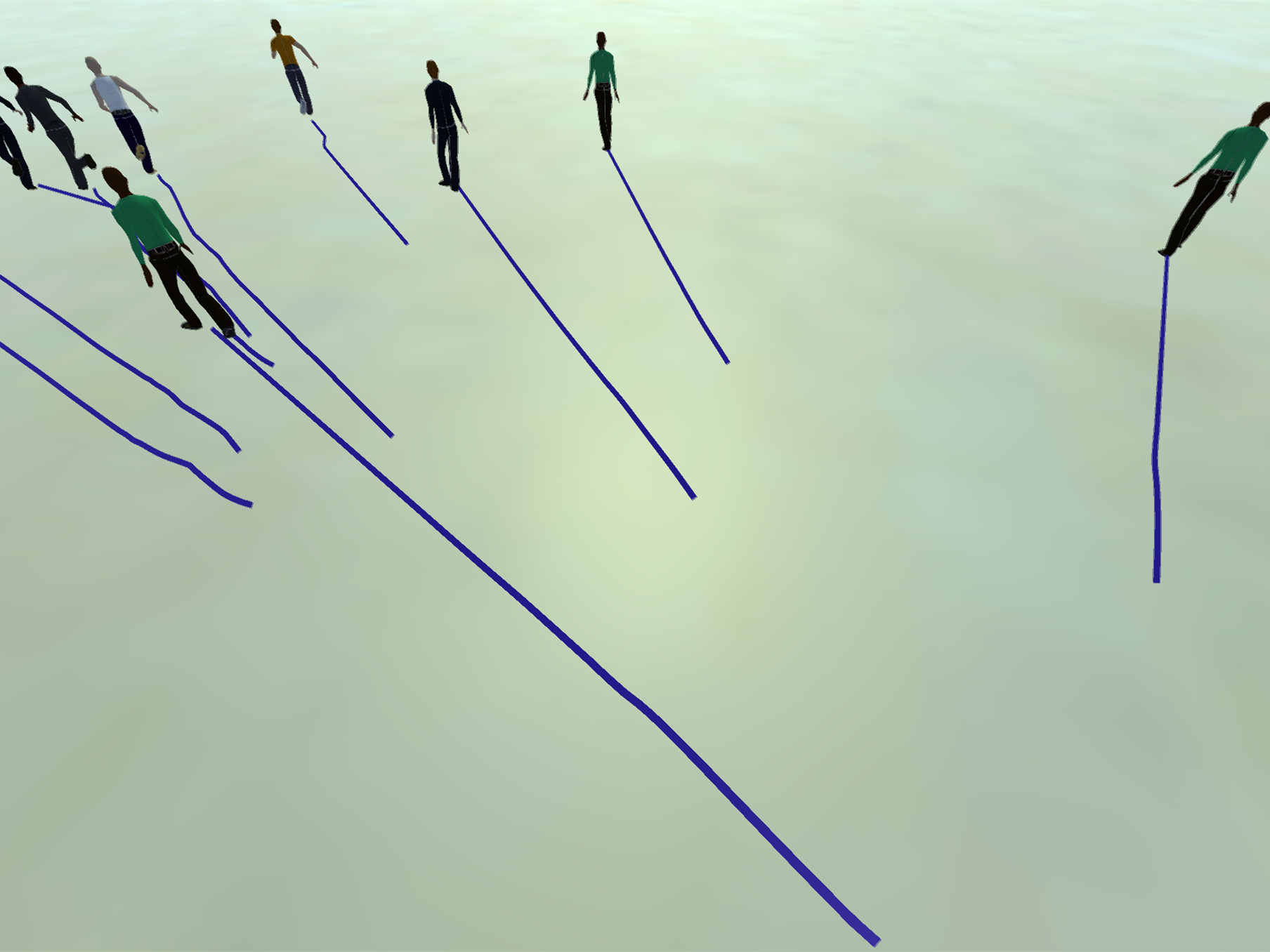}    \\
 (a)  & (b)    \\
 \includegraphics[width=3.5cm]{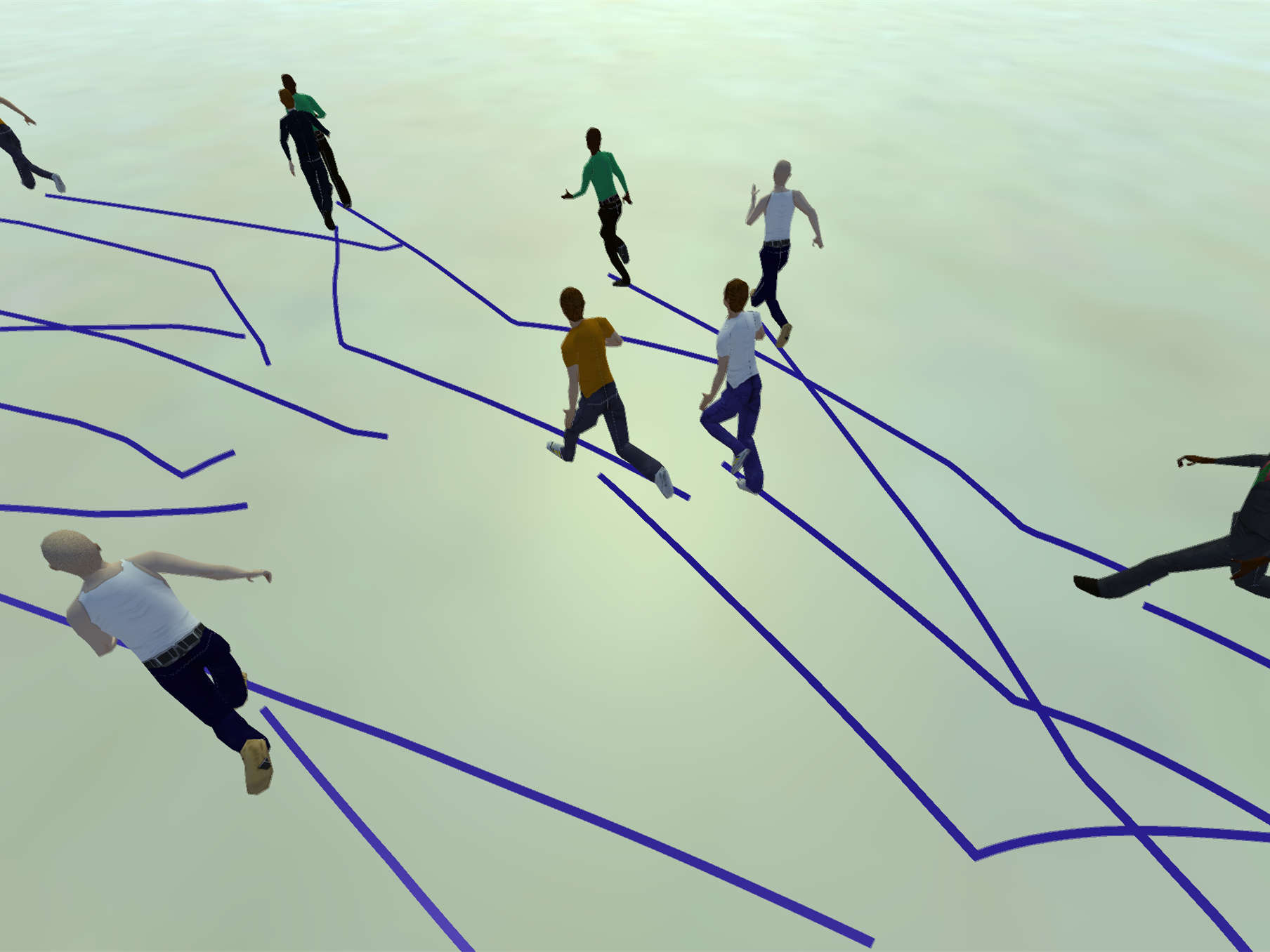} & \includegraphics[width=3.5cm]{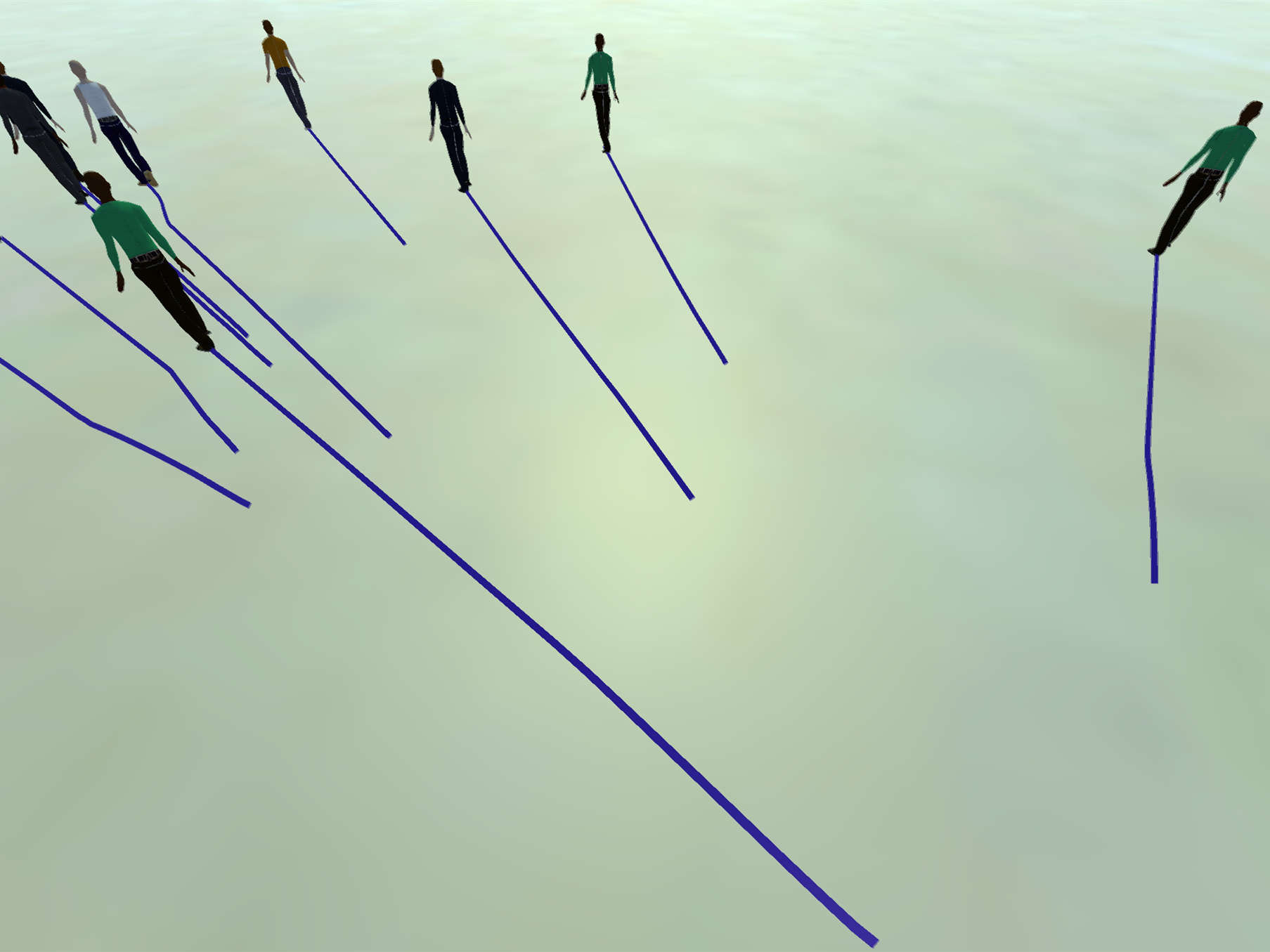}\\
 (c) & (d) \\
 \includegraphics[width=3.5cm]{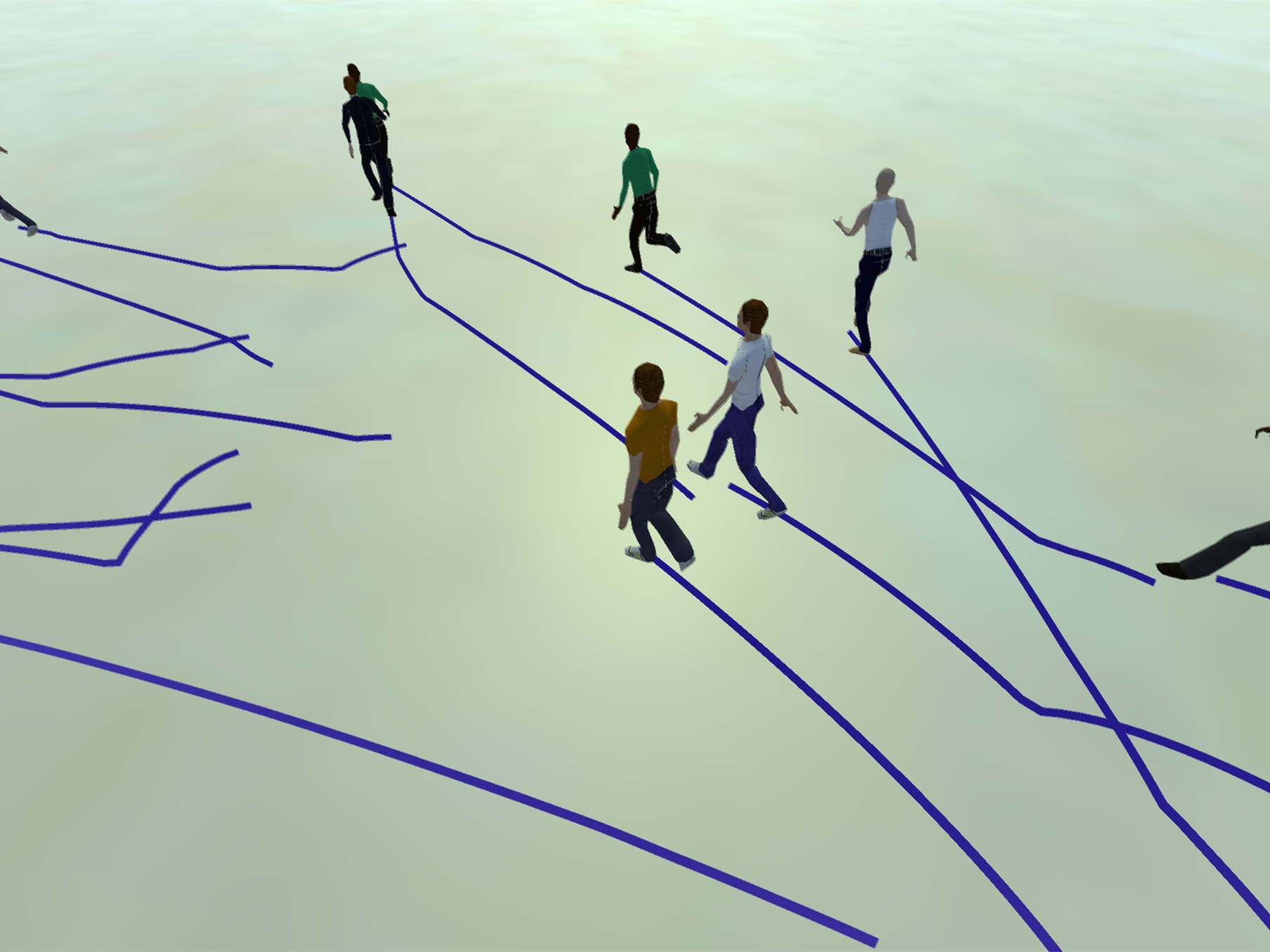} &  \includegraphics[width=3.5cm]{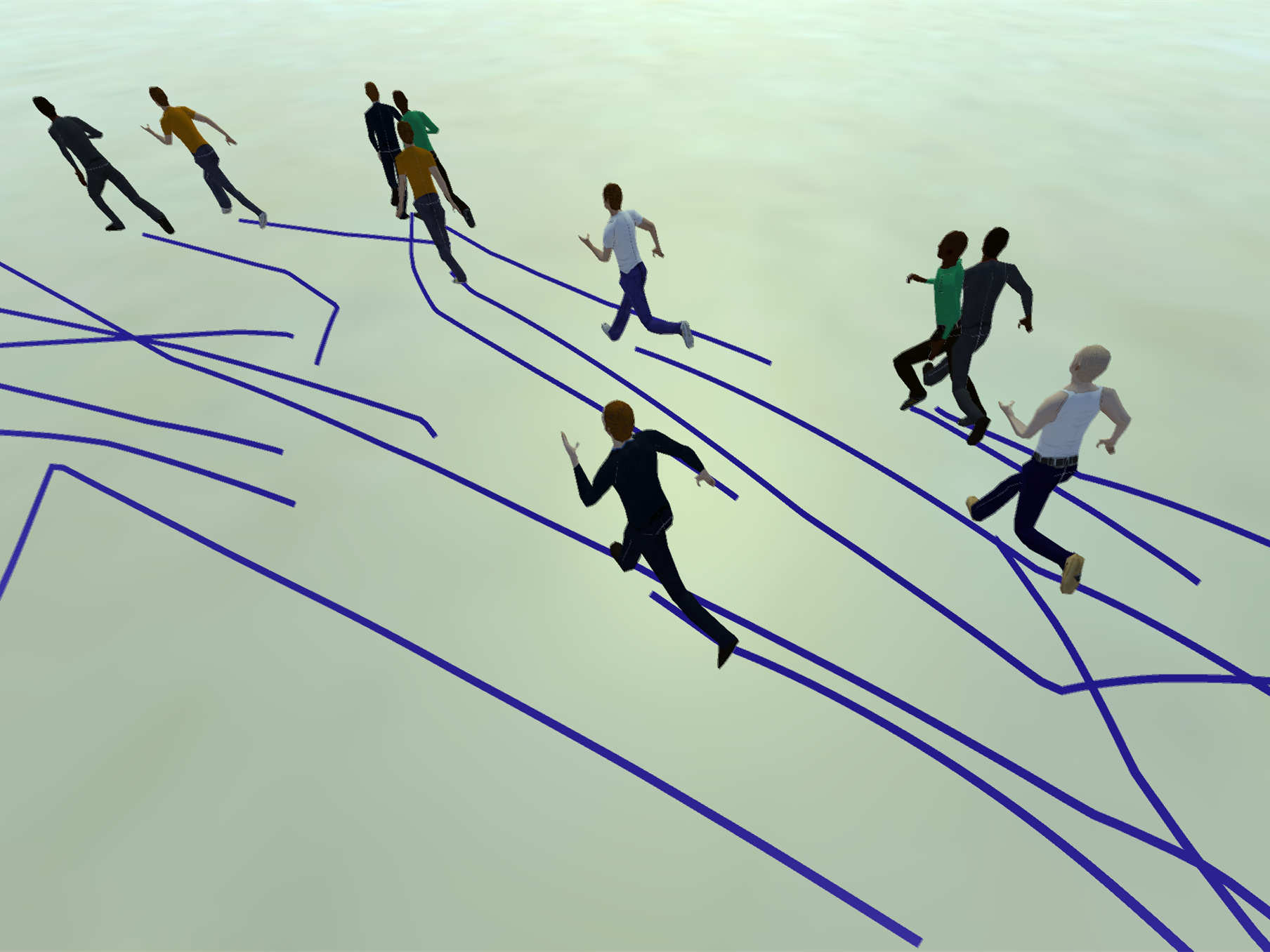} \\
 (e) & (f)  \\
\end{tabular} 
\caption{
Crowd simulation results by different models in the scenario of terrorist attacks on Kenya's shopping mall at the $214th$ frame: 
(a) Real-world scenario, (b) Neto model, (c) Neto-PS model, (d) Durupinar model, (e) Durupinar-PS model, and (f) our model.
} 
\label{fig2019060501} 
\end{figure} 

\begin{figure}[htbp] \subfigure[]
{\centerline{\includegraphics[width=6.2cm]{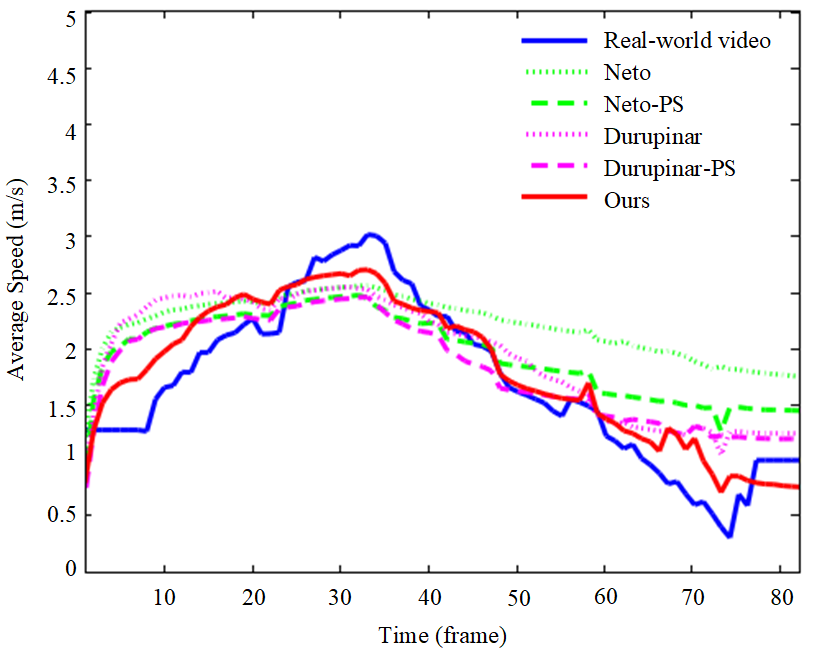}}} 
\subfigure[]
{\centerline{\includegraphics[width=6.2cm]{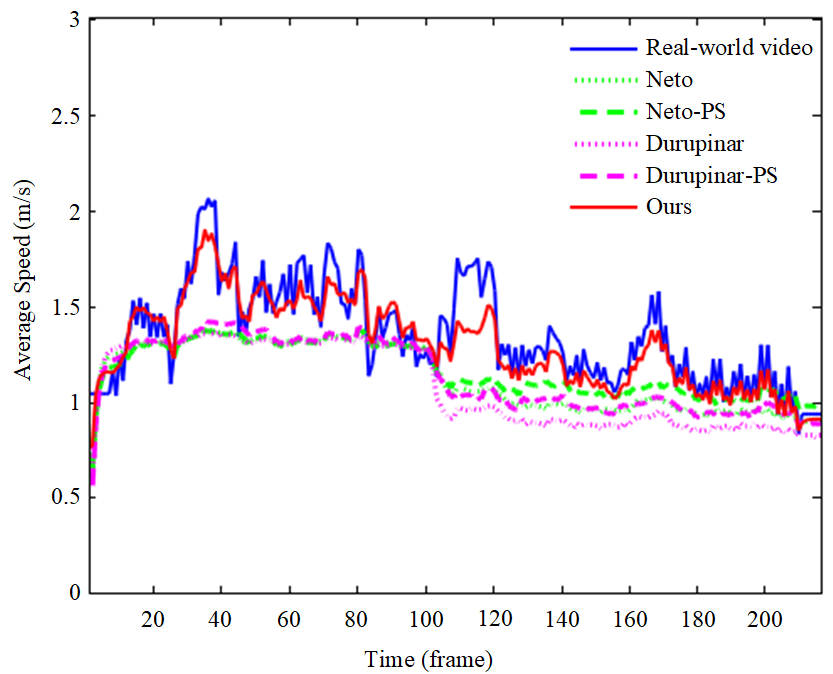}}} 
\caption{The average speeds of all the individuals at different timesteps for these different models in (a) the Square scenario and (b) 
the scenario of terrorist attacks on Kenya's shopping mall. 
The average speeds of our model at different timesteps are closer to the real scene than those of other models. 
The average speeds of the Neto-PS and Durupinar-PS models at different timesteps are closer to the real-world scenarios, than 
those of the original Neto and Durupinar models. 
 } 
\label{fig2019052102} 
\end{figure} 

\subsubsection{Comparisons with real-world emergency scenarios}
In this subsection, we compare the simulation results obtained by different methods with real-world videos, 
particularly real emergency incidents.

The Durupinar and Neto models don't consider physical strength consumption. The mechanism of physical strength consumption 
in this paper is integrated into these two models, which are denoted as Durupinar-PS and Neto-PS. In the real-world Square scenario, 
we compare our simulation results with the Durupinar, Durupinar-PS, Neto, and Neto-PS models (Figure \ref{fig2019052101}). 
We also compare our results in a real-world emergency scenario, which is related to terrorist attacks on Kenya's shopping mall (Figure \ref{fig2019060501}).
More details can be seen in the supplementary video. 
Figure \ref{fig2019052102} shows the average speeds of all the individuals at different timesteps for these different models. We find that 
the simulation results of the Durupinar-PS and Neto-PS are closer to the movements and behaviors in the real videos, than those of the original Durupinar and Neto models. 
These comparisons validate that our proposed mechanism of physical strength consumption (physical strength consumption calculation and 
the effect of physical strength consumption on emotion) can enhance the performance of existing crowd simulation models that are only based on emotion. 
Our model describes emotional changes in a comprehensive manner. Based on the James-Lange theory, we describe 
three stages individual emotions undergo (Section 3.3.2) 
and combine emotional contagion with the effect of physical strength consumption on panic. 
The comparisons with Durupinar-PS and Neto-PS models show that our model integrating emotion and physical strength consumption is 
better than other emotion-based crowd simulation models.


One piece of real-world video including both crowd and vehicles is chosen to simulate by our method. In this real scenario, 
we compare our simulation result with the Durupinar, Durupinar-PS, Neto, and Neto-PS models. 
In this paper, we mainly focus on emotions of crowds, especially panic emotions in emergencies. 
In particular, the drivers can express their panic through vehicles.
Moreover, in emergencies, the drivers may not follow the traffic rules.
In emergency scenarios, some behaviors of vehicles, such as sudden acceleration, intuitively demonstrate the drivers' panic. These behaviors can also 
cause the surrounding pedestrians to be panicked and thus act indirectly as emotional contagions. In common cases, 
when pedestrians are walking in front of vehicles, vehicles will slow down, change direction, or stop to avoid pedestrians. 
Vehicles and surrounding pedestrians influence each other in such traffic. Based on above analysis, in our model we 
treat vehicles as one kind of special large-sized agents with full physical strength and high moving speed. The radius of the 
vehicles is set to 3. 
The comparisons show that our simulation result conforms to the real-world video, and can 
enhance the performance of existing crowd simulations under the complex scenarios including both crowd and vehicles. 
More details can be seen in the supplementary video.

Table \ref{Entropy metric and spatial distance for three different scenarios} shows the values of entropy metric and spatial distance for 
the above three scenarios. Comparing with the Durupinar, Durupinar-PS, Neto, Neto-PS models, our model can generated more similar simulation results 
with real-world scenarios.

\begin{figure}[htbp]
\centering
\begin{tabular}{cc}
 \includegraphics[width=4cm]{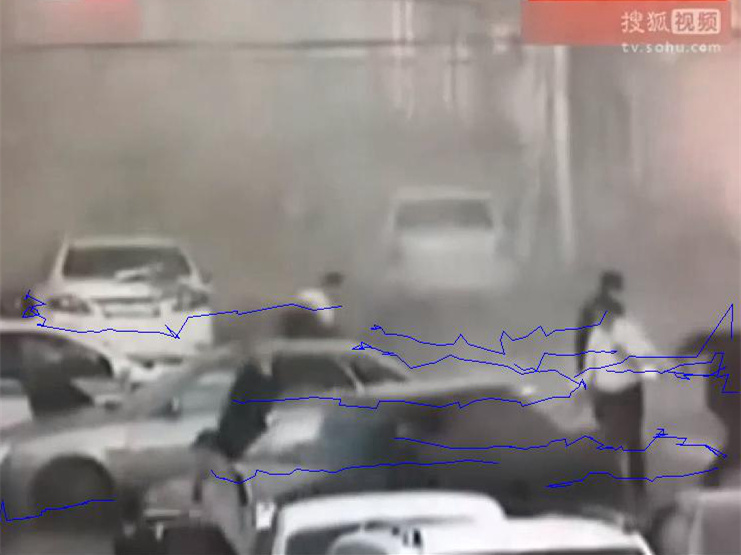} &  \includegraphics[width=4cm]{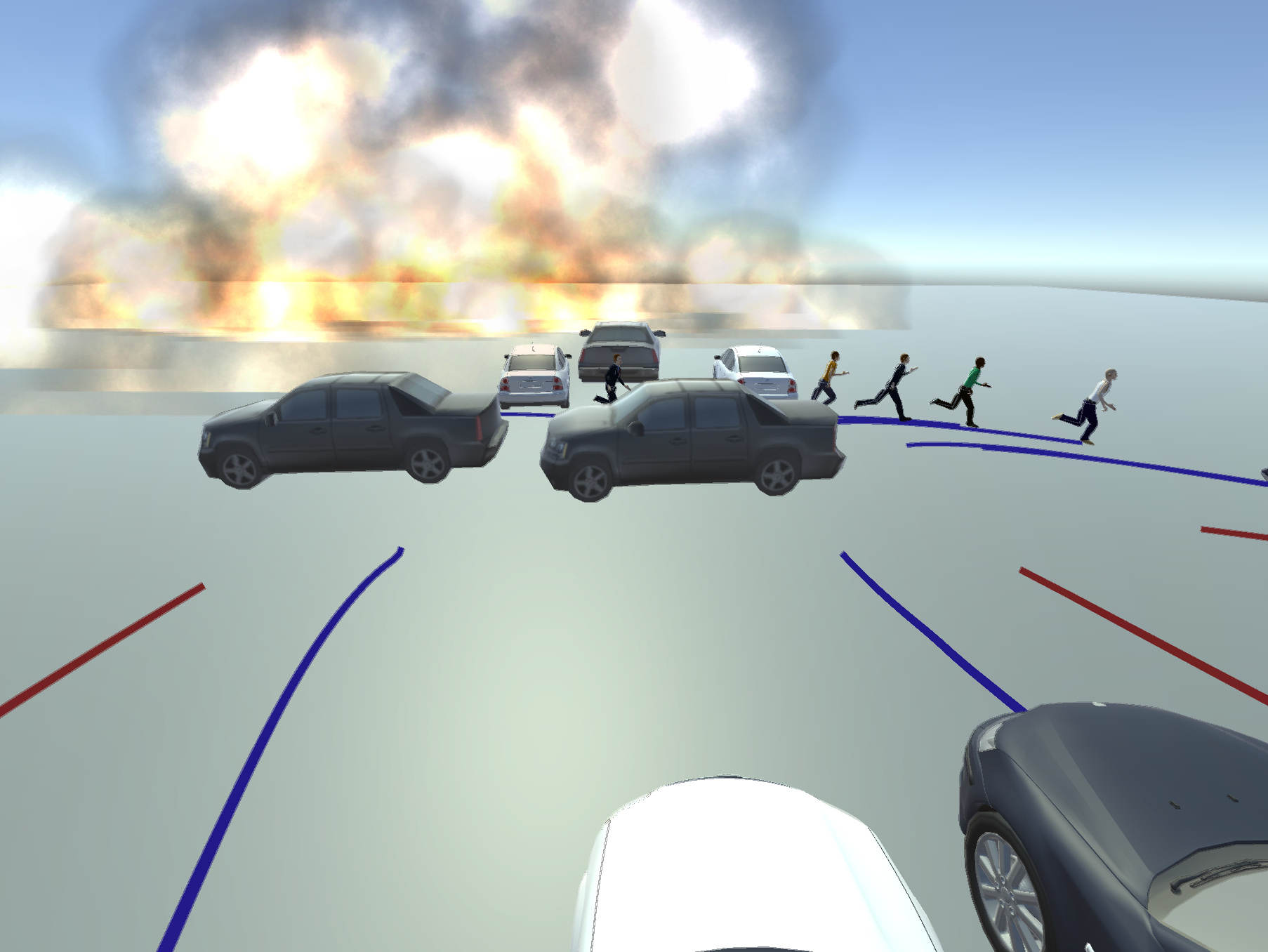}    \\
 (a)  & (b)    \\
 \includegraphics[width=4cm]{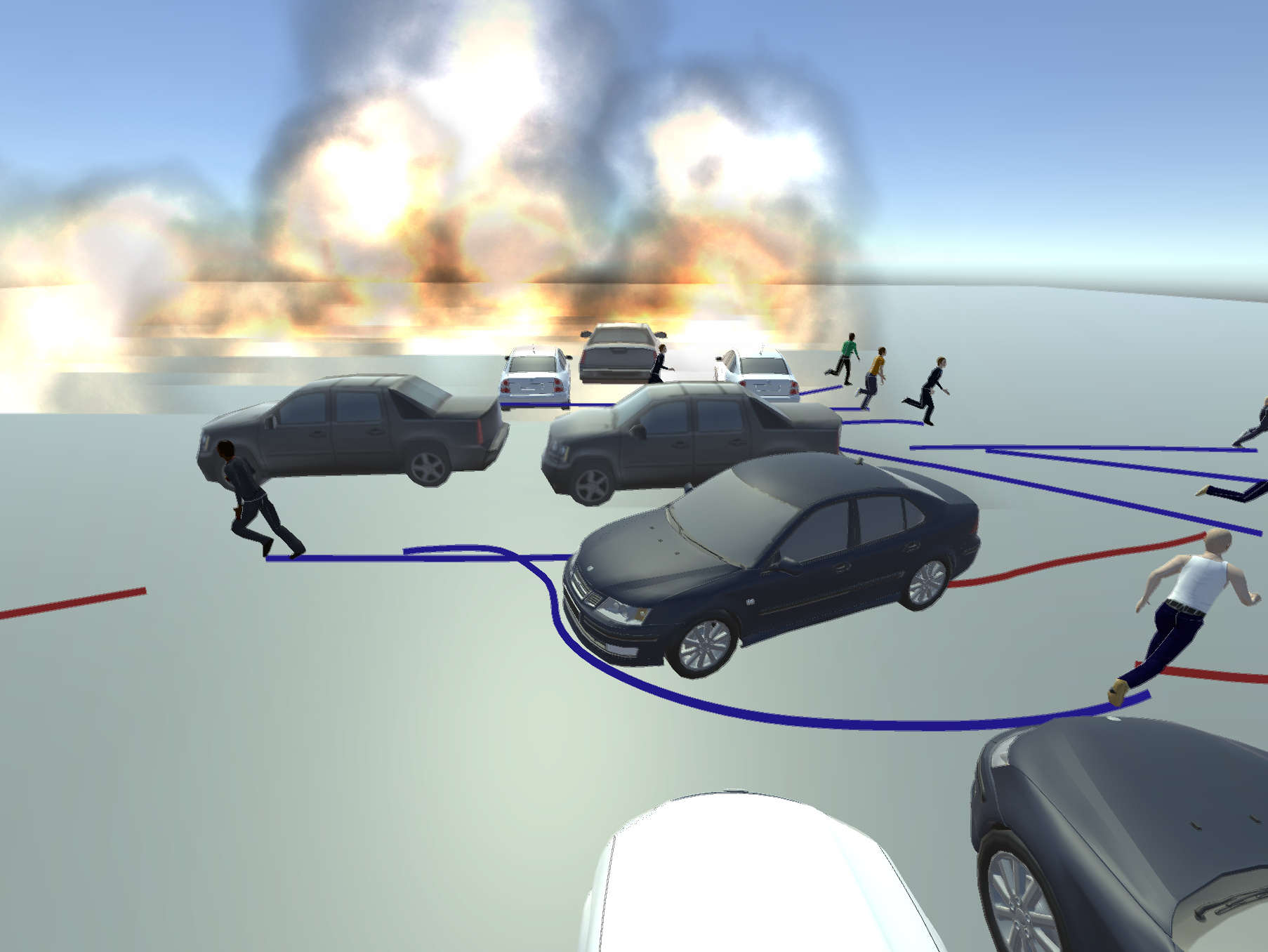} & \includegraphics[width=4cm]{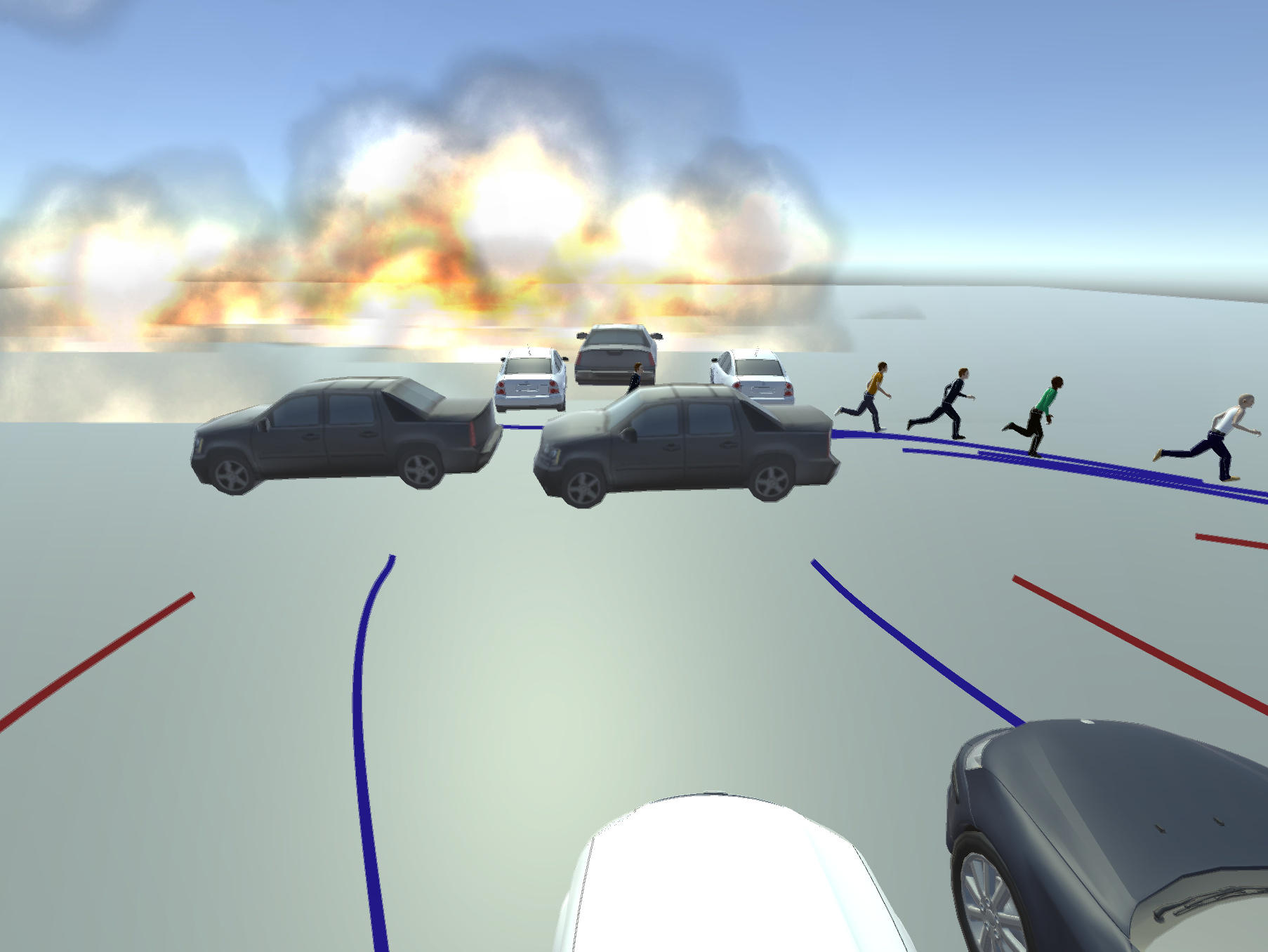}\\
 (c) & (d) \\
 \includegraphics[width=4cm]{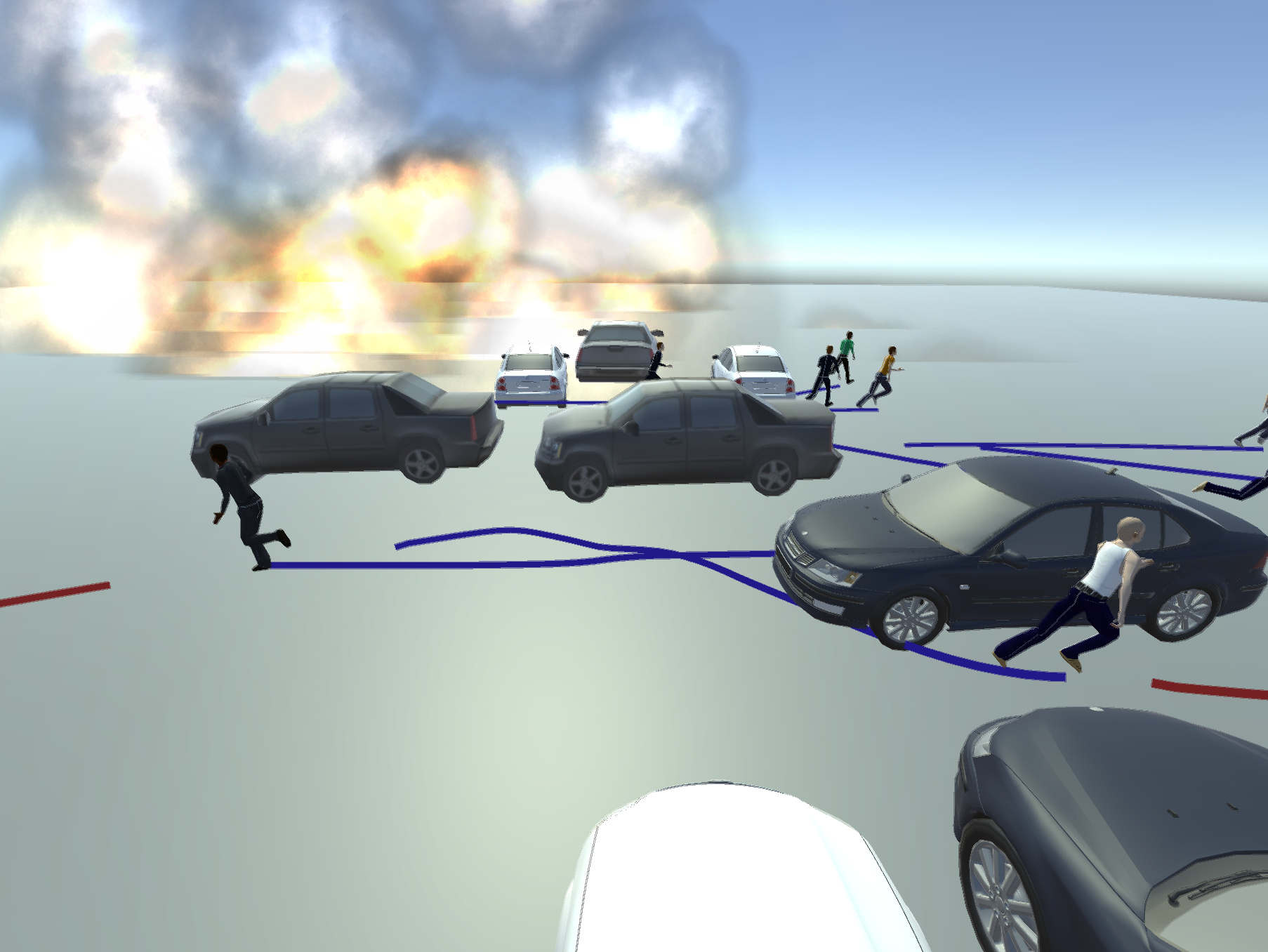} &  \includegraphics[width=4cm]{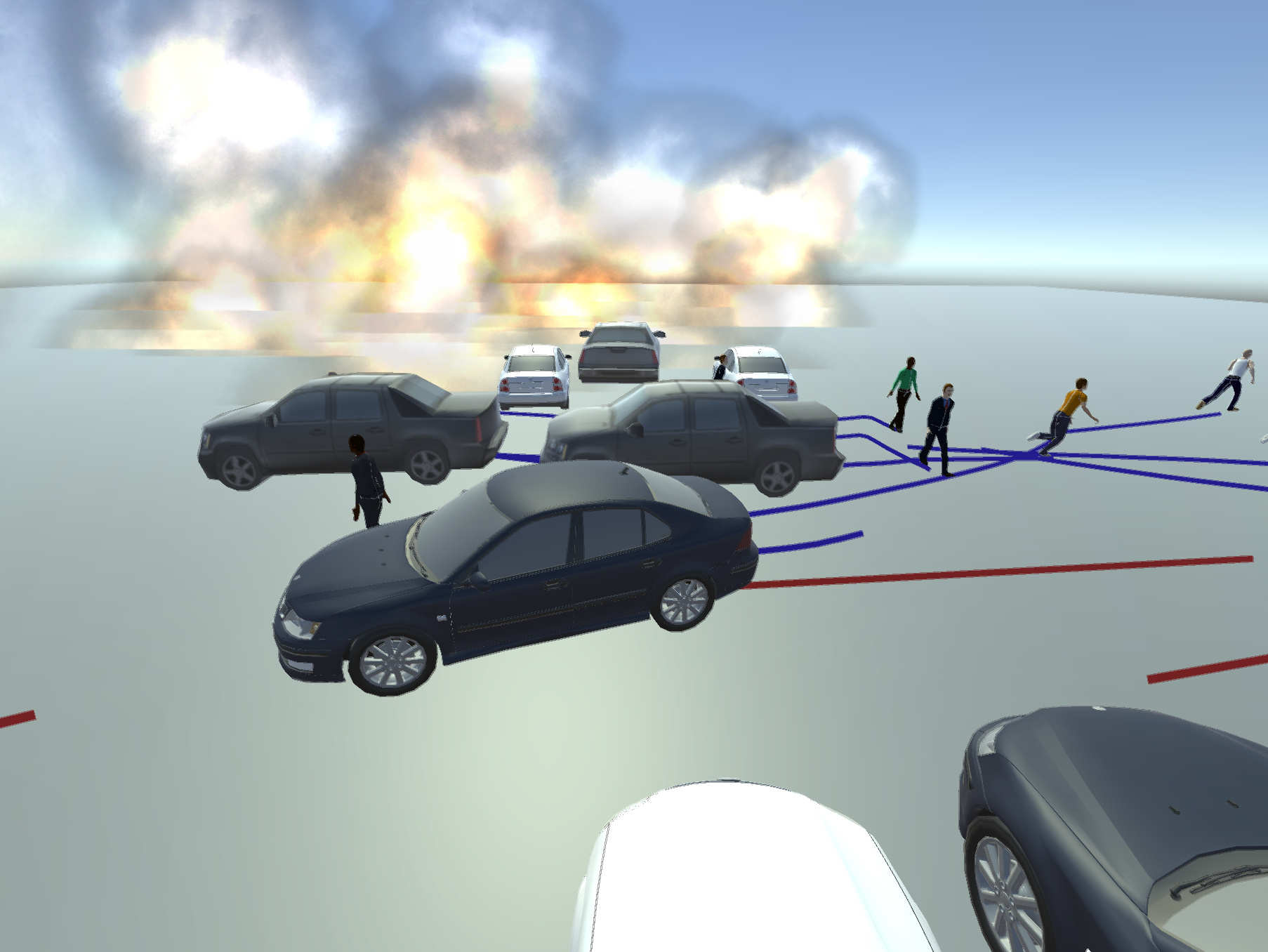} \\
 (e) & (f)  \\
\end{tabular} 
\caption{
Crowd simulation results by different models in the scenario including both crowd and vehicles at the $88th$ frame: 
(a) Real-world scenario, (b) Neto model, (c) Neto-PS model, (d) Durupinar model, (e) Durupinar-PS model, and (f) our model.
} 
\label{fig2019070601} 
\end{figure}

\begin{table}[htbp]
\scriptsize
\setlength{\belowcaptionskip}{10pt}
\renewcommand{\arraystretch}{1.7}
\caption{Entropy metric and spatial distance for different simulation algorithms on scenarios of Square, Crowd and vehicles, and Terrorist attacks on shopping mall. 
A lower value implies higher similarity with respect to the real-world data.   }
\label{Entropy metric and spatial distance for three different scenarios}
\begin{tabular}{|c|l|c|c|}
\hline
Scenario                                                                                       & \multicolumn{1}{c|}{Model} & Entropy metric & Spatial distance \\ \hline
\multirow{5}{*}{Square}                                                                        & Ours                       & 1.278777       & 0.210871         \\ \cline{2-4} 
                                                                                               & Durupinar                  & 3.292511       & 0.316492         \\ \cline{2-4} 
                                                                                               & Durupinar-PS               & 1.883470       & 0.238043         \\ \cline{2-4} 
                                                                                               & Neto                       & 3.415817       & 0.327066         \\ \cline{2-4} 
                                                                                               & Neto-PS                    & 1.568437       & 0.220118         \\ \hline
\multirow{5}{*}{Crowd and vehicles}                                                            & Ours                       & 3.740438       & 0.816685         \\ \cline{2-4} 
                                                                                               & Durupinar                  & 5.851273       & 2.402975         \\ \cline{2-4} 
                                                                                               & Durupinar-PS               & 5.785572       & 2.067541         \\ \cline{2-4} 
                                                                                               & Neto                       & 5.984731       & 2.346846         \\ \cline{2-4} 
                                                                                               & Neto-PS                    & 5.768604       & 2.070267         \\ \hline
\multirow{5}{*}{\begin{tabular}[c]{@{}c@{}}Terrorist attacks \\ on shopping mall\end{tabular}} & Ours                       & 1.060493       & 0.225969         \\ \cline{2-4} 
                                                                                               & Durupinar                  & 5.782041       & 0.761416         \\ \cline{2-4} 
                                                                                               & Durupinar-PS               & 3.809553       & 0.405739         \\ \cline{2-4} 
                                                                                               & Neto                       & 5.783371       & 0.755286         \\ \cline{2-4} 
                                                                                               & Neto-PS                    & 3.230872       & 0.382926         \\ \hline
\end{tabular}
\end{table}


\begin{figure}[htbp]
\centering
\begin{tabular}{cc}
 \includegraphics[width=4cm,height=2.5cm]{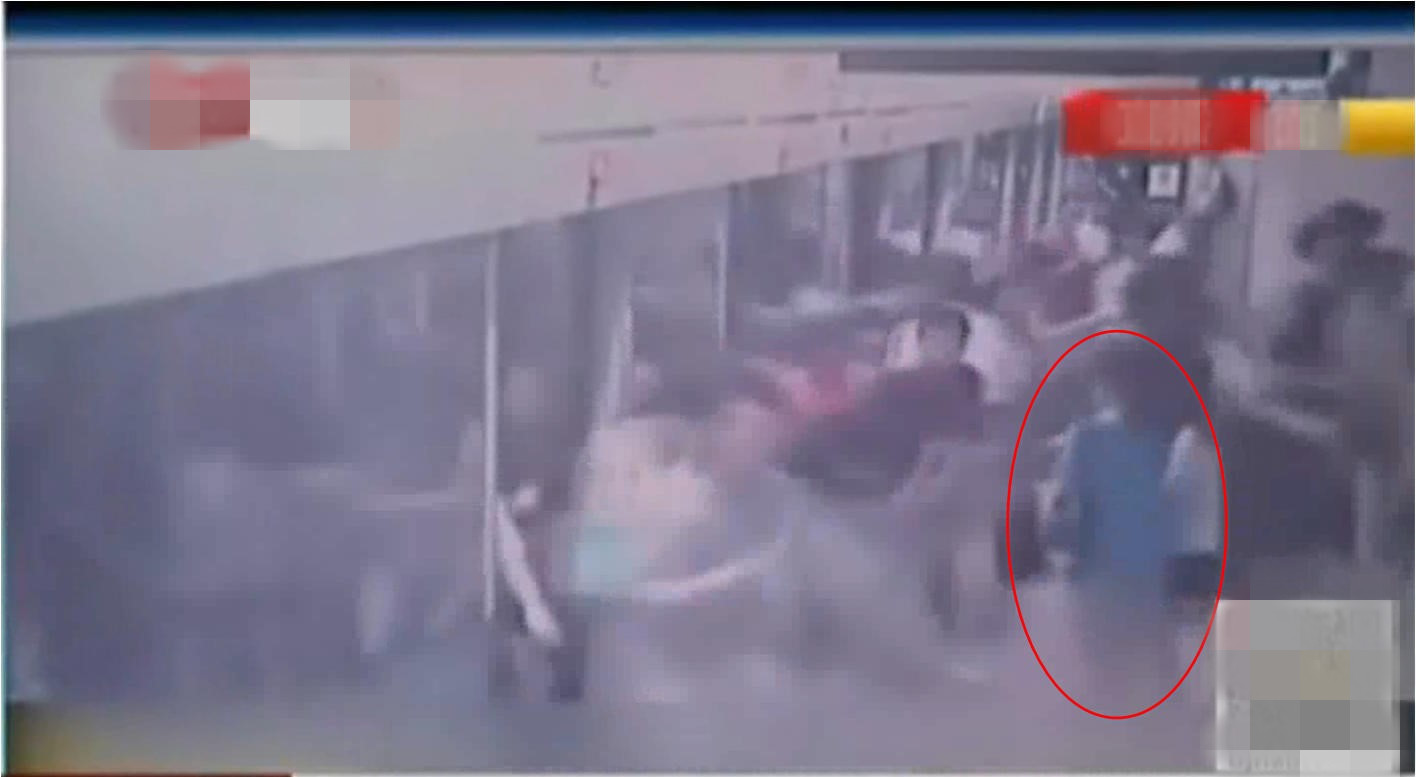} &  \includegraphics[width=4cm,height=2.5cm]{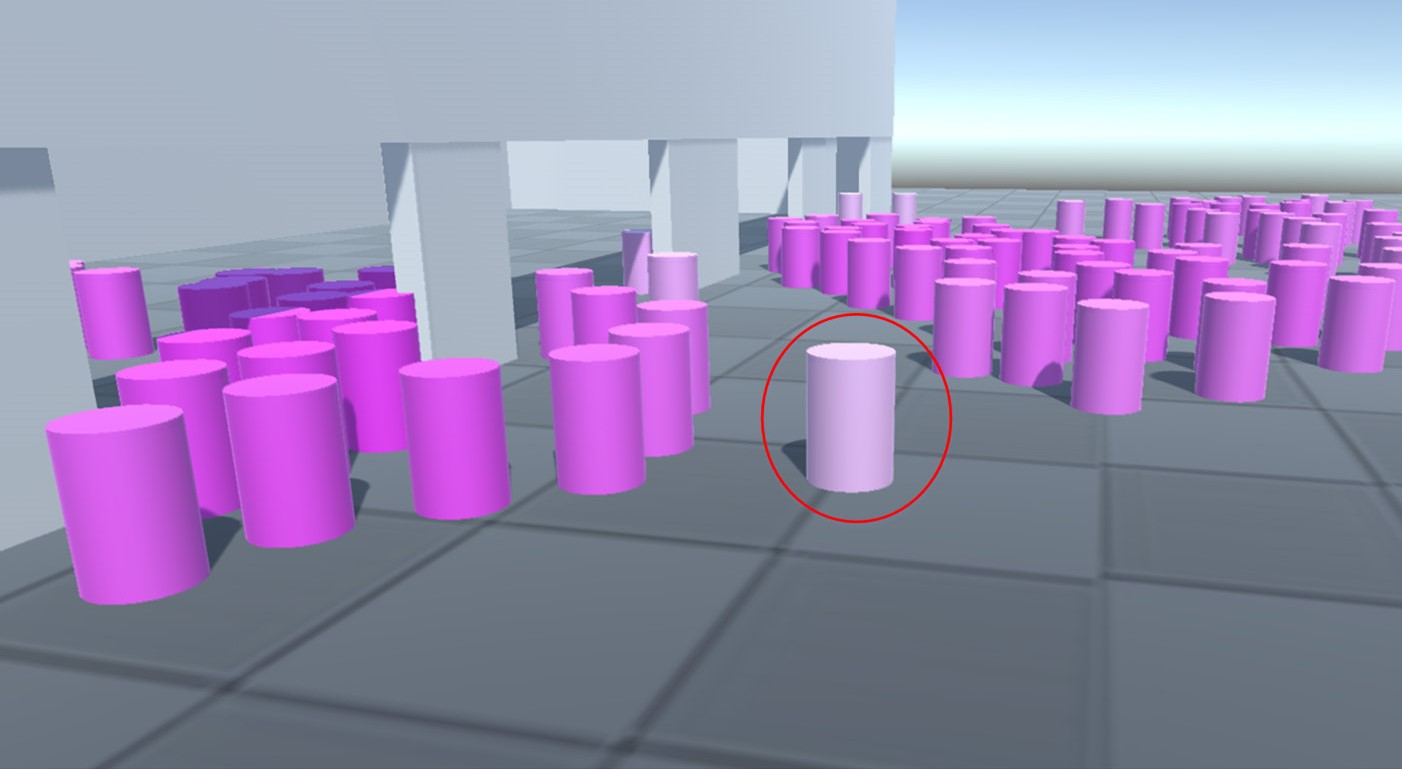} \\
 (a)   & (b)   \\
  \includegraphics[width=4cm,height=2.5cm]{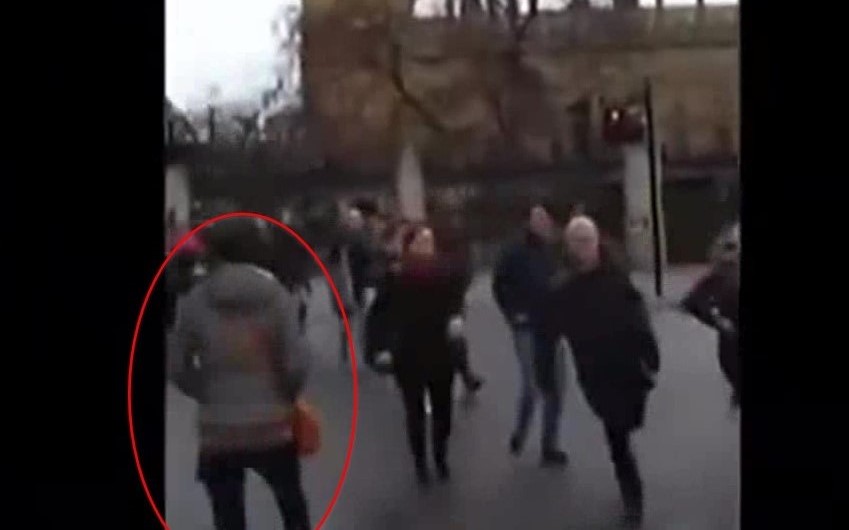} &  \includegraphics[width=4cm,height=2.5cm]{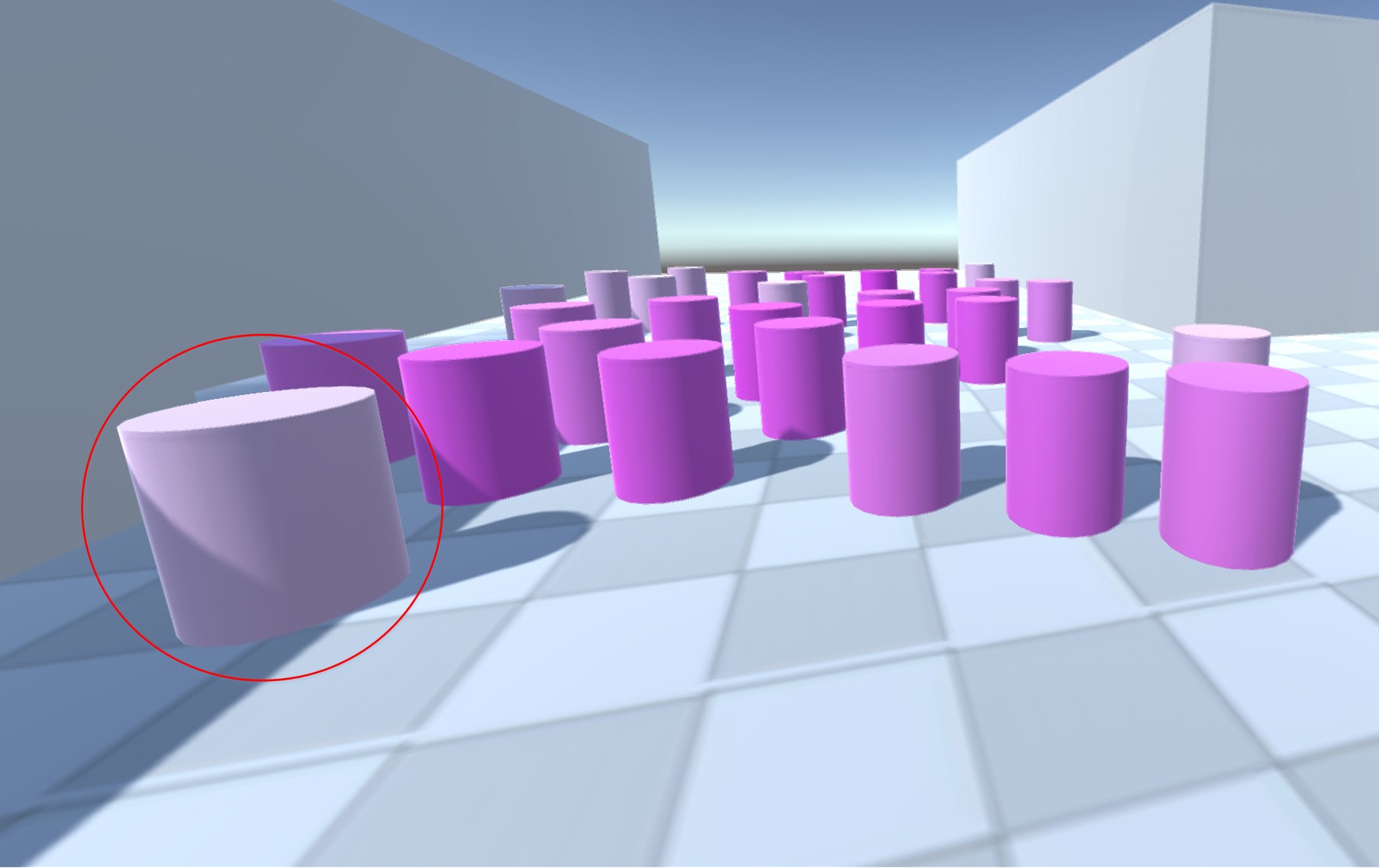} \\
 (c)   & (d)   \\
\end{tabular} 
\caption{
Comparisons between real-world videos and simulation results by our approach. 
(a) The mobile phone explosion incident on the subway in the Shanghai Metro Line 8; 
(c) the shooting incident at the British Parliament building on March 22, 2017; 
(b,d) our corresponding simulation results. 
At first, the individuals in the red circle aren't panicked. Because of emotional contagion, they are then 
influenced by the panicked crowd around them, become panicked, and run away from the hazards.
} 
\label{fig:25} 
\end{figure}

\begin{figure}[htbp] \begin{centering} 
  \centering 
  \includegraphics[width=7cm]{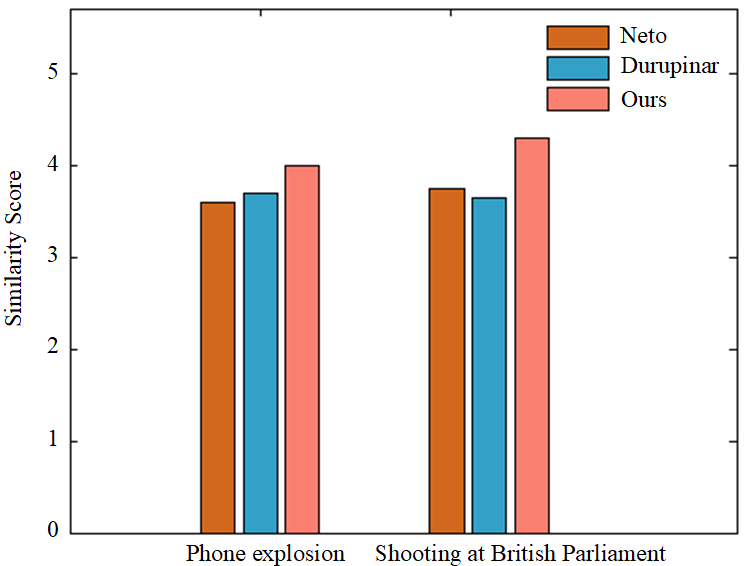}
  \centering
  \caption{ Comparison of similarity scores for movement states and the process of emotional contagion 
(higher values indicate greater similarity). A user study is performed and participants are asked to compare the movement states and processes of emotional contagion 
in the original videos with those in crowd simulation results achieved by different algorithms. }
  \label{fig:277}
  \end{centering} 
\end{figure}

We also take two real emergency incidents as examples to verify our proposed crowd simulation method. 
Our crowd simulation results of the scene after the mobile phone explosion on the subway in the Shanghai 
Metro Line 8 are presented in Figures \ref{fig:25}a and \ref{fig:25}b. Crowd simulation by our model of the shooting at the 
British Parliament building on March 22, 2017 is presented 
in Figures \ref{fig:25}c and \ref{fig:25}d. We show the spread of panic in both scenarios. The color of the 
cylinders represents the emotional intensity of the individuals. 
These two real-world videos have poor quality images. Even using manual methods, it is still difficult to track the positions 
of people in each frame accurately. Due to these objective limitations, we cannot measure the similarity between the simulated trajectories and the ground truth 
for these scenes in a direct way. Therefore, we use the dominant path and entropy metric to quantitively evaluate our simulation results. 
The dominant path is defined based on collectiveness of crowd movements and it can be treated as the movement trend of the crowd \cite{84,110}. 
We also calculate the spatial distance between the simulated trajectories and the ground truth for the scenarios.
From Table \ref{Entropy metric and spatial distance for two different scenarios} and Figure \ref{fig:277}, we can see that 
both the overall moving trend and the process of emotional contagion 
are similar to what is found in the recorded real-world crowd video clips.

\begin{table}[]
\scriptsize
\setlength{\belowcaptionskip}{10pt}
\renewcommand{\arraystretch}{1.3}
\caption{Entropy metric and spatial distance for different simulation algorithms on scenarios of Phone explosion and Shooting at British Parliament. 
 }
\label{Entropy metric and spatial distance for two different scenarios}
\centering
\begin{tabular}{|c|l|c|c|}
\hline
Scenario                                                                                   & \multicolumn{1}{c|}{Model} & Entropy metric & Spatial distance \\ \hline
\multirow{3}{*}{Phone explosion}                                                           & Ours                       & 1.565650       & 1.137298         \\ \cline{2-4} 
                                                                                           & Durupinar                  & 4.826474       & 1.523882         \\ \cline{2-4} 
                                                                                           & Neto                       & 2.857793       & 1.495380         \\ \hline
\multirow{3}{*}{\begin{tabular}[c]{@{}c@{}}Shooting at \\ British Parliament\end{tabular}} & Ours                       & 0.842232       & 0.749183         \\ \cline{2-4} 
                                                                                           & Durupinar                  & 2.246331       & 0.784255         \\ \cline{2-4} 
                                                                                           & Neto                       & 1.626734       & 0.771396         \\ \hline
\end{tabular}
\end{table}

\begin{figure}[htbp]
\centering
\begin{tabular}{cc}
 \includegraphics[width=4.1cm,height=2.5cm]{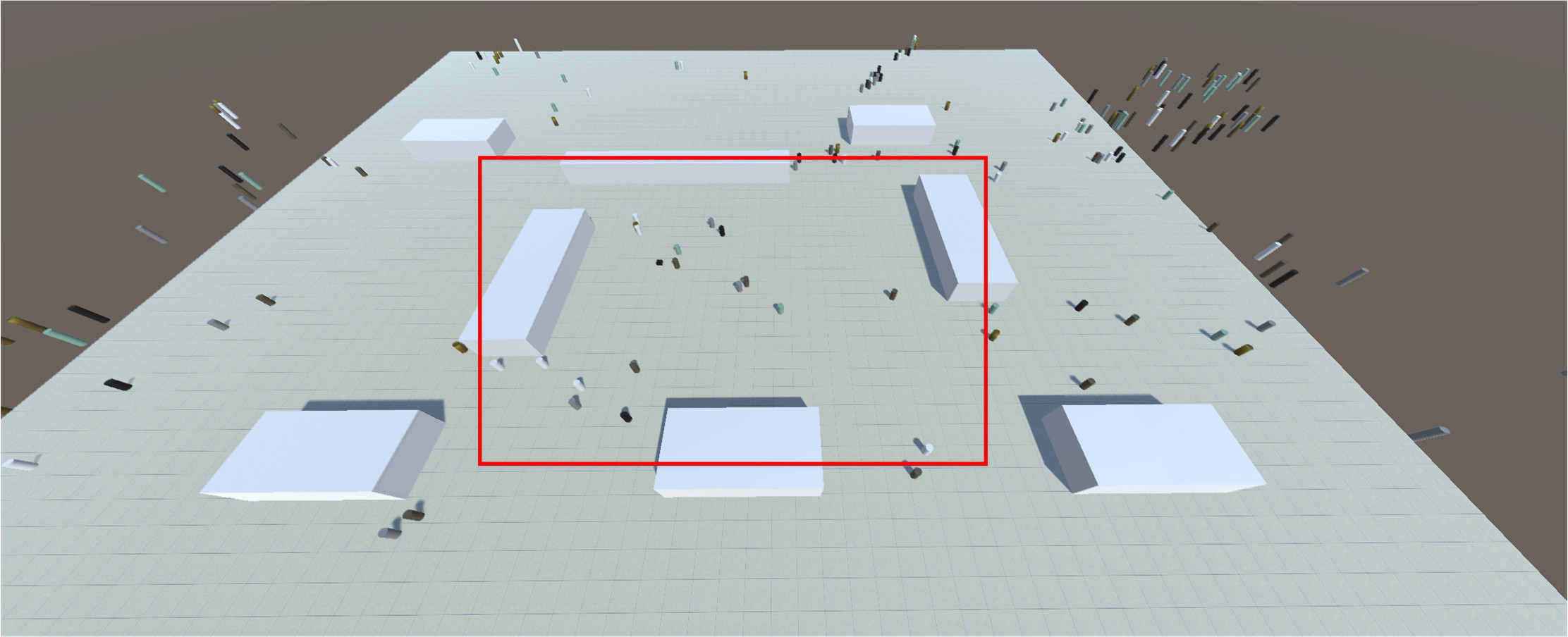} &  \includegraphics[width=4.1cm,height=2.5cm]{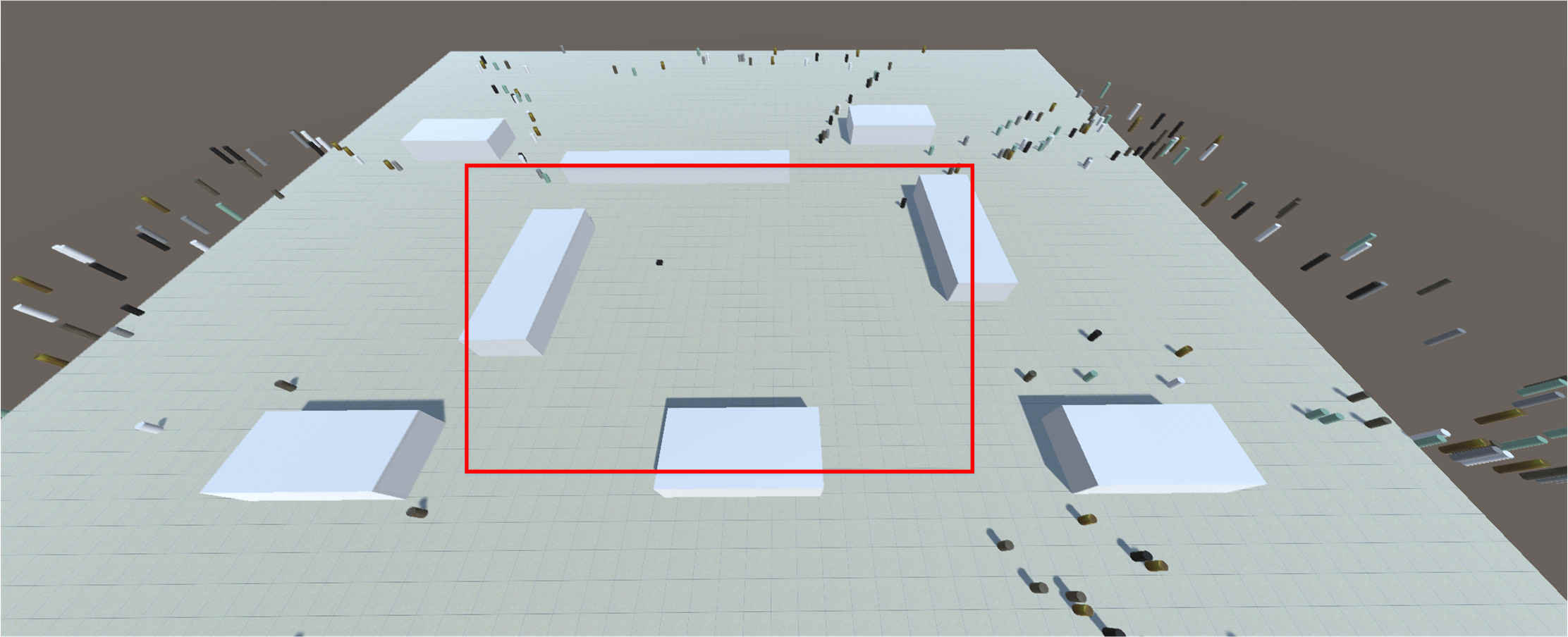} \\
 (a)  & (b)  \\
  \includegraphics[width=4.1cm,height=2.5cm]{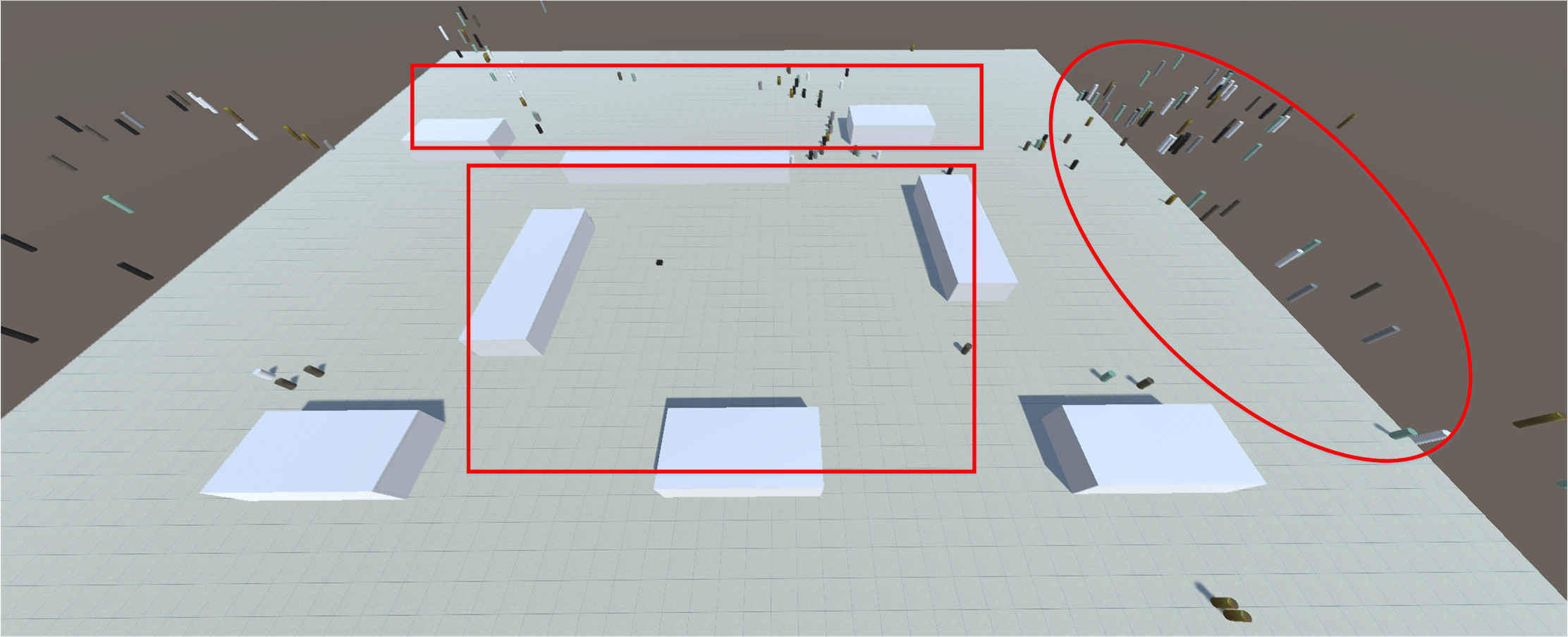} &  \includegraphics[width=4.1cm,height=2.5cm]{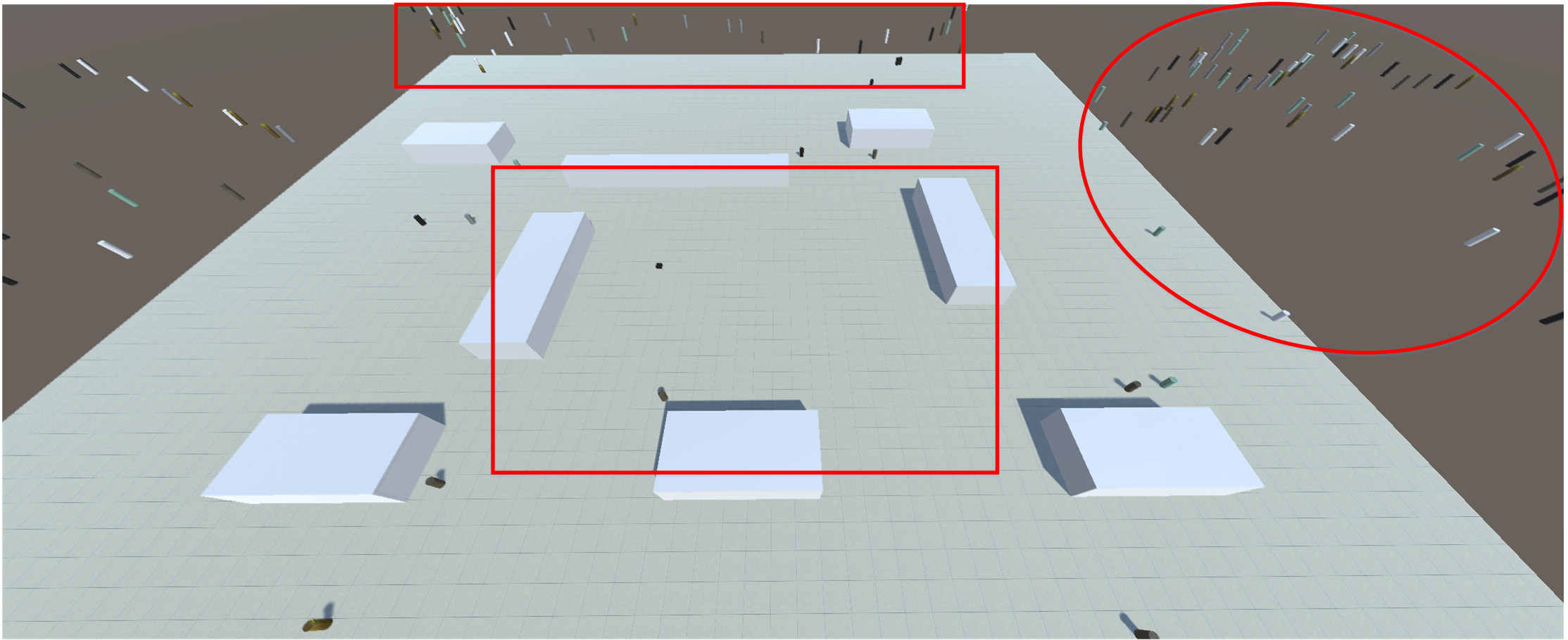} \\
 (c)  & (d)  \\
\end{tabular} 
\caption{
Crowd simulation results by different models in the virtual scenario at the $1000th$ frame: 
(a) Durupinar model, (b) Neto model, (c) Ours-Neto model, and (d) our model.
} 
\label{fig:1717} 
\end{figure} 

\subsubsection{Comparisons in virtual scenarios}

In the virtual scenario, we compare our simulation results with those 
of the Durupinar \cite{11}, Neto \cite{88}, and Ours-Neto models. 
The parameter values we used are described in Table \ref{2019318}.
In Figure \ref{fig:1717}a (the simulation result by the Durupinar model), 
the speeds of individuals are variable and their locations are scattered. Because of different thresholds and personality mechanisms, the Durupinar model can 
simulate heterogeneous crowd behaviors. However, there are too many 
individuals who are not affected by the panicked crowd and this result is unreasonable. 
In Figure \ref{fig:1717}b, individuals move much slower than the individuals in the simulation results by other models. 
The reason is that the emotion calculated by the Neto model is much smaller. 
Moreover, the individual movement is too regular, which is unsuitable for emergency situations. 
In Figure \ref{fig:1717}c (simulation result by our model with the same emotional contagion method as the Neto model) 
and Figure \ref{fig:1717}d (our simulation result), 
most of the individuals are affected by the hazard and run away from it.  
Because of physical strength consumption and personality factors, the speeds of individuals in our simulation result  
are more variable than those shown in the Ours-Neto model. Therefore, the simulation result by our model is more suitable for emergency situations than other models.

\subsection{Application of our model in various virtual scenarios} 
\label{app} 

\begin{figure}[htbp] \subfigure[]
{\centerline{\includegraphics[width=7cm]{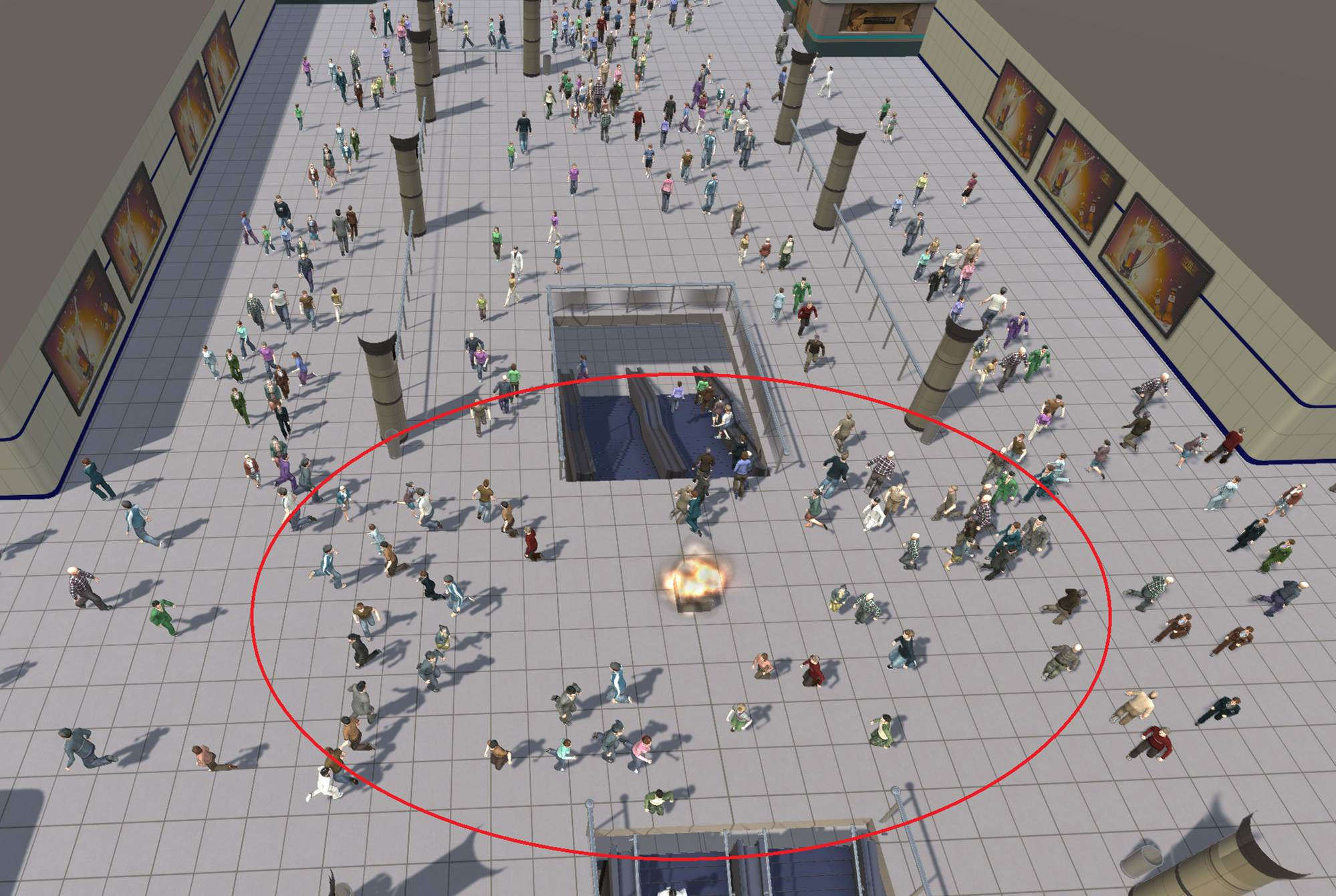}}} 
\subfigure[]
{\centerline{\includegraphics[width=7cm]{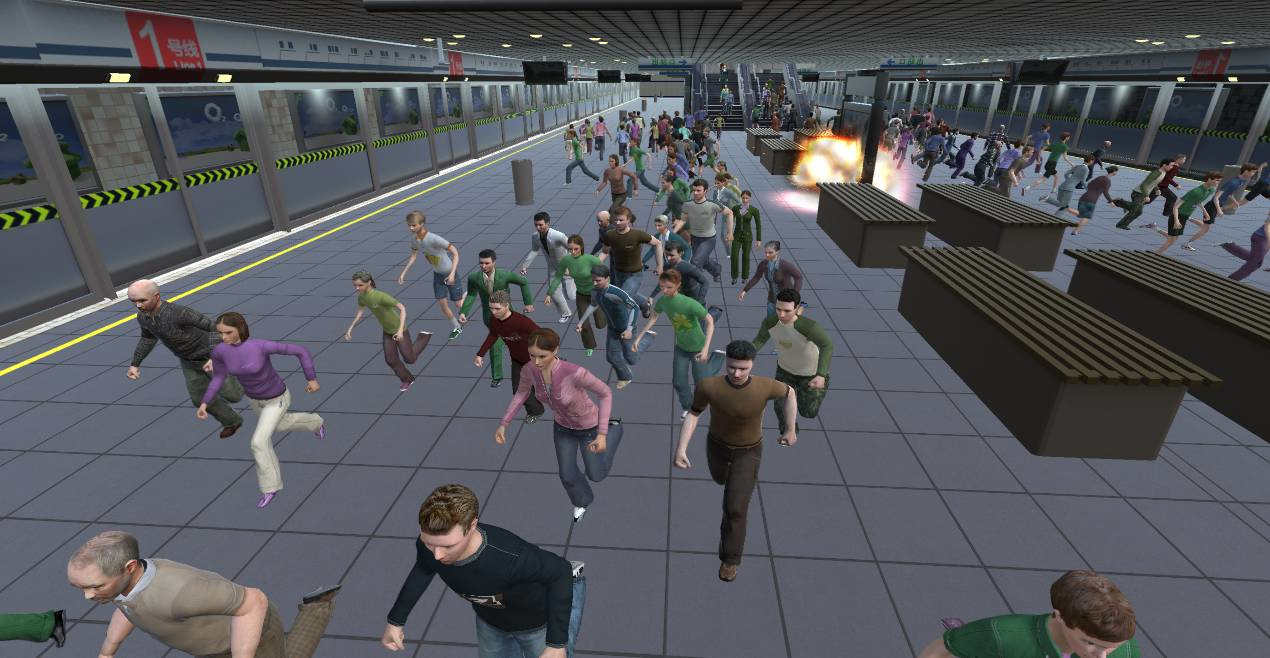}}} 
\caption{Crowd simulation results at a subway station: 
(a) the higher level of the subway station, (b) the lower level of the subway station. 
After the hazard occurs, the emotional contagion in our model begins to work. 
Although the direct impact of the hazard is limited, the hazardous area grows through emotional 
contagion among individuals and the number of individuals who run away from the hazard increases.
 } 
\label{fig:17} 
\end{figure}

\begin{figure}[htbp] \begin{centering} 
  \centering 
  \includegraphics[width=7cm]{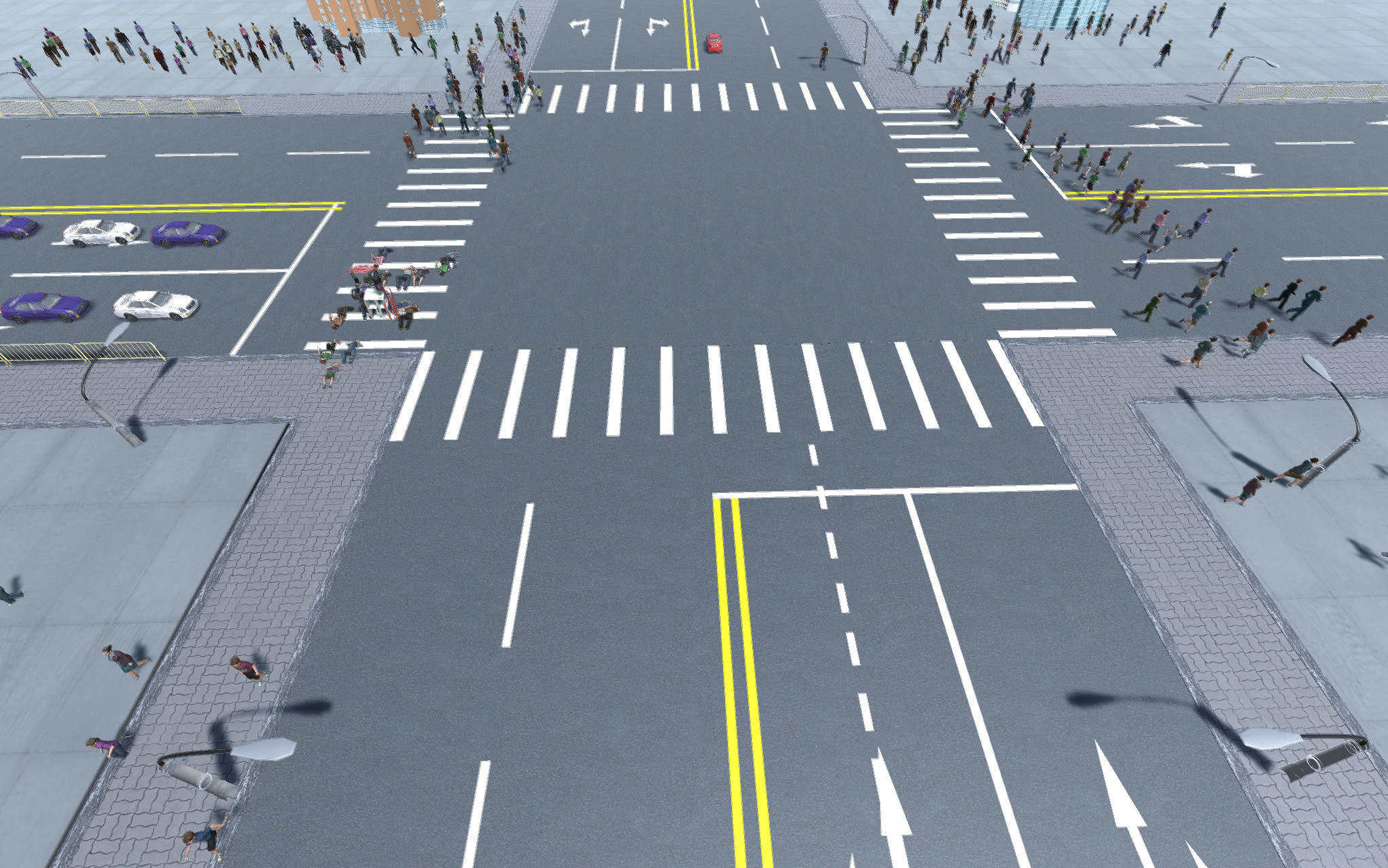}
  \centering
  \caption{Crowd simulation result at a crosswalk. 
At the lower left corner, a car explodes. Then individuals run away from the hazard.} 
  \label{fig:18}
  \end{centering} 
\end{figure}

Our model can be applied in different virtual scenarios. 
Subway stations and crosswalks are crowded and the probability of hazard occurrence in these scenarios is very high.
We simulate a hazard occurring in these scenarios and three examples are shown.
Figure \ref{fig:17}(a) shows crowd simulation at the higher level of the subway station. 
Figure \ref{fig:17}(b) shows crowd simulation at the lower level of the subway station. 
Figure \ref{fig:18} shows crowd simulation at a crosswalk.
We show each step of the process: hazard occurring, individuals running away from the hazard, emotional contagion spreading, and moving speed attenuating.
More details can be seen in the supplementary video. 
Our simulation results provide 
information about decision-making to deal with emergency situations.

\begin{figure}[htbp]
\centering
\begin{tabular}{cc}
 \includegraphics[width=3.8cm]{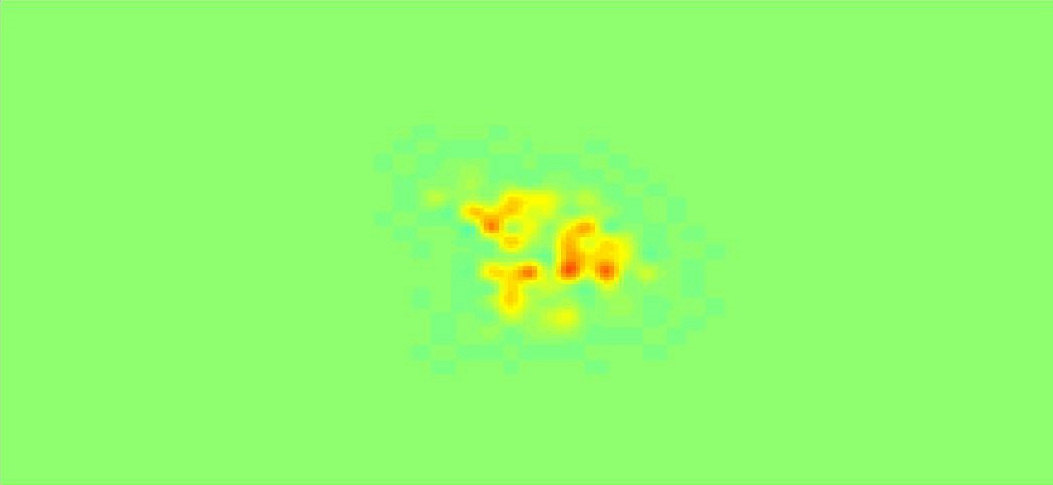} &  \includegraphics[width=3.8cm]{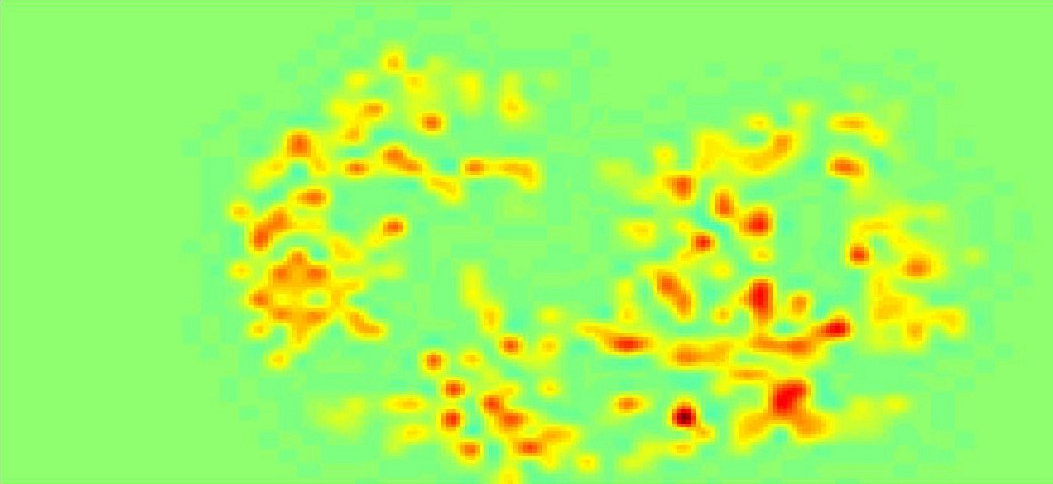} \\
 (a)  & (b)  \\
  \includegraphics[width=3.8cm]{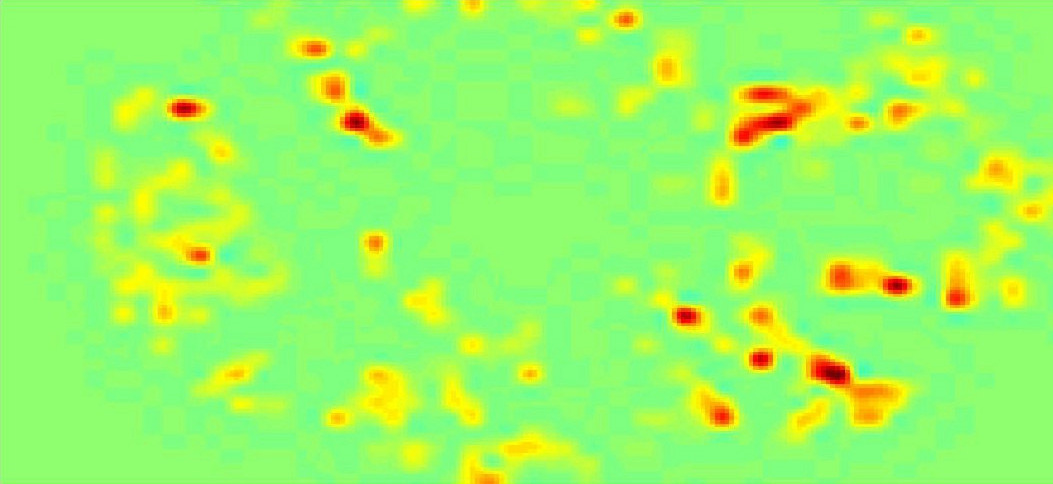} &  \includegraphics[width=3.8cm]{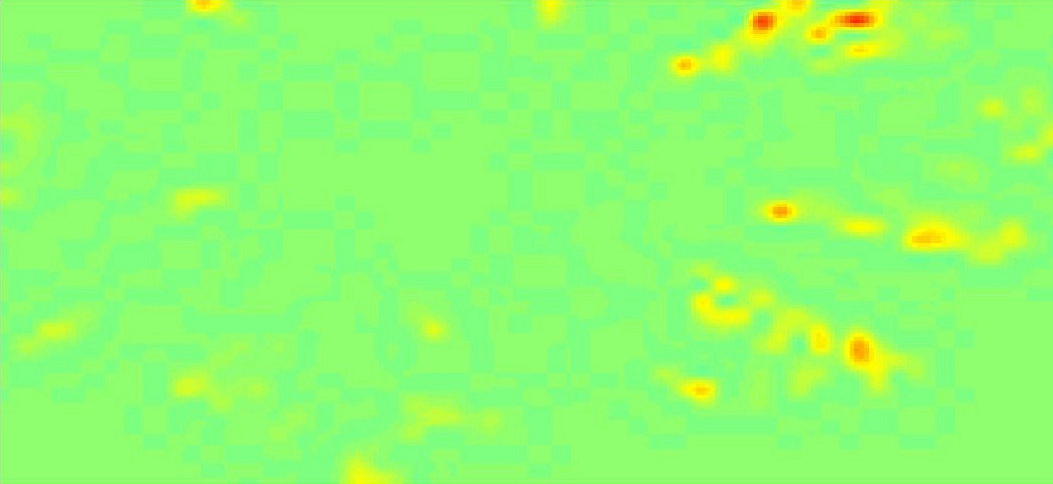} \\
 (c)  & (d)  \\
\end{tabular} 
\caption{Panic heat maps of the virtual scene: (a) heat map at the $13^{th}$ frame, (b) heat map at the $29^{th}$ frame, (c) heat map at the $57^{th}$ frame, (d) heat map at the $120^{th}$ frame. 
The red area is more panicked than the green area in the heat map. The deeper the color, the
more panicked the area. 
} 
\label{fig:13} 
\end{figure}

The heat maps of panic in the virtual scene are 
presented in Figure \ref{fig:13}. Although the direct impact of the hazard is limited, the panic area grows through the emotional contagion mechanism 
in our model. When individuals are far from the hazard, panic attenuates. 
As accidents may happen randomly in public places, we can take preventive action in advance and reduce loss by accurately predicting the panic area.

\begin{figure}[htbp]
\centering
\begin{tabular}{cc}
 \includegraphics[width=3.8cm]{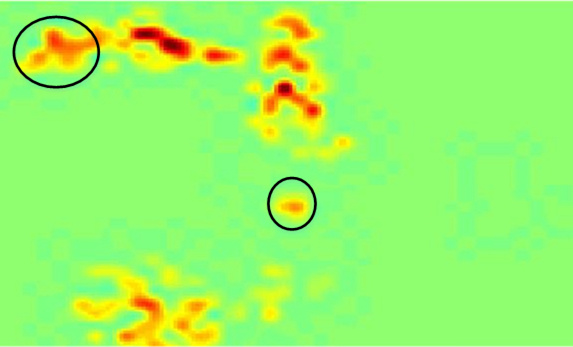} &  \includegraphics[width=3.8cm]{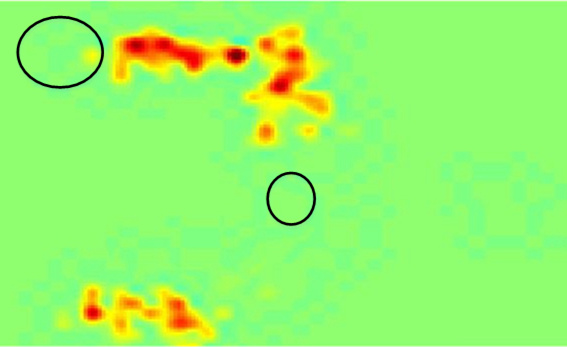} \\
 (a)  & (b)  \\
\end{tabular} 
\caption{\label{fig:19} The heat maps of panic at the $185^{th}$ frame of the crosswalk scenario: 
(a) heat map of the crowd simulation generated by our model and (b) heat map of the crowd simulation generated using the Durupinar model.
The red area is more panicked than the green area in the heat map. 
The deeper the color, the more panicked the area. 
We highlight the same area of the two simulation results. 
The individuals in our model simulation result are more panicked than 
the individuals in the Durupinar model simulation results.} 
\end{figure}

The panic heat maps generated from crowd simulations in the crosswalk scenario by our model and those generated using the Durupinar model are presented in Figure \ref{fig:19}. 
The individuals in the simulation results by our model are more panicked than 
those in the Durupinar model simulation results. 
The intensity of the panic calculated by the Durupinar 
model is lower than that calculated by our model. The reason is that our model considers not only emotional contagion 
among individuals, but also the impact of physical strength consumption on 
panic levels. Our model represents a comprehensive description of individual 
panic levels and is more conducive to the spread of panic than the Durupinar model. Therefore, the simulation results by 
our model are more reasonable for emergency situations.

\section{Conclusion and Limitations} 
\label{conclusion}

In contrast to traditional emotion-based crowd simulation models, 
we integrate physical strength consumption into our model. 
We not only present a panic level calculation, but also delineate 
the effect of physical strength consumption on panic. 
Finally, both physical strength consumption and panic determine the movement of each individual. 
Our proposed model is verified by simulations, and it is compared with real-world videos and previous approaches. 
Results have shown that our proposed model can reliably 
generate realistic group behaviors. 
It can also predict the changes of physical strength consumption and panic of a crowd in an
emergency situation.

However, our model has several limitations. Although our model can generate realistic crowd movements, 
the panic levels and physical 
strength consumption of the crowd in an emergency scene cannot be obtained directly. Our model can only infer them during the simulation. 
Thus, the initial state of our model is difficult to determine and it is usually time consuming to do so. 
In the future, we plan to use the latest wearable equipment to collect these data 
and provide a new method that can quickly and accurately determine the initial state.
Furthermore, at present, our model mainly focuses on emergency scenarios. In future work, we want to extend our model to a variety of general situations. 
We also plan to develop a coupled framework for crowd and traffic flow to generate realistic mixed traffic simulations 
and model the influence of these two flows on each other.


%


\ifCLASSOPTIONcaptionsoff
  \newpage
\fi

\bibliographystyle{IEEEtran}
\bibliography{bare_jrnl_compsoc}

\vspace{-15 mm}
\begin{IEEEbiography}[{\includegraphics[width=1in,height=1.25in,clip,keepaspectratio]{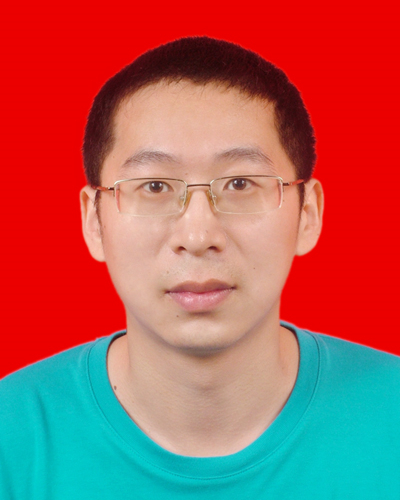}}]{Mingliang Xu}
received the Ph.D. degree in computer science and technology from the State Key Laboratory of CAD\&CG, Zhejiang University, Hangzhou, China.

He is a Full Professor with the School of Information Engineering, Zhengzhou University, Zhengzhou, China, where he is currently the Director of the Center for Interdisciplinary Information Science Research. He was with the Department of Information Science, National Natural Science Foundation of China, Beijing,
China, from 2015 to 2016. His current research interests include computer
graphics, multimedia, and artificial intelligence. He has authored over 60
journal and conference papers in the above areas, including the ACM
Transactions on Graphics, ACM Transactions on Intelligent Systems and
Technology, IEEE TRANSACTIONS ON PATTERN ANALYSIS AND MACHINE
INTELLIGENCE, IEEE TRANSACTIONS ON IMAGE PROCESSING, IEEE
TRANSACTIONS ON CYBERNETICS, IEEE TRANSACTIONS ON CIRCUITS
AND SYSTEMS FOR VIDEO TECHNOLOGY, ACM SIGGRAPH (Asia), ACM
MM, and ICCV.

Dr. Xu is the Vice General Secretary of ACM SIGAI China.
\end{IEEEbiography}
\vspace{-10 mm}
\begin{IEEEbiography}[{\includegraphics[width=1in,height=1.25in,clip,keepaspectratio]{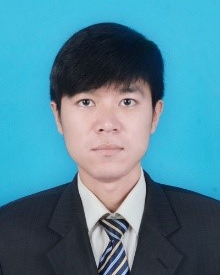}}]{Chaochao Li}
received the B.S. degree in computer
science and technology and the master's degree in
computer application technology from the School
of Information Engineering, Zhengzhou University,
Zhengzhou, China, where he is currently pursuing
the Ph.D. degree.
His current research interests include computer
graphics and computer vision.
\end{IEEEbiography}
\begin{IEEEbiography}[{\includegraphics[width=1in,height=1.25in,clip,keepaspectratio]{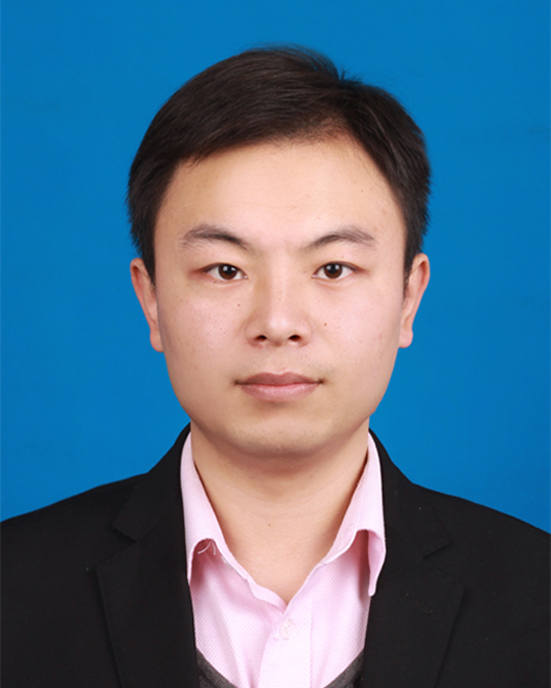}}]{Pei Lv}
received the Ph.D. degree from the State
Key Laboratory of CAD\&CG, Zhejiang University,
Hangzhou, China, in 2013.

He is currently an Associate Professor with
the School of Information Engineering, Zhengzhou
University, Zhengzhou, China. His current research
interests include video analysis and crowd simulation. He has authored over 20 journal and conference
papers in the above areas, including the IEEE
TRANSACTIONS ON IMAGE PROCESSING, IEEE
TRANSACTIONS ON CIRCUITS AND SYSTEMS FOR
VIDEO TECHNOLOGY, and ACM Multimedia.
\end{IEEEbiography}
\begin{IEEEbiography}[{\includegraphics[width=1in,height=1.25in,clip,keepaspectratio]{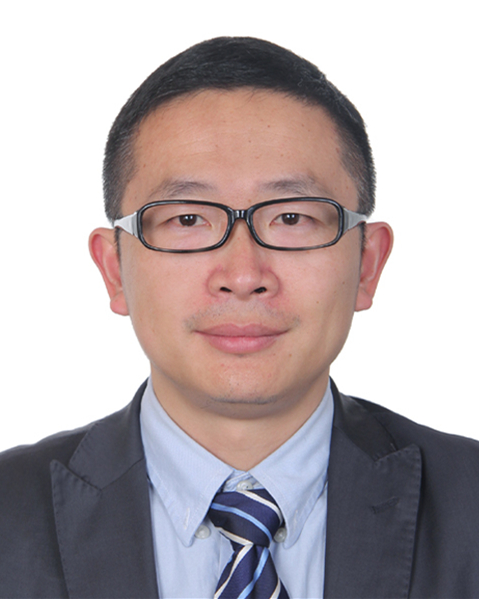}}]{Wei Chen}
is a professor at the State Key Lab of CAD\&CG, Zhejiang University. His research interests are visualization and visual analysis, and he has published more than 30 IEEE/ACM Transactions and IEEE VIS papers. He actively served as guest or associate editors of IEEE Transactions on Visualization and Computer Graphics, IEEE Transactions on Intelligent Transportation Systems, and Journal of Visualization. For more information, please refer to http://www.cad.zju.edu.cn/home/chenwei/.
\end{IEEEbiography}
\begin{IEEEbiography}[{\includegraphics[width=1in,height=1.25in,clip]{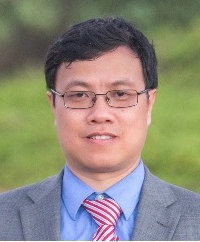}}]{Zhigang Deng}
is a Full Professor of Computer Science at University of Houston (UH). He is also the Director of Graduate Studies at UH's Computer Science Department and the Founding Director of the UH Computer Graphics and Interactive Media (CGIM) Lab. He earned a Ph.D. in Computer Science at the Integrated Media System Center (NSF ERC) and the Department of Computer Science at the University of Southern California in 2006. He completed a B.S. degree in Mathematics from Xiamen
University (China), and an M.S. in Computer Science from Peking University (China).His interests are in the broad areas of computer graphics, computer animation, human computer interaction, virtual human modeling and animation, and visual computing for biomedical/healthcare informatics. He is the
recipient of ACM ICMI Ten Year Technical Impact Award, UH Teaching Excellence Award, Google Faculty Research Award, UHCS Faculty Academic Excellence Award, and NSFC Overseas and Hong Kong/Macau Young Scholars Collaborative Research Award. Besides being the CASA 2014 conference general co-chair and the SCA 2015 conference general co-chair, he currently serves as an Associate Editor of several journals including Computer Graphics Forum and Computer Animation and Virtual Worlds Journal.
\end{IEEEbiography}
\begin{IEEEbiography}[{\includegraphics[width=1in,height=1.25in,clip]{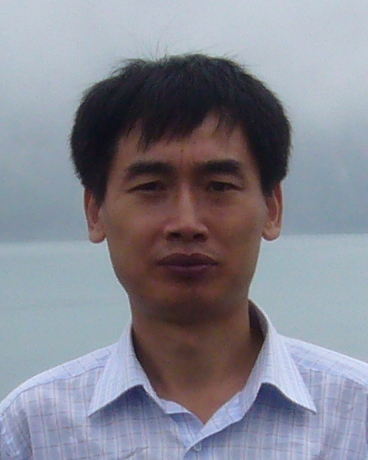}}]{Bing Zhou}
received B.S. and M.S. degrees from
Xi'an Jiaotong University, Xi'an, China, in 1986
and 1989, respectively, and a Ph.D. from
Beihang University, Beijing, China, in 2003, all in
computer science.

He is currently a Professor with the School
of Information Engineering, Zhengzhou University,
Zhengzhou, China. His current research interests
include video processing and understanding, surveillance, computer vision, and multimedia applications.
\end{IEEEbiography}
\begin{IEEEbiography}[{\includegraphics[width=1in,height=1.25in,clip,keepaspectratio]{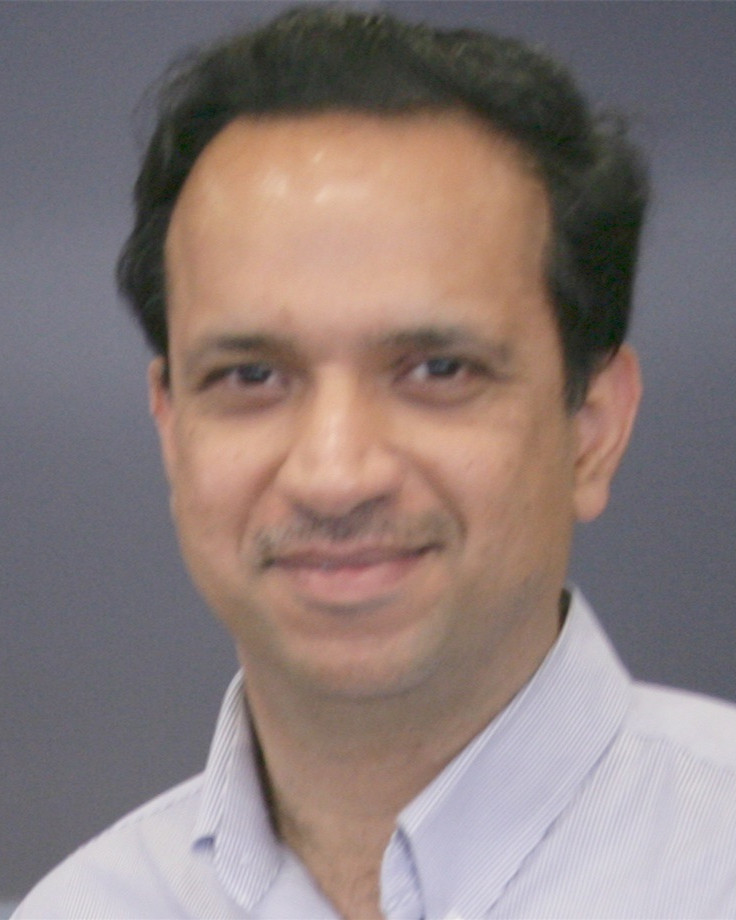}}]{Dinesh Manocha}
is the Paul Chrisman Iribe Chair in Computer Science \& Electrical and Computer Engineering at the University of Maryland College Park. He is also the Phi Delta Theta/Matthew Mason Distinguished Professor Emeritus of Computer Science at the University of North Carolina - Chapel Hill. He has won many awards, including Alfred P. Sloan Research Fellow, the NSF Career Award, the ONR Young Investigator Award, and the Hettleman Prize for scholarly achievement. His research interests include multi-agent simulation, virtual environments, physically-based modeling, and robotics. His group has developed a number of packages for multi-agent simulation, crowd simulation, and physics-based simulation that have been used by hundreds of thousands of users and licensed to more than 60 commercial vendors. He has published more than 500 papers and supervised more than 35 PhD dissertations. He is an inventor of 9 patents, several of which have been licensed to industry. His work has been covered by the New York Times, NPR, Boston Globe, Washington Post, ZDNet, as well as DARPA Legacy Press Release. He is a Fellow of AAAI, AAAS, ACM, and IEEE and also received the Distinguished Alumni Award from IIT Delhi. He was a co-founder of Impulsonic, a developer of physics-based audio simulation technologies, which was acquired by Valve Inc. See \url{http://www.cs.umd.edu/~dm}
\end{IEEEbiography}







\end{document}